\documentclass[10pt,a4paper]{article}
\usepackage{amssymb,amsmath,graphicx,enumerate}
\usepackage[paperheight = 11 in, paperwidth = 8.5 in,
top    = 0.75 in,
bottom = 0.75 in,
left   = 0.75 in,
right  = 0.75 in]{geometry}
\usepackage{float}
\usepackage{hyperref}
\usepackage{booktabs}
\usepackage{arydshln}
\usepackage{changepage}
\usepackage{wrapfig}
\usepackage[linewidth=1pt]{mdframed}
\usepackage[font=footnotesize,labelfont=bf,
justification=justified,
format=plain]{caption} 
\usepackage{multicol}
\usepackage{multirow}
\usepackage[dvipsnames]{xcolor}
\colorlet{cyan}{green!60!blue}
\colorlet{actual_purple}{red!50!blue}
\usepackage{booktabs}
\usepackage{arydshln}
\usepackage{changepage}
\usepackage{pdflscape}
\usepackage[round, numbers]{natbib}
\usepackage{cancel}
\usepackage[normalem]{ulem}

\usepackage{enumitem}
\usepackage{titlesec}
\titlespacing\section{0pt}{8pt plus 4pt minus 2pt}{8pt plus 2pt minus 2pt}
\titlespacing\subsection{0pt}{8pt plus 4pt minus 2pt}{8pt plus 2pt minus 2pt}
\titlespacing\subsubsection{0pt}{8pt plus 4pt minus 2pt}{8pt plus 2pt minus 2pt}
\newcommand{\supplementarysection}{%
  \setcounter{figure}{0}
  \let\oldthefigure\thefigure
  \renewcommand{\thefigure}{S\oldthefigure}
}

\usepackage{authblk}
\title{\large Mechanisms of thrombin inhibition by protein S and the TFPI$\alpha$-fVshort-protein S complex\vskip-1.5ex}
\author[1]{\small Alexander G. Ginsberg}
\author[2]{\small Josefin Ahnstr\"{o}m}
\author[2]{\small James T.B. Crawley}

\author[3, 4, 5]{\small Karin Leiderman}
\author[3]{\small Dougald M. Monroe}
\author[6]{\small Keith B. Neeves}
\author[7]{\small Suzanne F. Sindi}
\author[1, 8]{\small Aaron L. Fogelson}
\affil[1]{\footnotesize Department of Mathematics, University of Utah, Salt Lake City, UT, USA}
\affil[2]{\footnotesize Centre for Haematology, Imperial College London, London, UK}
\affil[3]{\footnotesize UNC Blood Research Center, University of North Carolina at Chapel Hill, Chapel Hill, North Carolina, USA}
\affil[4]{\footnotesize Computational Medicine Program, University of North Carolina at Chapel Hill, Chapel Hill, North Carolina, USA}
\affil[5]{\footnotesize Department of Biochemistry and Biophysics, University of North Carolina at Chapel Hill, Chapel Hill, North Carolina, USA}
\affil[6]{\footnotesize Department of Bioengineering, University of Colorado Denver, Anschutz Campus, Aurora, Colorado, USA}
\affil[7]{\footnotesize Department of Applied Mathematics, University of California Merced, Merced, California, USA}
\affil[8]{\footnotesize Department of Biomedical Engineering, University of Utah, Salt Lake City, UT, USA }
\date{}

\begin{document}

\maketitle

\begin{abstract}
Protein S (PS) is an important anticoagulant implicated in both bleeding and thrombotic disorders, making it a promising drug target. The anticoagulant function of PS arises in part because PS enhances the anticoagulant function of tissue factor pathway inhibitor alpha (TFPI$\alpha$). PS has been proposed to circulate in the bloodstream together with TFPI$\alpha$ and a truncated form of factor V (fV-short) in the trimolecular complex, TFPI$\alpha$-PS-fV-short, which we call protein S complex (PSC). PSC has been proposed to strongly inhibit thrombin production by enhancing the ability of TFPI$\alpha$ to inhibit clotting factor Xa up to 100-fold and by localizing to platelet membranes, limiting fXa activity shortly after coagulation starts.

Yet, exactly how PS functions with TFPI$\alpha$ as an anticoagulant remains poorly understood. To investigate, we extend an experimentally validated mathematical model of blood coagulation to include PSC and free PS (not a part of PSC) in the plasma, as well as free PS and TFPI$\alpha$ in platelets. We find that shortly after coagulation initiation, PSC strongly inhibits thrombin production. We find that the (unknown) magnitude of the enhanced affinity of PSC binding to inhibit fXa critically regulates PSC's impact on thrombin production. We find that under flow, PSC can unexpectedly accumulate on platelets to concentrations $\sim$ 50 times higher than in the plasma. We also find that PSC limits thrombin production by occupying fV-specific binding sites on platelets.

Our results show that changes in PSC can dramatically impact severity of pathological bleeding disorders. For the east Texas bleeding disorder, elevated PSC concentrations eliminate thrombin bursts, leading to bleeding. With fV deficiency, reducing PSC rescues thrombin production in severe fV deficiency and returns thrombin production due to mild fV deficiency to normal. Finally, thrombin production in severe hemophilia A can be substantially improved by blocking PSC's anticoagulant function.

\begin{center}
    \textbf{Significance statement}
\end{center}

Blood coagulation is critical for an effective clotting response. Protein S (PS) is an important anticoagulant implicated in pathological bleeding and thrombosis. PS binds with TFPI$\alpha$ and factor V short to form the protein S complex (PSC) which enhances anticoagulant function by incompletely understood means. To investigate, we extend an experimentally validated mathematical model of blood coagulation and platelet deposition under flow by adding new proteins, including PSC, and their reactions. We find that PSC accumulates on platelets far more than expected, and that determining the affinity for PSC binding with fXa is necessary to understand PSC's anticoagulant properties.   We show that deficiencies in PSC can rescue thrombin production in several bleeding disorders, including severe hemophilia A.

\end{abstract}

\section{Introduction}

The body halts bleeding by forming blood clots. First, platelets aggregate at the site of an injury \cite{guyton2011guyton_and_hall, brass2019hemostatic, scridon2022platelets_in_hemostasis, jackson2007platelet_aggregation}. However, the platelet aggregate is unstable and can detach from the injury site \cite{jackson2007platelet_aggregation} or leak \cite{guyton2011guyton_and_hall}. In the second step of clotting known as coagulation, a polymer mesh of fibrin forms and stabilizes the platelet aggregate \cite{guyton2011guyton_and_hall, brass2019hemostatic, marder2012hemostasis_book}.
    
Protein S (PS) is an important inhibitor of blood coagulation. It can help prevent pathological blood clots \cite{gierula2020anticoagulant} that constrict blood flow or break off into the bloodstream as emboli that cause strokes and heart attacks \cite{koupenova2017thrombosis, lichota2020thrombosis}, but it can also contribute to excess bleeding \cite{dahlback2023natural}. Thus, PS is a promising target for drugs that could treat excess clotting or uncontrolled bleeding \cite{eladnani2025protein_s_therapeutic_target_review, prince2018targeting, eladnani2025protein_s_drug_target_hem_a, wilson2024protein_s_antibody}.

The need to understand how PS carries out its inhibitory functions is therefore compelling. While PS has long been known to have anticoagulant functions \cite{di1977protein_s_discovery, d2008protein_s_discovery}, it was much more recently discovered to enhance the anticoagulant properties of tissue factor pathway inhibitor alpha (TFPI$\alpha$) \cite{hackeng2006protein}. Our understanding of how PS and TFPI$\alpha$ function in concert as an anticoagulant remains incomplete.

TFPI$\alpha$ regulates activated factor X (fXa) \cite{ahnstrom2024tissue}, a protein critical for coagulation \cite{guyton2011guyton_and_hall, marder2012hemostasis_book, gierula2020anticoagulant}. When an injury to a blood vessel exposes tissue factor (TF) in the subendothelium of the vessel wall, the TF pathway of coagulation is initiated. TF binds to the plasma protein factor VIIa to form the subendothelium-bound enzyme complex TF:fVIIa, which, in turn, triggers a network of clotting factor activation reactions by activating factor X (fX) to fXa.

FXa is essential for further coagulation reactions. Along with activated clotting factor V (fVa), fXa forms the fVa:fXa enzyme complex, known as prothrombinase (PRO), on the surfaces of activated platelets. PRO in turn cleaves prothrombin into thrombin, which cleaves the abundant plasma protein fibrinogen into fibrin monomers which then polymerize to form the polymer mesh that stabilizes the clot \cite{guyton2011guyton_and_hall, marder2012hemostasis_book, gierula2020anticoagulant}. Without fXa, the reactions leading to fibrin mesh formation do not occur.
    
Once initiated, the TF pathway can produce large bursts of thrombin \cite{guyton2011guyton_and_hall, marder2012hemostasis_book} in which thrombin concentration grows from sub-pM to as much as 100-200 nM in a matter of minutes. However, TFPI$\alpha$ can significantly delay thrombin bursts or prevent them altogether by reducing the concentration of fXa available to form PRO \cite{reglinska2014shbg}. TFPI$\alpha$ can further delay thrombin bursts by reducing the amount of the TF:VIIa complex available to produce fXa \cite{dahlback2023natural}. TFPI$\alpha$ can also delay thrombin bursts by reducing the concentration of partially-activated forms of fV (which we refer to as fVh) available to produce an altered form of platelet-bound prothrombinase (PROh, a complex of fVh and fXa) \cite{mast2022tfpi} that may be especially important during early coagulation.

\begin{figure}[h!]

    \includegraphics[width =\textwidth]{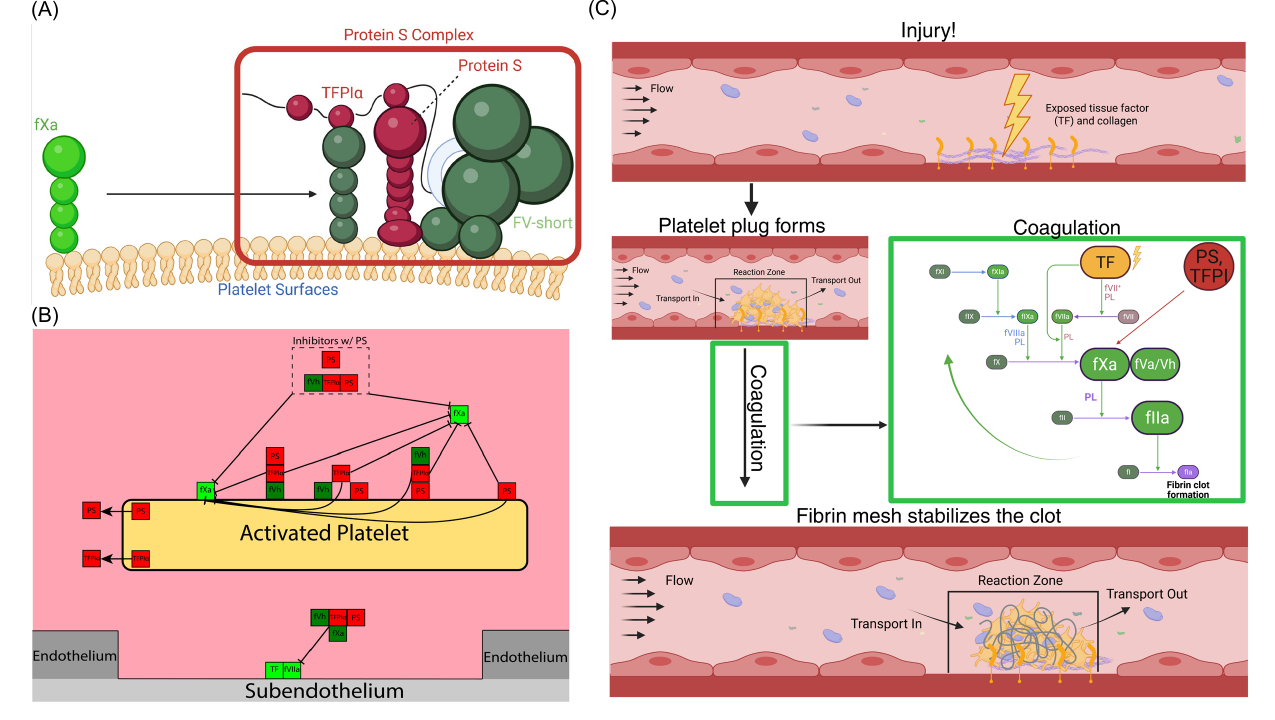}

    \caption{\footnotesize
		 (A): PSC, consisting of TFPI$\alpha$ bound via its Kunitz 3 (K3) domain to the sex-hormone-binding globulin (SHBG) of PS, with the negatively-charged C-terminus tail of TFPI$\alpha$ bound to the positively charged truncated B domain of fVh, binds and inhibits fXa via the Kunitz 2 (K2) domain of TFPI$\alpha$. The figure shows this happening on a platelet surface. (B): Two main species involving PS are hypothesized to have an anticoagulant role: free PS and PSC (PS:TFPI$\alpha$:fVh.) Free PS can bind to platelet membranes and to activated factor Xa (fXa) regardless of whether fXa is membrane-bound or not. PSC can likewise bind to platelets, either via its PS or its fVh (but not its TFPI$\alpha$); and PSC can bind to fXa or fXam via its TFPI$\alpha$. Fluid-phase PSC bound to fXa can also bind TF:fVIIa on the subendothelium exposed by injury. Red = anticoagulant; light green = procoagulant; dark green = procoagulant but inhibited. (C): An injury to a blood vessel wall occurs, exposing collagen (shown as purple strands) and tissue factor (TF, shown as thin, yellow proteins). As platelets from upstream (blue blobs) contact the exposed collagen in the reaction zone, they activate (spiky yellow spheres), attracting more platelets to the injury site, and thereby forming a platelet plug. At the same time, clotting factor fVIIa contacts TF, initiating a cascade of activation of procoagulant clotting factors (green), eventually cleaving fibrinogen into fibrin, which in turn forms a stabilizing mesh over the platelet plug. Anticoagulants (red) such as protein S (PS) and tissue factor pathway inhibitor (TFPI) regulate the coagulation reactions. Panels (A) and (C) created in BioRender.\normalsize
         }
         \label{figure:intro}

    \end{figure}

TFPI$\alpha$ reduces the availability of the three previously discussed species--fXa, the TF:fVIIa complex, and fVh--by binding to them and thus preventing their binding to other clotting factors \cite{mast2022tfpi}. TFPI$\alpha$ consists of 3 Kunitz domains and a C-terminal tail (Fig. \ref{figure:intro}). Kunitz 1 is important for binding and inhibiting TF:fVIIa \cite{girard1989kunitz_2}, Kunitz 2 binds and inhibits fXa \cite{girard1989kunitz_2}, Kunitz 3 binds to PS \cite{ahnstrom2012kunitz_3, ndonwi2010kunitz}, and the C-terminal tail is critical for binding fVh \cite{gierula2023c_terminal}.

PS contains an N-terminal Gla domain by which it binds strongly to negatively-charged phospholipid membranes.  Consequently, it has been proposed that PS helps to localize TFPI$\alpha$ to activated platelet surfaces \cite{gierula2020anticoagulant, gierula2023c_terminal}, facilitating TFPI$\alpha$'s binding to the relatively high concentrations of fXa found there (also see \cite{hackeng2006protein, wood2014protein_s_cofactor, reglinska2014shbg}). 

FV consists of 6 domains A1-A2-B-A3-C1-C2 \cite{dahlback2017novel} and binds to activated platelets by its C1 and C2 domains \cite{macedo1999crystal_fv_c2}.  Upon activation of fV to fVa by thrombin, the B domain is completely excised, allowing fVa to bind to fXa to form prothrombinase.  The B domain contains a basic region and an acidic region \cite{dahlback2017novel}.  In the presence of phospholipids, fXa cleaves the B-domain so as to remove the basic region but leave in place the acidic region.  TFPI$\alpha$ binds strongly to this partially activated form of fV, and thus can be localized to activated platelet surfaces together with partially activated fV. Likewise in fV-short, a fV variant present at low concentrations in plasma ($\sim 0.2$ nM), the basic region is missing but the acidic region remains \cite{vincent2013east_texas}. Further, platelets contain stores of partially-proteolyzed forms of fV which retain the acidic region \cite{wood2013tfpi} and which are released upon platelet activation. All three of these modified forms of fV can bind the C-terminal tail of TFPI$\alpha$ (see \cite{wood2013tfpi}) because they contain the acidic but not basic region of the B-domain. All three of these modified forms of fV have at least partial fXa cofactor activity \cite{dahlback2017novel} because of their structure. Further, there is no evidence that suggests that these three forms of fV would have markedly different fXa cofactor activity from one another or from fVa (see \cite{petrillo2021tfpi_five_short}). We therefore collectively refer to these partially activated forms of fV which are missing the basic region but retain the acidic region of their B domain\cite{dahlback2022hydrophobic, dahlback2023natural, gierula2023c_terminal} as ``fVh'' and do not distinguish between them for the remainder of this paper.

Not only can TFPI$\alpha$ bind PS or fVh, it can bind them simultaneously. In fact, fVh, TFPI$\alpha$, and PS are proposed to circulate at a concentration of 0.2 - 0.5 nM in a trimolecular complex that we call the protein S complex (PSC)  \cite{vincent2013east_texas, dahlback2023natural, ahnstrom2024tissue}.  Because of the relatively strong binding between fVh and TFPI$\alpha$ and the comparatively high concentrations of free PS in the plasma ($\sim$ 150 nM for PS versus $\sim$ 0.2 nM for fVh and TFPI$\alpha$), it has been hypothesized that essentially all fVh and TFPI$\alpha$ in the plasma circulate as constituents of PSC \cite{dahlback2022hydrophobic}. 

PSC has several potential advantages over TFPI$\alpha$ in inhibiting thrombin production. While TFPI$\alpha$ alone cannot bind to platelet surfaces, PSC can bind to them by its PS and/or its fVh (see \cite{gierula2023c_terminal}). Further, PS and fVh may enhance TFPI$\alpha$'s ability to inhibit fXa up to 100-fold \cite{dahlback2018factor, gierula2020anticoagulant}, although the dissociation constant ($\text{K}_{\text{D}}$) describing the binding of TFPI$\alpha$ in PSC with fXa is unknown \cite{dahlback2022hydrophobic}. Due to the low concentration of PSC in the plasma but strong affinity for fXa, it is hypothesized that PSC limits fXa production and activity during times when coagulation activity is low \cite{dahlback2022hydrophobic}, for example early during coagulation (before a thrombin burst has occurred) or following minor injuries that expose little TF on the subendothelium.

In the past few years, it has been hypothesized that fVh regulates the concentration of PSC in the plasma \cite{ahnstrom2024tissue, dahlback2023natural}  by extracting TFPI$\alpha$ from the endothelium \cite{dahlback2017novel}. Then, since fVh and TFPI$\alpha$ occur in the plasma at much lower concentrations than PS, they are the limiting factors in determining PSC concentration. In forming a large trimolecular complex, it is hypothesized that TFPI$\alpha$ is protected from renal filtration. Therefore, if fVh production were elevated, for example, the hypothesis would predict that the PSC concentrations in the plasma would also be elevated. So, PSC may play a critical role under pathological fV conditions (see \cite{dahlback2023natural, ahnstrom2024tissue}).

Evidence for this hypothesis comes from an east Texas family (see e.g., \cite{dahlback2023natural, ahnstrom2024tissue}) in which 22 individuals showed a similar moderately severe bleeding disorder \cite{kuang2001east_texas_discovery}. FV-short concentrations in affected family members were $\sim$ 10 times the standard fV-short concentration: $1.7$-$8.4$ nM in affected individuals \cite{vincent2013east_texas} vs $\sim 0.2$ nM in normal plasma \cite{dahlback2023natural}. Affected family members concurrently showed elevated levels of TFPI$\alpha$ in complex with fV-short.  Because TFPI$\alpha$ and fV-short are believed to circulate as constituents of PSC, it is likely that affected family members had similarly ($\sim$ 10-fold) elevated plasma concentrations of PSC \cite{dahlback2023natural, ahnstrom2024tissue}. 
 
More  evidence supporting this hypothesis lies in fV deficiencies which lead to unexpectedly mild bleeding disorders \cite{duckers2008low_plasma_factor_v, dahlback2017novel} considering fV's critical role in coagulation. In particular, severe fV deficiencies with plasma concentrations 1\% of the normal concentration (20 nM) lead to mild bleeding, while mild fV deficiencies (10\% normal concentration) do not lead to excess bleeding. FV deficiencies also correlate with less TFPI$\alpha$ \cite{duckers2008low_plasma_factor_v}, suggesting that the loss of fV's procoagulant activity may be balanced by the loss of TFPI$\alpha$'s anticoagulant functions \cite{dahlback2017novel} due to a reduction in PSC \cite{ahnstrom2024tissue}.

To investigate the role of PS and in particular of PSC, we extend an existing, experimentally validated, mathematical model of blood coagulation (see \cite{kuharsky2001surface, fogelson2006coagulation, fogelson2012blood, link2020mathematical, miyazawa2023i, miyazawa2023ii, link2018local, stobb2024mathematical}) to include PSC and free PS (PS not included in PSC or bound to C4 binding protein) in the plasma, as well as free PS and TFPI$\alpha$ released locally by platelets at sites of clot formation. In section \ref{section:methods} (Methods), we review our previous mathematical models of the coagulation reactions (\ref{subsection:methods:review}), then we detail the model extensions to include protein S (\ref{subsection:methods:model_extensions}). We begin the Results section (section \ref{section:results}) by describing the strong impact of PS on thrombin production (\ref{subsection:ps_psc_impact}), followed by demonstrating that PSC only matters shortly after the initiation of coagulation (\ref{subsection:psc_only_matters_shortly_after_the_initiation}). We then describe mechanisms by which PSC inhibits thrombin production (Section \ref{subsection:psc_mechanisms}), namely that PSC significantly accumulates in the injury zone, inhibiting fXa (\ref{subsection:psc_significantly_accumulates}); that the particular binding affinity for PSC with fXa is critical for thrombin inhibition (\ref{subsection:particular_binding_affinity_for_psc_to_fxa}); that PSC competes with fVam and fVhm to bind and inhibit fXa (\ref{subsection:psc_competes_with_fvh_to_bind_fxa}); and that PSC competes for fV binding sites on platelets (\ref{subsection:psc_competes_with_fv_for_binding_sites_on_platelets}).  In section \ref{subsection:extreme_conditions}, we discuss the role of PSC in pathological clotting conditions, including conditions such as the east Texas bleeding disorder where PSC may be in excess (\ref{subsection:excess_psc}); factor V deficiencies where blood may also be PSC deficient (\ref{subsection:factor_v_deficiencies}; and hemophilia, which may be treated by targeting PSC (\ref{subsection:hemophilia_a}). We conclude the results section by describing the role of free PS (Section \ref{subsection:free_ps_overall} in terms of its competition with PSC for PS-specific binding sites on platelets and in terms of its (purported) ability to bind fXa, taking a closer look at binding between free PS and fXa in section \ref{subsection:free_ps_fxa_binding_affinity}. We discuss the context for and implications of our results in section \ref{section:discussion}, including future directions in section \ref{subsection:discussion:future_directions} and implications for bleeding disorders in section \ref{subsection:discussion:bleeding_disorders}.

\section{Methods}

\label{section:methods}

The model used in this paper substantially extends models that we have published \cite{kuharsky2001surface, fogelson2006coagulation, fogelson2012blood, link2020mathematical, miyazawa2023i, miyazawa2023ii} that describe the dynamic interplay of TF-initiated coagulation and platelet deposition on a small vascular injury under flow. The reason for the current extension is to include reactions and mechanisms of protein S-related inhibition that have recently received much interest in the research community. In this section, we first review the features of the previously published models and then introduce the changes to the model needed to incorporate the new biology.

\subsection{Review of previous mathematical model of the coagulation reactions}
\label{subsection:methods:review}

We consider a vessel the size of an arteriole or venule. In the vessel, a small vascular injury has exposed the subendothelium on part of its wall. As a result, TF and collagen are exposed to flowing blood, triggering coagulation and platelet deposition, respectively. We focus on events in a thin boundary layer above the injury, a region we call the ``reaction zone'' whose initial height ($\approx$ 1-2 $\mu$m) can be estimated by a boundary layer analysis accounting for the near-wall flow velocity and the diffusivities of the relevant protein and cellular species \cite{kuharsky2001surface}.

Within the reaction zone, we consider all species, protein and cellular, to vary over time but to be uniformly distributed in space, so we describe each species by its concentration that is a function of time but not space. We therefore use an ordinary differential equation (ODE) to express the rate of change of each concentration due to reactions and transport into and out of the reaction zone. The overall model consists of a set of coupled nonlinear ODEs for all the species. 

The model includes the number densities of three populations of platelets: unactivated fluid-phase platelets, activated platelets bound to the subendothelium, and activated platelets bound to other
platelets in the thrombus.  The model includes mechanisms by which
platelets become activated and bind to the subendothelium or other platelets. The model also includes mechanisms by which fluid-phase platelets move with the flow. Bound platelets, however, are stationary: they do not
move with the flow.

\begin{figure}[h!]
    \includegraphics[width = \textwidth]{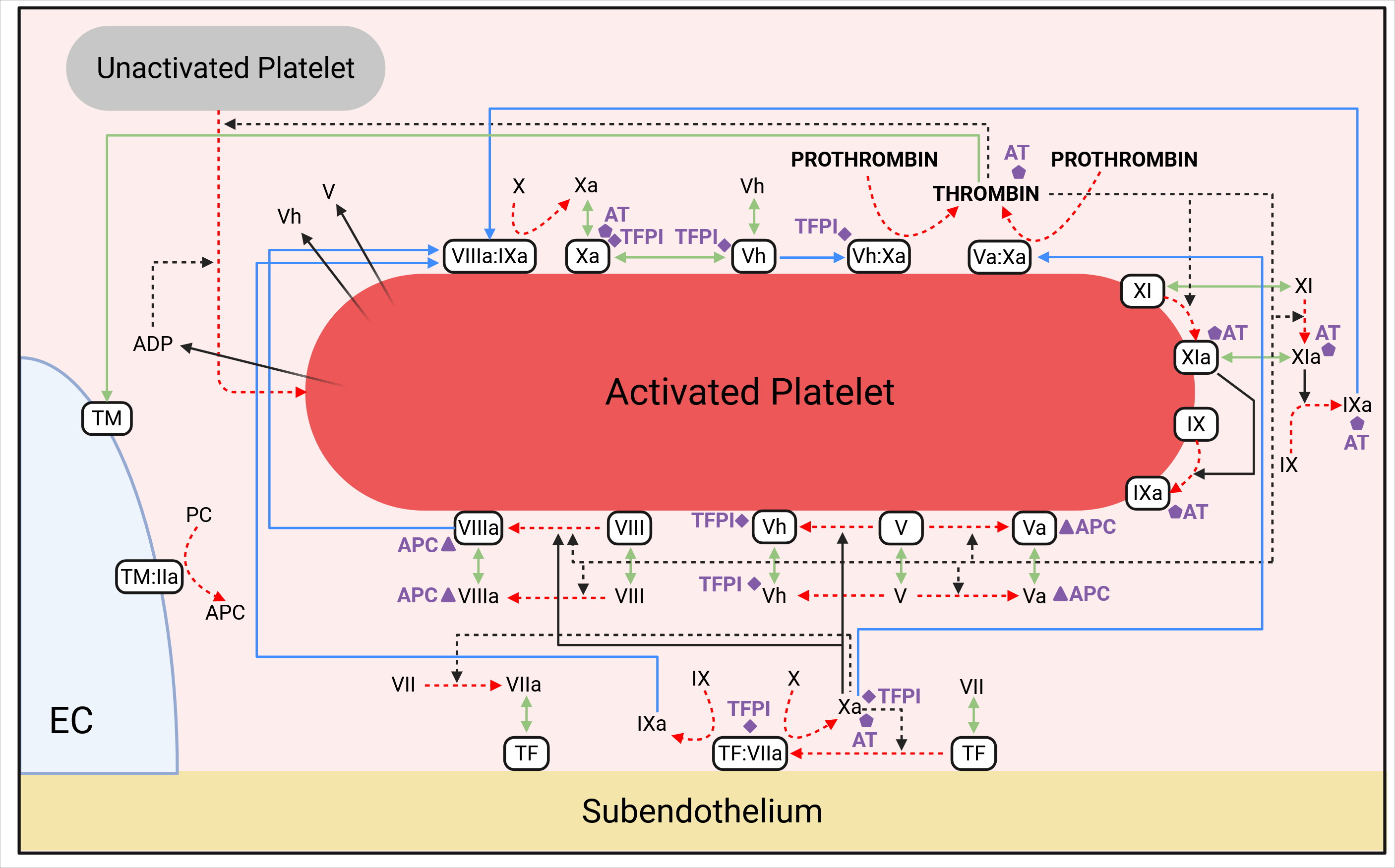}

    \caption{\footnotesize \textbf{Diagram of coagulation reactions before the extension adding PS species.} Proteins involved in the coagulation interact in the plasma, on activated platelets surfaces, on the subendothelium, or on endothelium cells. Platelets are activated by thrombin, adenosine diphosphate (ADP), or (not shown) contact with the subendothelium. Red dashed arrows indicate the enzymatic conversion of an unactivated species to an activated species. Black dashed arrows indicate feedback acceleration of reactions, whereas solid black arrows indicate feedforward acceleration, and solid black arrows with a fading tail indicate release of species from the platelet upon activation. Solid green lines and solid blue lines respectively indicate binding to a cell surface and transport of proteins in the fluid. Proteins in rounded boxes are bound to cell surface, and a purple protein next to a second protein indicates that the purple protein inhibits the second protein by binding to it. Roman numerals indicate the corresponding clotting factor, e.g., ``X'' indicates clotting factor X, whereas ``Xa'' indicates activated clotting factor X, and ``A:B'' indicates a complex of species A and B. TM = thrombomodulin, PC = protein C, APC = activated protein C, TF = tissue factor, AT = antithrombin, TFPI = tissue factor pathway inhibitor $\alpha$, AT = antithrombin. Modified from \cite{stobb2024mathematical} using BioRender. \normalsize}
    \label{supplemental_figure:reaction_diagram_excluding_protein_s_species}
\end{figure}

As for protein species, the model includes concentrations of zymogens, pro-cofactors and the corresponding enzymes and active cofactors of the TF pathway (see Fig. \ref{supplemental_figure:reaction_diagram_excluding_protein_s_species}), as well as those of the major chemical inhibitors of coagulation. It distinguishes clotting proteins in part by their chemical identity. Because,  the initiation of coagulation occurs on the SE surface whereas critical reactions in the amplification phase of coagulation occur on activated platelet surfaces, the model further distinguishes proteins by whether they are in the plasma, bound to the SE, or bound to an activated platelet surface. The model allows enzymes activated on the SE to reach the surface of a platelet by unbinding from the SE, moving through the plasma, and then binding to a receptor on an activated platelet. Only species in the plasma move with the fluid; the others are stationary while they are bound to a surface. While in the plasma, species may be carried downstream by the flow, removing them from the RZ.

Because activated platelet surfaces provide
the sites on which critical procoagulant and inhibitory reactions occur, platelets play a prominent role in the model. Since each activated platelet has a limited ability to support these reactions, the increasing availability of additional platelet surfaces as platelets accumulate in the reaction zone strongly influences the progression of the coagulation reactions. Platelets also have an anticoagulant role in that each platelet's adhesion to the injured vessel wall covers a portion of the subendothelium and blocks access to TF:fVIIa on that part of the subendothelium, physically inhibiting the enzymatic activity of TF:fVIIa. The progressive increase in platelet adhesion blocks increasing fractions of the total TF:fVIIa activity.

Coagulation interactions before the extension to include PS species are shown in Fig. \ref{supplemental_figure:reaction_diagram_excluding_protein_s_species} and outlined here:
\begin{enumerate}[noitemsep,topsep=0pt]
    \item FVII and fVIIa bind to TF on the subendothelium. FXa activates fVII in plasma or when it is bound to TF. FXa binds TF:fVII directly from the plasma without having to first bind to the subendothelium.

    \item TF:fVIIa binds and activates fIX and fX from the plasma. 

    \item Thrombin in plasma or on an activated platelet surface and fXa on an activated platelet surface activate the cofactors fV and fVIII. While thrombin converts fV to activated fVa, fXa converts fV to partially activated fVh. Thrombin converts fVh to fVa. 
    
    \item Upon activation, platelets release a prescribed number of fV and fVh molecules into the plasma.

    \item Platelet-bound fVIIIa and fIXa bind to form the tenase complex fVIIIa:fIXa which activates platelet-bound fX, producing platelet-bound fXa.
    
    \item Platelet-bound fVh and fVa bind platelet-bound fXa to form the prothrombinase complexes fVh:fXa (``PROh'') and fVa:fXa (``PRO''), respectively. The prothrombinase complexes can activate platelet-bound prothrombin into thrombin which is immediately released into the plasma.

    \item FIX is activated by fXIa in plasma and on an activated platelet surface.  FXI is activated by thrombin in plasma and on an activated platelet surface.

    \item The model includes the chemical inhibitors/inactivators antithrombin (AT), activated protein C (APC) and TFPI$\alpha$.
    
    \begin{enumerate}[noitemsep,topsep=0pt]
        \item By binding to fIXa, fXa, fXIa, and thrombin, AT permanently inactivates them.
        \item APC binds to fVa and fVIIIa in the plasma and on an activated platelet surface, permanently inactivating them. However, APC cannot bind to fVa or fVIIIa already incorporated in prothrombinase (fVa:fXa) or tenase (fVIIIa:fIXa) on an activated platelet surface. APC is produced in the endothelial zone, adjacent to the reaction zone, in the direction transverse to the flow, by a complex of thrombin and endothelial-cell-bound thrombomodulin. To bind to thrombomodulin and activate protein C, thrombin must diffuse from the reaction zone to the endothelial zone; while to affect coagulation dynamics, APC must in turn diffuse from the endothelial zone into the reaction zone.

        \item TFPI$\alpha$ in plasma binds to fXa and then TFPI$\alpha$:fXa can inhibit subendothelium-bound TF:fVIIa. TFPI$\alpha$ also binds to and inhibits fXa and blocks the cofactor activity of fVh in the plasma or on an activated platelet surface, as well as to the fVh in PROh on an activated platelet surface.
    \end{enumerate}
    
    \item The activity of TF:fVIIa decreases as platelet deposition on the subendothelium increases because an adherent platelet physically blocks access to the subendothelium surface to which it is attached.
\end{enumerate}

\medskip
In the following we use notation fVam to denote fVa bound to the surface membrane of activated platelets.  Similarly, fVhm denotes fVh bound to platelet surfaces. Other proteins bound to platelets are notated similarly.

\subsection{Model extensions to include protein S}
\label{subsection:methods:model_extensions}
We extend our mathematical model by incorporating several new protein species involving protein S (PS) and their corresponding reactions, as depicted in Fig. \ref{figure:intro}. The new species are: the protein S complex (PSC), PSC bound to fXa, PS that is not part of PSC (free PS), free PS bound to fXa, and the respective platelet-membrane-bound forms of each of these species. The corresponding reactions are the binding/unbinding to/from platelets and of PSC and PS to/from fXa that allow those species to form.

We assume that PSC circulates in the bloodstream, and is carried to and is removed from the reaction zone by flow. We do not consider PSC formation in the model due to incomplete kinetic information in the literature about this process. So, while the constituents of PSC are present in our model, reactions in which they form PSC are not. 
    
A PSC molecule in the reaction zone can bind to an activated platelet surface by its PS to PS-specific binding sites, by its fVh to fV-specific binding sites, or by both its PS and fVh. PSC cannot bind to an activated platelet surface by its TFPI$\alpha$ since such an interaction is not known to occur with a physiologically relevant binding affinity. PSC can bind directly to and inhibit fXa, regardless of whether the PSC and/or fXa are in plasma or bound to an activated platelet surface. Hence, the model includes the 8 product species, consisting of a fXa or activated platelet surface-bound fXa (fXam) bound to PSC or to one of the three activated platelet surface-bound PSC species (collectively referred to as PSCm). Fluid-phase PSC:fXa can carry out further anticoagulant functions, binding and inhibiting to TF:fVIIa. However, both fluid-phase PSC and PSC:fXa can be carried away by flow.  We assume that all TFPI$\alpha$ circulates only as part of PSC; no TFPI$\alpha$ circulates in the plasma. The latter is a departure from our previous model \cite{miyazawa2023i, miyazawa2023ii}.

We assume that free PS circulates in the plasma and is carried to and removed from the reaction zone by flow, and that, upon activation, a platelet immediately releases a specified number of free PS molecules as well as a specified number of TFPI$\alpha$, fV, and fVh molecules. Because PS \cite{schwarz1985identification_of_ps_in_platelets} and fVh \cite{wood2013tfpi} are both located in $\alpha$-granules, whereas TFPI$\alpha$ is not \cite{maroney2007tfpi_not_in_alpha_granules}, it is unlikely that PSC is present in platelets. The free PS in the reaction zone can bind to PS-specific binding sites on an activated platelet surface. Free PS can bind to and inhibit fXa, regardless of whether free PS and/or fXa are in plasma or membrane bound.  The model includes activated platelet surface-bound free PS (PSm) and the four species in which fXa or fXam is bound to free PS or PSm.

\subsubsection{Concentrations, protein copy numbers, and rate constants}
Table \ref{table:values} gives the plasma concentrations, protein copy numbers of molecules released by platelets upon activation, and the rate constants we use to incorporate protein S into the model.  We set the dissociation constant ($\text{K}_{\text{D}}$) for PSC-fXa binding to TF:fVIIa to that for TFPI$\alpha$-fXa binding to TF:fVIIa \cite{miyazawa2023i} based on reports \cite{dahlback2018factor} that PSC is not better than TFPI$\alpha$ at inhibiting TF-fVIIa. Further, when a literature value of a $\text{K}_{\text{D}}$ is available but not individual binding ($\text{k}^+$) and unbinding ($\text{k}^-$) rates, we set $\text{k}^- = 1$/sec and estimate $\text{k}^+$ as $\text{k}^+ = \text{k}^-/\text{K}_{\text{D}}$. In some cases, such as for the exceptionally small $\text{K}_{\text{D}}$ values for PSC binding to fXa, this would yield $\text{k}^+ > 10^9$/Msec exceeding the diffusion rate limit. In that case, we set $\text{k}^- = 0.01$/sec. In particular, to ensure that the binding rates of TFPI$\alpha$, free PS species, and PSC species with fVh and fXa are all on the same scale, we set all the corresponding values of $\text{k}^{-}$ to 0.01/s.

    	\begin{table}[h!]
		\begin{minipage}{\textwidth}\centering
			\caption[Concentrations and binding/unbinding rates.]{ \textbf{Concentrations and binding/unbinding rates.} \small 
				The table presents all new quantities we add to the existing mathematical model. The columns from left to right describe the category of the quantity; the particular protein species to which the quantity pertains; the name of the quantity; observed literature values of the quantity; any inferences we make using the quantity; the corresponding baseline values in the model; and lastly all references for the literature value. $\text{K}_{\text{D}}$ indicates a dissociation constant. $\text{k}^{+}$ and $\text{k}^{-}$ refer respectively to the binding and unbinding rates of the particular species to which the quantity pertains and were extrapolated from the corresponding $\text{K}_{\text{D}}$ provided by the cited paper. $\dagger$: PS and fVh have been shown to have no effect on fXa-bound TFPI$\alpha$ (unpublished work from J. Ahnstr\"om). 
                }

			\label{table:values}
			\small
            \resizebox{\textwidth}{!}{%
			\begin{tabular}{c c c c c c c}
            
				\toprule[2pt]
				\toprule[1pt]
				\rule{0pt}{3ex} 
				\multirow{1}{6em}{\centering \textbf{Quantity Type}} & \multirow{1}{5em}{\centering \textbf{Species}} & \multirow{1}{6em}{\centering\textbf{Quantity}} & \multirow{1}{6em}{\centering\textbf{Value}} & \multirow{1}{6em}{\centering \textbf{Inferred Quantity}} & \multirow{1}{4em}{\centering \textbf{Baseline Value}} & \multirow{1}{4em}{\centering\textbf{Source}}\\
				
				& & & & & &\\
				[1ex] \midrule[1.5pt] \rule{0pt}{3ex}\multirow{5}{5em}{\centering Platelet} & \multirow{3}{5em}{\centering PS} & PS binding sites per platelet & 400/1100 & same & 1100 & \citep{stavenuiter2013platelet}/\citep{mitchell1987cleavage} \\
				[0.5ex]\rule{0pt}{3ex}
				
				& & PS copies per platelet & 5300 & same & 5300 &  \citep{burkhart2012first}\\
				[1ex]\cdashline{2-7}\rule{0pt}{3ex}

				& \multirow{1}{5em}{\centering TFPI}  & TFPI$\alpha$ copies per platelet & 1200 & same & 1200 &  \citep{burkhart2012first}\\
				[1ex]\cmidrule[0.5pt]{1-7}\rule{0pt}{3ex}

				\multirow{18}{6em}{\centering Binding / unbinding} & \multirow{2.5}{8em}{\centering TFPI$\alpha$ binding to fVh} & \multirow{3}{10em}{\centering $\text{K}_{\text{D}}$} & \multirow{3}{5em}{\centering $2.1$ nM } & k$^{+}$ & $4.8 \cdot 10^{6}$ nM & \multirow{2.5}{5em}{\centering  \citep{gierula2023c_terminal} } \\
				[0.5ex]\rule{0pt}{3ex}
				
				& &  &  & k$^{-}$ & 0.01 nM &  \\
				[0.5ex]\cdashline{2-7}
				\rule{0pt}{3ex} 

                & \multirow{2.5}{8em}{\centering TFPI$\alpha$ binding to fXa} & \multirow{3}{5em}{\centering $\text{K}_{\text{D}}$} & \multirow{3}{5em}{\centering 5 nM }& k$^{+}$ & $2 \cdot 10^{6}$ nM &  \multirow{2.5}{5em}{\centering $\sim$  \cite{santamaria2017factor}, \cite{ahnstrom2012identification}, \cite{hackeng2006protein}, \cite{huang1993kinetics} } \\
				[0.5ex]\rule{0pt}{3ex}
				
				& &  &  & k$^{-}$  & 0.01 nM& \\
				[0.5ex]\cdashline{2-7}
				\rule{0pt}{3ex} 
				
				 & \multirow{2.5}{6em}{\centering Free PS binding fXa} & \multirow{3}{8em}{\centering $\text{K}_{\text{D}}$ } & \multirow{3}{5em}{\centering No binding \cite{ndonwi2008ps_needs_fxa} - 19 nM \cite{heeb1994protein}} & k$^{+}$ & $10^5$ & 
                 \\
				[0.5ex]\rule{0pt}{3ex}

				& &  & & k$^{-}$ & $0.01$ &  \\
				[0.5ex]\cdashline{2-7}
				\rule{0pt}{3ex}

				& \multirow{2.5}{8em}{\centering PSC species binding fXa} & \multirow{3}{5em}{\centering $\text{K}_{\text{D}}$} & \multirow{3}{5em}{\centering $\sim$ 0.01 - 0.5 nM }& k$^{+}$ & $2\cdot 10^8$ nM &  \multirow{2.5}{5em}{\centering \cite{gierula2020anticoagulant} citing \cite{dahlback2017novel}} \\
				[0.5ex]\rule{0pt}{3ex}
				
				& & & & k$^{-}$ & $0.01$ nM  &    \\
				[0.5ex]\cdashline{2-7}
				\rule{0pt}{3ex} 
				
				& \multirow{3}{8em}{\centering PS binding to platelets} & \multirow{3}{5em}{\centering $\text{K}_{\text{D}}$} & \multirow{3}{5em}{\centering 10 nM }& k$^{\text{on}}$ & $10^8$ & \multirow{2.5}{5em}{\centering \citep{veer1999regulation}   } \\
				[0.5ex]\rule{0pt}{3ex}
				
				& &  & & k$^{\text{off}}$ & $1$  & \\
				[0.5ex]\cdashline{2-7}
				\rule{0pt}{3ex}

				&  \multirow{3}{9em}{\centering PSC bound to fXa binding tissue factor - fVIIa} & \multirow{3}{5em}{\centering $\text{K}_{\text{D}}$} & \multirow{3}{5em}{\centering 0.11 nM} & k$^{+}$ & $10^7$  &   \multirow{2.5}{5em}{\centering See \cite{miyazawa2023i} and $\dagger$. }\\
				[0.5ex]\rule{0pt}{3ex}
				
				& &  &  & k$^{-}$ & $0.0011$ nM &  \\
				[1ex]\cmidrule[0.5pt]{1-7}\rule{0pt}{3ex}

				\multirow{4}{7em}{\centering Nonzero upstream and initial concentrations} & \multirow{1}{5em}{\centering PSC} & Upstream/initial concentration & 0.2 - 0.5 nM & same & 0.5 nM & \citep{dahlback2023natural} \\
				[0.5ex]
				\rule{0pt}{3ex} 

                & \multirow{1}{5em}{\centering Free PS} & Upstream/initial concentration & $\sim$150 nM & same & $\sim$150 nM & \citep{gierula2020anticoagulant}\\
				[0.5ex]
				\rule{0pt}{3ex}
				
				& \multirow{1}{6em}{\centering Zymogen fV} & Upstream/initial concentration & 16 - 20 nM & same &  20 nM & \citep{duckers2008low_plasma_factor_v}\\
				[0.5ex]
				\bottomrule[1pt]
				\bottomrule[2pt]
			\end{tabular}\normalsize
            }
		\end{minipage}
        \vskip-3ex\text{}
	\end{table}

\subsubsection{Example differential equations that illustrate the model extension}
We have added 19 new ODEs to the model and modified numerous others to include Protein S in the model (see Supplemental Material).  Here, as examples, we discuss two of the new ODEs. Equation \ref{equation:ode_for_psc} describes the concentration of fluid-phase PSC. The five labeled terms in the equation describe the flow-mediated delivery and removal of PSC from the reaction zone; the binding and unbinding of PSC via its protein S to binding sites on activated platelets specific for protein S; the binding and unbinding of PSC via its fVh to binding sites on activated platelets specific for fV, fVh, and fVa; the binding of PSC to fluid-phase fXa; and the binding of PSC to platelet-bound fXam (that is fXa bound to a fX/fXa-specific binding site on an activated platelet). The binding and unbinding reactions are assumed to be governed by mass action kinetics. The binding reactions depend on the concentrations of available binding sites, i.e., ones not already bound to another protein.  Here, $p_{\text{PS}}^{\text{avail}}$ and $p_5^{\text{avail}}$ are respectively the concentrations of available PS and fV binding sites on activated platelets in the reaction zone.

    \hskip-3ex
	\resizebox{\textwidth}{!}{%
    \begin{minipage}{\textwidth}
    \begin{align}	
        \label{equation:ode_for_psc}
		\frac{d[\text{PSC}]}{dt}	&= \underbrace{\text{k}_{\text{flow}}\left([\text{PSC}]^{\text{up}} - [\text{PSC}]\right)}_{\text{1. Transport by flow}} \underbrace{-\text{k}_{\text{PS}}^{\text{on}}[\text{PSC}] p_{\text{PS}}^{\text{avail}} + \text{k}_{\text{PS}}^{\text{off}}[\text{PSm:TFPI:fVh}]}_{\text{2. Binding/unbinding to from platelets via PS}} ~\underbrace{~-~\text{k}_{\text{PS:TFPI:fVhm}}^{\text{on}}[\text{PSC}] p_{\text{5}}^{\text{avail}} + \text{k}_{\text{PS:TFPI:fVhm}}^{\text{off}}[\text{PS:TFPI:fVhm}]}_{\text{3. Binding/unbinding to from platelets via fVh}} \notag \\
		&\hskip2.5ex \underbrace{-\text{k}^{+}_{\text{PSC:fXa}}[\text{PSC}] [\text{fXa}] + \text{k}^{-}_{\text{PSC:fXa}}[\text{PSC:fXa}]}_{\text{4. Binding/unbinding to/from fXa}} ~\underbrace{-\text{k}^{+}_{\text{PSC:fXam}}[\text{PSC}][\text{fXam}] ~+~ \text{k}^{-}_{\text{PSC:fXam}}[\text{PSC:fXam}]}_{\text{5. Binding/unbinding to/from fXam}}
	\end{align}
    \end{minipage}
    }\vskip1ex
	
The rate of change of the concentration [PS] of fluid-phase free protein S is described in Eq.\ \ref{equation:ode_for_free_ps}.  The first four labeled terms in this ODE describe flow-mediated delivery and removal of PS from the reaction zone; the binding and unbinding of PS to protein-S-specific binding sites on activated platelets; the binding/unbinding of PS to fluid-phase fXa; and the binding/unbinding of PS to platelet-bound fXam. Note that PSC and PS compete for the protein-S-specific binding sites on activated platelet surfaces. The final term in Eq. \ref{equation:ode_for_free_ps} describes the rate of release of PS from platelets as platelets are activated.  This rate is the product of the prescribed number $n_{ps}$ of PS molecules stored in each platelet and the rate at which the concentration of activated platelets changes.  $\text{PL}^{\text{as}}$ and $\text{PL}^{\text{av}}$ refer, respectively, to the concentrations of subendothelium-bound activated platelets and activated platelets in the thrombus that are not directly bound to the subendothelium.
    
    \begin{minipage}{\textwidth}
    \begin{align} 
        \label{equation:ode_for_free_ps}
		\frac{d}{dt}\left[  \text{PS}  \right]&=  \underbrace{\text{k}_{\text{flow}} ([\text{PS}]^{\text{up}} - [\text{PS}] )}_{\text{1. Removal/replenishment by flow}} \hskip1ex ~~\underbrace{- \text{k}_{\text{PS}}^{\text{on}}[\text{PS}] \cdot p_{\text{PS}}^{\text{avail}} + \text{k}_{\text{PS}}^{\text{off}}[\text{PSm}]}_{\text{2. Binding/unbinding to/from platelets}}   
		\hskip1ex \underbrace{- \text{k}_{\text{PS:fXa}}^{+} [\text{PS}] \cdot [\text{fXa}] + \text{k}_{\text{PS:fXa}}^{-}[\text{PS:fXa}]}_{\text{3. Binding/unbinding to/from fXa}} \notag\\ 
		&\hskip2.5ex \underbrace{- \text{k}_{\text{PS:FXam}}^{+} [\text{PS}]\cdot [\text{fXam}] + \text{k}_{\text{PS:fXam}}^{-}[\text{PS:fXam}]}_{\text{4. Binding/unbinding to/from fXam}} ~~\underbrace{+ n_{ps} \cdot \frac{d}{dt} ( [\text{PL}^{\text{as}}] + [\text{PL}^{\text{av}}] )}_{\text{5. Release from platelets of free PS}}
	\end{align} 
    \end{minipage}

\normalsize
\section{Results:}
\label{section:results}

To determine the effect of PS species on thrombin production, we ran simulations with PS-related parameters varied around the baseline values given in Table \ref{table:values}. For example, to understand the impact of the binding affinity between PSC and fXa, we varied the corresponding $\text{K}_{\text{D}}$ near its baseline value (0.05 nM), while holding all other parameters fixed at their respective baselines. All simulations were run with baseline values for PS-related parameters unless otherwise specified. Because different injuries may lead to different tissue factor exposure, we varied the density of tissue factor $[\text{TF}]_\text{s}$ exposed on the subendothelium, generally from 0 - 5 fmol/cm$^2$ or from 0 - 15 fmol/cm$^2$. 

To quantify the impact of PS species on thrombin production, we examine the concentration of thrombin in the reaction zone over time, as would be done in thrombin generation assays \cite{tripodi2020thrombin_review, depasse2021thrombin_review, baglin2005thrombin_measurement, castoldi2011thrombin_generation}. From time-courses of [Thrombin] over 40 minutes, we compute several metrics: the thrombin concentration at 10 minutes ([Thrombin]$_{10}$), and the thrombin lag time ($t_{\text{lag}}$). We use [Thrombin]$_{10}$ as a marker of the progress of coagulation and $t_{\text{lag}}$ to delineate the start of the thrombin burst \cite{castoldi2011thrombin_generation}. Lag time is typically defined as the time from coagulation initiation until [Thrombin] reaches substantial levels \cite{tripodi2020thrombin_review, depasse2021thrombin_review, baglin2005thrombin_measurement, castoldi2011thrombin_generation}, which we take to be 1 nM, following our previous practice, e.g., \cite{miyazawa2023i, stobb2024mathematical}.

\subsection{PS significantly impacts thrombin production} 
\label{subsection:ps_psc_impact}
To calibrate the effect of PS, we first ran simulations with plasma [TFPI$\alpha$]=0.5 nM and without PS. We see that (Fig. \ref{supplemental_figure:tfpi}) TFPI$\alpha$ alone has little effect on thrombin production across a range of [TF]$_s$ values. Thrombin time-courses and metrics $t_{lag}$ and [Thrombin]$_{10}$ hardly change. TFPI$\alpha$ mainly binds fVh early and PROh later, but it never binds enough of either to impact thrombin production.

\begin{figure}[h!]

        \includegraphics[width = \textwidth]{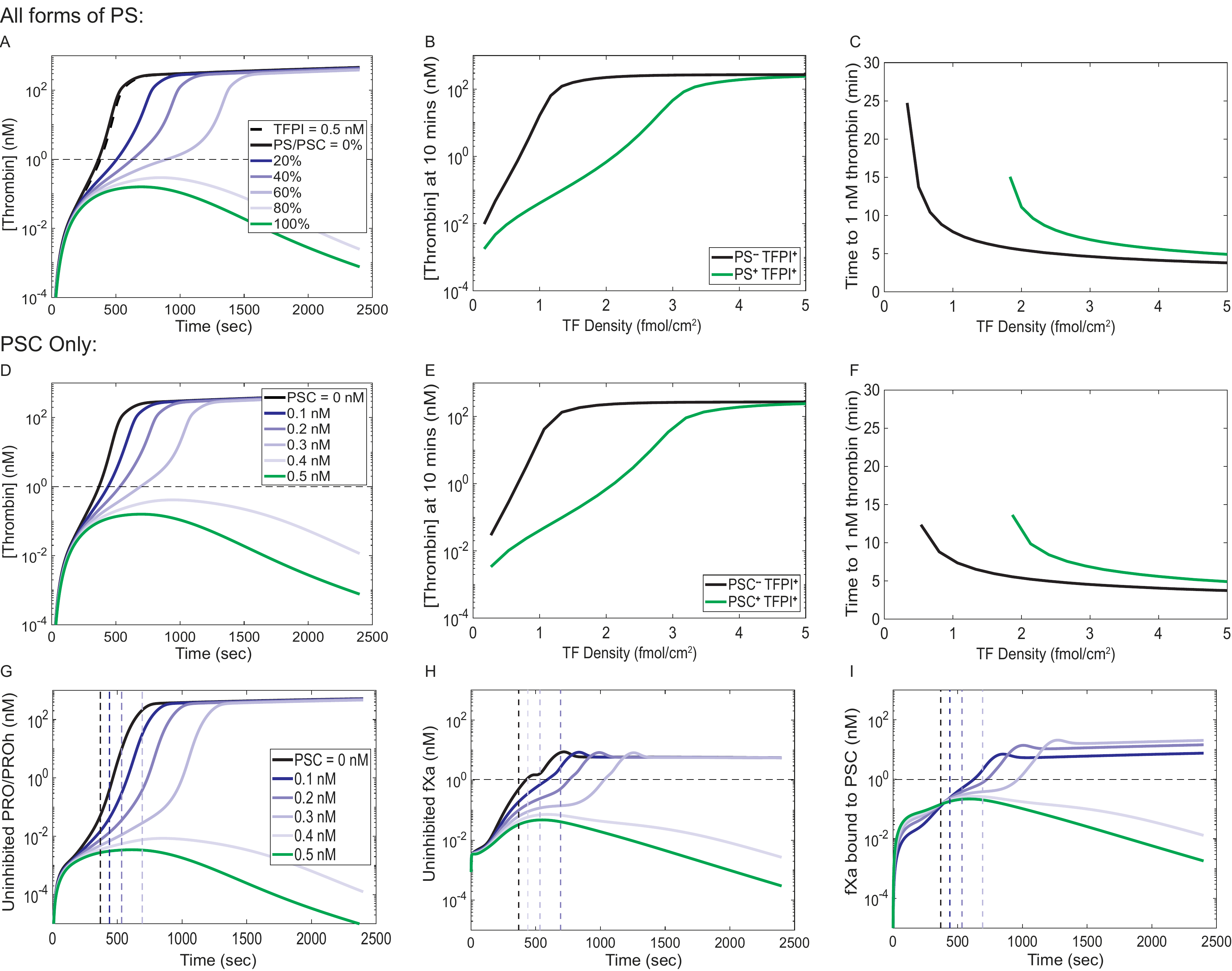}

        \vskip1ex
              \caption{\footnotesize (A) Increasing the concentration of protein S species in plasma and platelets from 0 to baseline ($\sim$0.5 nM PSC and 160 nM free PS in the plasma, and 5300 free PS molecules in each platelet) delays or prevents a thrombin burst for [TF]$_s$ = 1.5 fmol/cm$^2$. Introducing PS leads to (B) significantly lower [Thrombin]$_{10}$ and (C) higher $t_{lag}$ across tissue factor levels. (D) For [TF]$_s$ = 1.5 fmol/cm$^2$, increasing plasma [PSC] from 0 to baseline delays thrombin production for lower plasma [PSC] and prevents the thrombin burst for higher [PSC]. For a range of [TF]$_s$, higher plasma [PSC] leads to (E) lower [Thrombin]$_{10}$ and (F) higher $t_{lag}$ or eliminates the thrombin burst altogether.  Increasing [PSC] significantly delays (G) the rise of uninhibited prothrombinase (both PRO and PROh) and (H) total uninhibited fXa, while (I) total fXa bound to PSC goes up sooner. Vertical dashed lines in (G) - (I) denote the times at which [Thrombin] reaches 1 nM. For all panels, the shear rate is 100 1/s and $\text{K}_{\text{D}} = 0.05$ nM for PSC binding with fXa.
        \normalsize}
        \label{figure::ps_psc_impact}

    \end{figure}

In contrast, when we replace plasma TFPI$\alpha$ with PSC (0.5 nM), and include free PS in the plasma (150 nM) and in platelets (copy number 5300 per platelet), we see a strong inhibitory effect. Fig. \ref{figure::ps_psc_impact}A shows that for [TF]$_s$= 1.5 fmol/cm$^2$ and PS levels between 0 and baseline, higher PS levels shift the time-course of thrombin to the right and sufficiently high levels prevent the thrombin burst altogether. Panels B and C show the impact of baseline levels of PS species on thrombin metrics for a range of [TF]$_s$ levels. PS prevents a thrombin burst for [TF]$_s$ $\lessapprox 2$ fmol/cm$^2$, increases $t_{lag}$ and reduces [Thrombin]$_{10}$ for [TF]$_s$ between $\sim$2 - 4 fmol/cm$^2$, and has little impact for [TF]$_s > 4$ fmol/cm$^2$. We discuss reasons underlying the strong inhibitory effect of PSC compared to the weak inhibitory effect of TFPI$\alpha$ in the absence of PSC in more detail in section \ref{section:discussion}.

Inhibition due to PS species in these simulations is due largely to PSC. Fig. \ref{figure::ps_psc_impact} D - F show the impact of PSC on thrombin time-courses and thrombin metrics as PSC is varied but free PS (plasma and platelet stores) remains at baseline. The overall impact is similar to that seen in panels A - C. However, the smaller shifts to the right in panel D than in panel A suggest that the net effect of free PS is to \emph{reduce} the effect of inhibition by PSC (see Section \ref{subsection:free_ps_overall}). In contrast, in panels E - F, we see that the thrombin metrics vs [TF]$_s$ in the absence of PSC (black curves) are shifted slightly to the right of the corresponding curves in B and C, suggesting that free PS is slightly inhibitory in the absence of PSC (see Section \ref{subsection:free_ps_fxa_binding_affinity}).

High-level mechanisms by which PSC inhibits thrombin production become clear upon examination of key steps in the coagulation reaction network (Fig. \ref{figure::ps_psc_impact} G - I). Panel G shows that PSC concentrations leading to a delay in the thrombin burst similarly delay prothrombinase formation. This delay corresponds to a delay in the rise of uninhibited fXa (panel H) which is in turn due to increased inhibition of fXa by PSC (panel I) early during coagulation. For simulations without a thrombin burst, the higher concentrations of PSC lead to sufficiently low availability of fXa early during coagulation and thus insufficient prothrombinase formation for a thrombin burst to occur.

\subsection{PSC only matters shortly after the initiation of coagulation}
\label{subsection:psc_only_matters_shortly_after_the_initiation}

\begin{figure}[h!]

        \includegraphics[width = \textwidth]{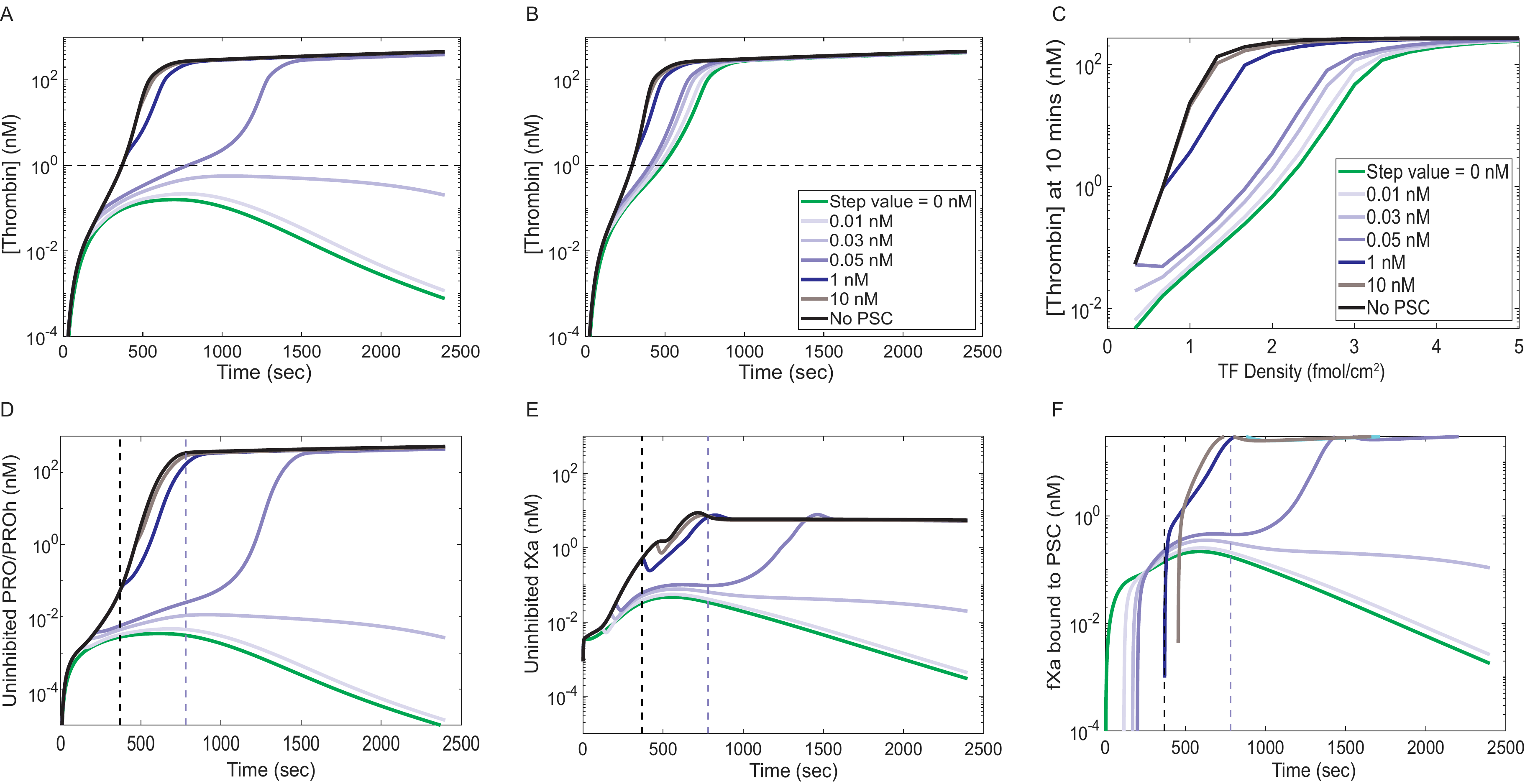}
        \vskip1ex
        \caption{\footnotesize In all panels, plasma PSC is absent until [Thrombin] reaches the specified ``step value'', at which time [PSC] is immediately set to 0.5 nM in upstream plasma. Panels (A) and (B) show that PSC impacts the time courses of thrombin most when it is added before [Thrombin] reaches 1 nM, for [TF]$_s = 1.5$ and $2.5$ fmol/cm$^2$, respectively. Panel (C) shows similar results for [Thrombin]$_{10}$ across a wide range of [TF]$_s$ values. For [TF]$_s=1.5$ fmol/cm$^2$, increasing the delay in adding PSC significantly accelerates the rise of (D) uninhibited prothrombinase (both PRO and PROh), and (E) total uninhibited fXa, while (F) total fXa bound to PSC rises later. Vertical dashed lines in (D) - (F) denote the times at which [Thrombin] reaches 1 nM. \normalsize}
        \label{figure:psc_only_matters_shortly_after_the_initiation}

	\end{figure}

Indeed, we find that PSC matters only shortly after coagulation begins, while [Thrombin] $<$ 1 nM. We infer this from simulations in which we initiate coagulation with no PSC in the plasma, and add PSC to the plasma at its baseline concentration (0.5 nM) only when [Thrombin] reaches a specified value. We compare the results with those obtained when PSC is present from the beginning of coagulation or absent for the entire simulation. Across a range of $[\text{TF}]_{\text{s}}$, adding PSC when [Thrombin] reaches 10 nM produces thrombin metrics only slightly different than when PSC is absent altogether (Fig. \ref{figure:psc_only_matters_shortly_after_the_initiation} A - C). In contrast, early addition of PSC has a significant and sometimes profound effect on later thrombin production. For $[\text{TF}]_\text{s}$=1.5 fmol/$\text{cm}^2$, adding PSC when thrombin reaches 1 nM or 0.5 nM delays the subsequent rapid rise in thrombin by about one minute.  Adding PSC when thrombin reaches 0.1 nM to 0.05 nM delays thrombin reaching 1 nM, slows substantially thrombin's rise from 1 nM to 10 nM, but has little effect on the speed of its rise after [Thrombin] reaches 10 nM. Adding PSC when thrombin reaches 0.03 nM (or sooner) prevents a thrombin burst completely. For $[\text{TF}]_\text{s}$=2.5 fmol/$\text{cm}^2$ (panel B), PSC at 0.5 nM does not prevent a thrombin burst even if present from the start, but the earlier PSC is added to the plasma, the greater the delay until the burst.  Panel C shows that when present early PSC affects thrombin generation for $[\text{TF}]_\text{s} \lessapprox$ 3 fmol/$\text{cm}^2$, while for higher $[\text{TF}]_\text{s}$ it has only a small effect even if present from the start.

The sequence of reactions underlying the effects of added PSC on thrombin are shown in Fig. \ref{figure:psc_only_matters_shortly_after_the_initiation} D - F. Adding PSC to plasma when thrombin is $\sim$ 0.01-1.0 nM reduces prothrombinase concentration (panel D), either slowing or leading to a net decrease in the prothrombinase concentration. The formation of prothrombinase requires the presence of uninhibited platelet-bound fXa and we see in panel E that the concentration of this species drops quickly after PSC is added to the plasma.  In situations that produce a thrombin burst, the decrease is temporary but contributes to the delay in prothrombinase formation. In panel F, we see that the concentration of fXa bound to and inhibited by PSC increases sharply with the addition of PSC to the plasma, reducing the availability of uninhibited fXa.

\subsection{Mechanisms by which PSC inhibits thrombin production}
\label{subsection:psc_mechanisms}
Simulations suggest that three specific mechanisms underlie the inhibitory role of PSC. (i) PSC accumulates in the injury zone reaching concentrations much higher than in plasma (Section \ref{subsection:psc_significantly_accumulates}). (ii) PSC binds fXa tightly, potentially 50 - 100 fold more tightly than does TFPI$\alpha$ alone (Section \ref{subsection:particular_binding_affinity_for_psc_to_fxa}). This allows PSC to successfully compete with fV/fVh to bind fXa species (Section \ref{subsection:psc_competes_with_fvh_to_bind_fxa}). (iii) PSC occupies fV-specific binding sites on platelets, competing with fV, fVh, and fVa and thus reducing prothrombinase formation and increasing removal of procoagulant fV species by blood flow (Section \ref{subsection:psc_competes_with_fv_for_binding_sites_on_platelets}).

\subsubsection{PSC significantly accumulates in the injury zone, inhibiting thrombin}
\label{subsection:psc_significantly_accumulates}

\begin{figure}[h!]\centering

        \includegraphics[width = \textwidth]{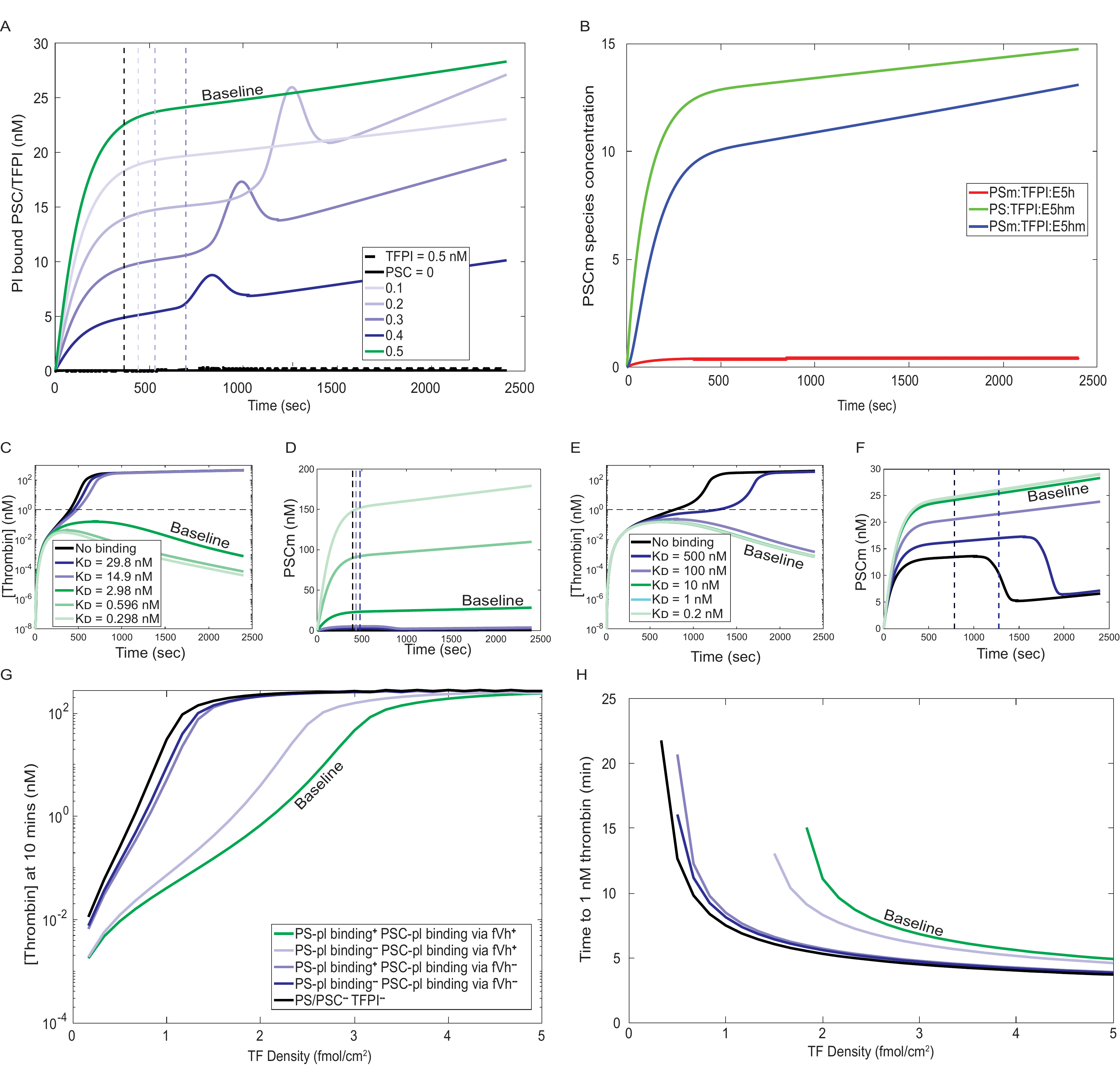}

        \caption{\footnotesize
        For [TF]$_s$=1.5  fmol/cm$^2$,   (A): Increasing plasma PSC levels from 0 towards physiological levels ($\sim$ 0.2 - 0.5 nM) leads to increasing concentrations of PSC bound to platelets, much higher than the concentrations of platelet-bound forms of TFPI$\alpha$ (TFPI-fVhm) available to inhibit fXa if TFPI$\alpha$ were to replace PSC. (B) PSC mainly binds to platelets via fVh. (C) Varying the $\text{K}_{\text{D}}$ for PSC binding to platelets by fVh around its physiological value ($\text{K}_{\text{D}} = 2.98$ nM, dark green) significantly impacts thrombin time-courses and (D) PSC levels. (E) Varying the $\text{K}_{\text{D}}$ for PS (for both free PS and PSC) binding to platelets around its expected physiological value (10 nM, dark green) modestly impacts thrombin time-courses and (F) platelet-bound PSC levels. For a wide range of [TF]$_s$, removing the ability of PSC (and free PS) to bind to platelets by PS has a mild effect on thrombin levels, whereas (G-H) removing the ability of PSC to bind to platelets by its fVh makes PSC's inhibitory functionality be comparable to the case where it cannot bind to platelets at all but still more anticoagulant than the case where PSC and TFPI$\alpha$ are altogether absent. In panels \textit{A}, \textit{D}, and \textit{F}, vertical dashed lines indicate the time at which [Thrombin] reaches 1. nM In all panels, shear rate is 100 1/s.  \normalsize}
		\label{figure:psc_significantly_accumulates}
	\end{figure}

\vskip1ex
Unlike TFPI$\alpha$ alone, PSC can bind to platelets, either by its PS, its fVh, or by both. To investigate the accumulation of PSC on platelets in the injury zone, we ran simulations for a variety of plasma PSC concentrations, a variety of $\text{K}_{\text{D}}$ for binding of PS to platelets, a variety of $\text{K}_{\text{D}}$ for binding of PSC to platelets via its fVh, and a variety of [TF]$_s$ values. Fig. \ref{figure:psc_significantly_accumulates} A-F show results for
$[\text{TF}]_\text{s}$ 1.5 fmol/cm$^2$.  In panel A, we see 
that PSC accumulates in the reaction zone to concentrations (5 - 30 nM) much higher than does TFPI$\alpha$ (0 - 1 nM) and much higher than the PSC concentration in plasma (0.5 nM). 

Interestingly, simulations suggest that PSC attaches to platelets mainly by its fVh rather than its PS. Fig. \ref{figure:psc_significantly_accumulates} B shows that for 0.5 nM plasma [PSC], PSC is mostly bound to platelets by its fVh (12 - 15 nM), second by both its fVh and its PS (10 - 13 nM), and hardly at all by its PS alone (0 - 0.5 nM). Panel C shows that relatively small changes to $\text{K}_{\text{D}}$ for the binding of PSC to platelets by its fVh significantly alter thrombin production, preventing or allowing a thrombin burst. In particular, if the $\text{K}_{\text{D}}=2.98$ nM (dark green), its baseline value, or lower, there is no thrombin burst, and reducing $\text{K}_{\text{D}}$ reduces [Thrombin] at each time. Increasing $\text{K}_{\text{D}}$ from $14.9$ nM (light purple) shifts thrombin time-courses to the left. The effects of these relatively small changes in $\text{K}_{\text{D}}$ are likely due to dramatic changes in [PSC] bound to platelets (changing from $\sim 5$ nM to near $\sim 200$ nM, panel E). Panels G - H show that across $[\text{TF}]_\text{s}$ values, preventing PSC from binding to platelets by its fVh eliminates nearly all inhibitory functionality of PSC.  

In contrast, binding of PSC to platelets by its PS only modestly increases PSC accumulation and has relatively little impact on thrombin production except for a limited range of [TF]$_{\text{s}}$. Fig. \ref{figure:psc_significantly_accumulates} E - F show the impact of altering the binding strength between PS and platelets on the time-courses of the thrombin and membrane-bound PSC concentrations, respectively. With a physiological $\text{K}_{\text{D}} \sim$ 10 nM (panel F dark green curve) for PS binding to platelets, the membrane-bound PSC concentration is $\sim$double that when PS does not bind to platelets at all (black curve). The extra PSC binding that results from reducing the $\text{K}_{\text{D}}$ from 500 to 100 nM can prevent a thrombin burst (panel E), and further reducing the $\text{K}_{\text{D}}$ moderately reduces [Thrombin] at each time. However, by comparing thrombin metrics with and without the ability of PS to bind to platelets, Panels G - H show that PS binding to platelets can prevent thrombin bursts only for a limited range of $[\text{TF}]_\text{s}$ ($\sim 1.5$ - 2 fmol/cm$^2$), and otherwise only moderately increases $t_{\text{lag}}$ and decreases [Thrombin]$_{10}$. PSC binding to platelets by PS has little impact on thrombin production because PS binds to platelets less strongly than does fVh (see Table \ref{table:values}) and, more importantly, PSC must compete for PS-specific platelet binding sites with the $\sim$ 300 times more concentrated free PS in the plasma. Increasing the number of PS-specific binding sites on platelets (Fig. \ref{supplemental_figure:ps_binding_sites_on_platelets}) increases the concentration of PSC bound to platelets via its PS (panel B) and moderately reduces thrombin production (panel A). See Section \ref{subsection:free_ps_overall} for more details on competition between free PS and PSC for PS-specific binding sites on platelets.

\subsubsection{The particular binding affinity for PSC to fXa is critical for thrombin inhibition}
\label{subsection:particular_binding_affinity_for_psc_to_fxa}

\begin{figure}[h!]

        \includegraphics[width = \textwidth]{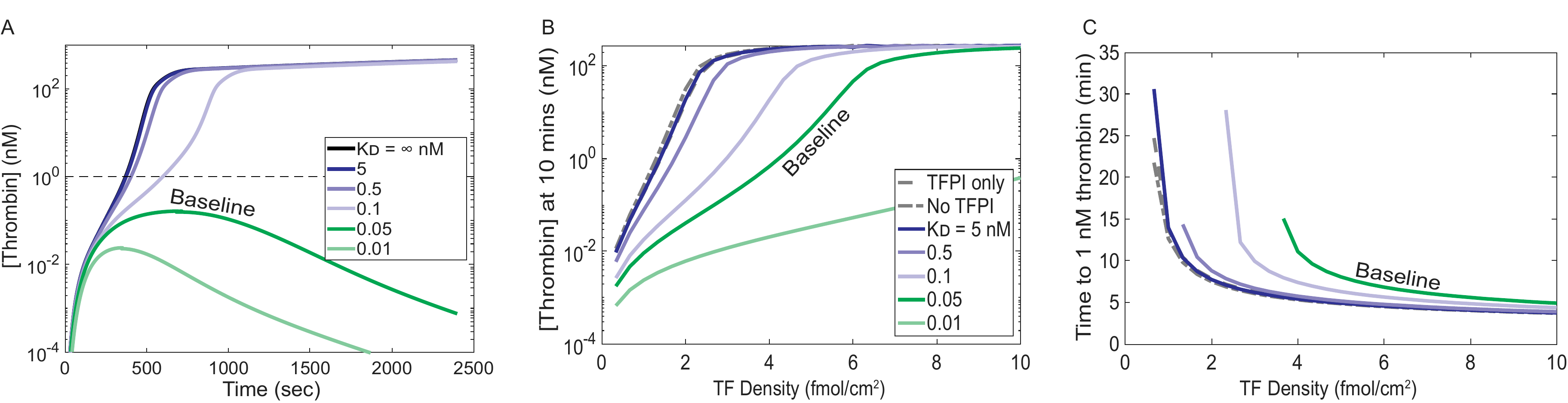}
        \caption{\footnotesize \textbf{Enhanced binding affinity of TFPI$\alpha$ in PSC to fXa inhibits thrombin.} (A) For [TF]$_s$ = 1.5 fmol/cm$^2$, decreasing the $\text{K}_{\text{D}}$ for binding of PSC to fXa from infinity to physiological levels ($\text{K}_{\text{D}}$ 0.1 - 0.05 nM) delays thrombin time-courses for larger $\text{K}_{\text{D}}$ and prevents a thrombin burst entirely for smaller $\text{K}_{\text{D}}$. Across a range of [TF]$_s$ values,  higher binding affinity for PSC to fXa (B) leads to lower [Thrombin]$_{10}$ and (C) increases $t_{lag}$ or eliminates the thrombin burst altogether (C).  The shear rate is 100 1/s in all panels. \normalsize}
		\label{figure:psc-e10_binding_affinity}
	\end{figure}

Simulations (Fig. \ref{figure:psc-e10_binding_affinity}) show that relatively small changes to the binding affinity between PSC and fXa can significantly impact thrombin time-courses and thrombin metrics. Panel A shows that for $[\text{TF}]_\text{s}$ = 1.5 fmol/cm$^2$, PSC with a $\text{K}_{\text{D}}$ comparable to that of TFPI$\alpha$ alone impacts thrombin time-courses as if it did not bind fXa at all. On the other hand, reducing the $\text{K}_{\text{D}}$ to near physiological levels (K$_{\text{D}}$ = 0.1 - 0.05 nM) prevents the thrombin burst and strongly inhibits thrombin across $[\text{TF}]_\text{s}$. In particular, without enhanced PSC-fXa binding compared to TFPI$\alpha$-fXa binding, PSC has little impact on thrombin metrics, whereas reducing $\text{K}_{\text{D}}$ to near physiological levels dramatically reduces [Thrombin]$_{10}$ (panel C), prevents a thrombin burst for $[\text{TF}]_\text{s}$ between $\sim 0.5$ and 4.5 fmol/cm$^2$, and otherwise greatly increases the lag time (panel D). In fact, reducing $\text{K}_{\text{D}}$ to 0.01 nM prevents a thrombin burst for all tested $[\text{TF}]_\text{s}$.

\subsubsection{PSC competes with fVam and fVhm to bind and inhibit fXa}
\label{subsection:psc_competes_with_fvh_to_bind_fxa}
To inhibit fXa, PSC must bind fXa before either fVam or fVhm does. Fig. \ref{figure:psc_competes_with_fvh_to_bind_fxa} shows the competition between PSC and fVam/fVhm to bind fXa. Panels A - C show that shortly after coagulation initiation, PSCm (panel C) is available in a much higher concentration than fVam / fVhm (panel A) to bind and inhibit a limited concentration of fXam (panel B). For tested nonzero values of PSC, shortly after coagulation initiation [PSCm] reaches $\approx$ 3 - 25 nM while concentrations of fVam/fVhm species hardly exceed 1 nM. However, for plasma [PSC] sufficiently low that thrombin bursts occur, [fVam] and [fVhm] increase to over 100 nM shortly after the thrombin burst, outcompeting PSC for fXam, resulting in most fXam becoming part of PRO/PROh. In contrast, for higher plasma [PSC] (panel E), PSC binds only slightly higher concentrations of fXa, but sufficiently slows and decreases the availability of fXa for forming PRO/PROh that a thrombin burst never occurs.

\subsubsection{PSC competes for factor V binding sites on platelets}
\label{subsection:psc_competes_with_fv_for_binding_sites_on_platelets}

\begin{figure}[h!]

        \includegraphics[width = \textwidth]{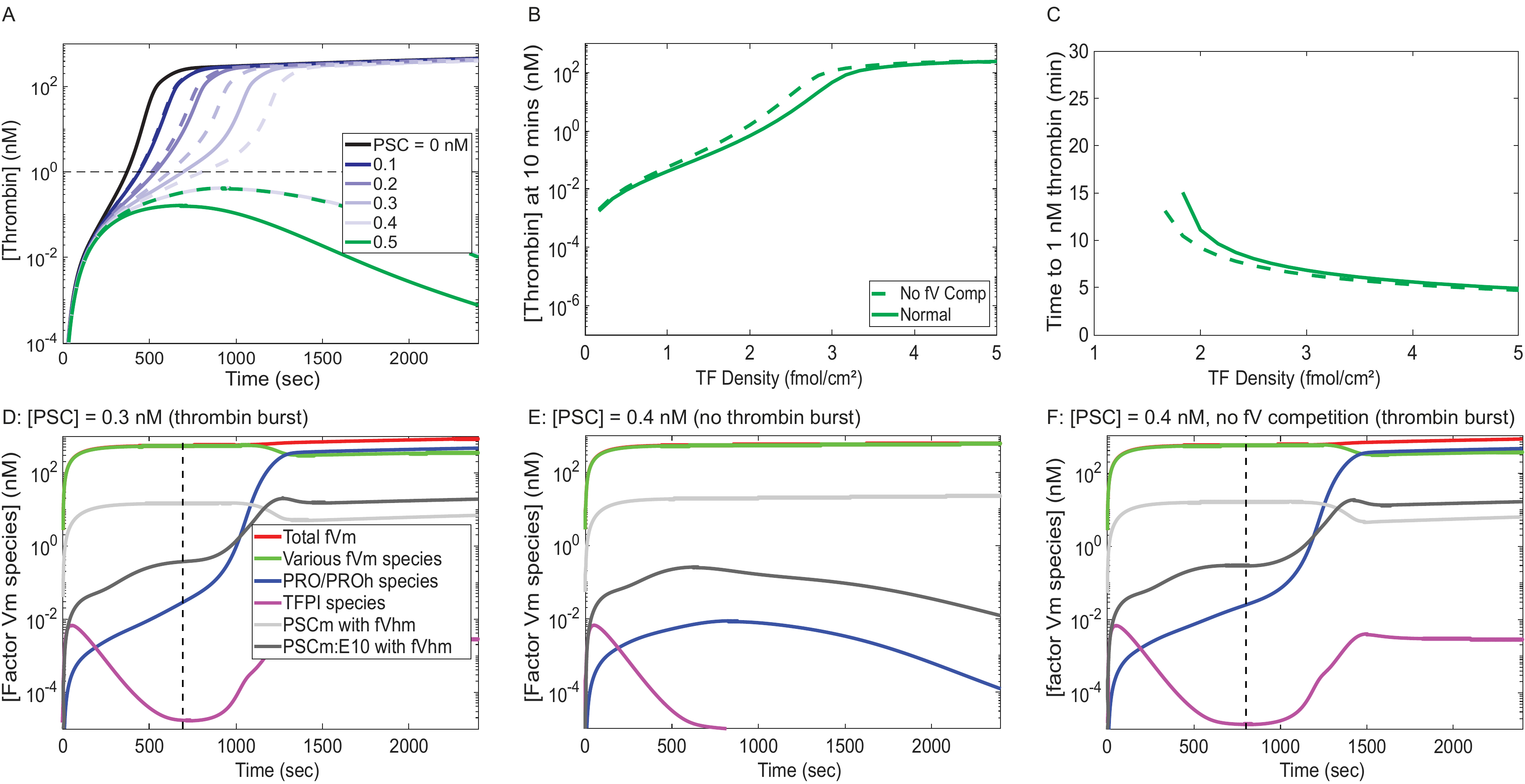}
        \vskip1ex
        \caption{\footnotesize \textbf{Indirect inhibition of fV by PSC due to competition for fV binding sites on platelets.} Upon allowing fV species to bind to platelets at binding sites already occupied by PSC, (A) thrombin time-courses accelerate (dashed curves) for [TF]$_s$ = 1.5 fmol/cm$^2$ compared to thrombin time-courses where fV species cannot bind to platelets at binding sites already occupied by PSC (solid curves), (B) [Thrombin]$_{10}$ moderately increases, and (C) lag times moderately decrease. (D) For [PSC]= 0.3 nM, PSC species out compete PRO/PROh species for fV-specific binding sites on platelets throughout the simulation  until the thrombin burst is well under way. (E) In the absence of a thrombin burst as when [PSC] = 0.4 nM, PSC out competes fV species for fV-specific binding sites throughout the simulation.  Panel (F) shows that for [PSC] = 0.4 nM, removing the inhibitory effect of PSC competition for fV-specific binding sites by allowing fV to bind to such sites already occupied by PSC restores a thrombin burst. In (D) - (F), solid green lines include fVm, fVam, fVm:fXam, fVm bound to membrane-bound thrombin, fVhm, fVhm:fXam, and APC:fVhm; solid pink lines include TFPI$\alpha$:fVh, TFPI$\alpha$:PROh, fXa bound to TFPI$\alpha$ in turn bound to fVh or fVhm; solid dark-gray lines include PSC whose fVh is membrane bound; solid light-gray lines include PSC bound to fXam where the fVh in the PSC is membrane bound.  Shear rate is 100 1/s in all panels. In (D) and (F) the vertical dashed line indicates the time at which [Thrombin] reaches 1 nM.\normalsize}
		\label{figure:psc_competition_for_fv_binding_sites}
	\end{figure}

In addition to competing to bind fXa, PSC and fV species compete to occupy fV-specific binding sites on platelets. Thus, PSC can inhibit the binding of fV species to platelet surfaces by occupying binding sites that would otherwise be available to them, reducing the concentrations of platelet-bound fV species, and thus inhibiting prothrombinase formation and thrombin production. We find that PSC-fV competition for fV-specific platelet binding sites only mildly inhibits thrombin production. Fig. \ref{figure:psc_competition_for_fv_binding_sites} shows the impact of this competition on thrombin production by comparing thrombin time-courses and thrombin metrics for baseline parameters with those from experiments where we remove the inhibitory effect of PSC bound to fV-specific binding sites on platelets by allowing fV species to bind at fV-specific binding sites even if those sites are already occupied by PSC.

Figures \ref{figure:psc_competition_for_fv_binding_sites} A - C show that for low [PSC], PSC-fV competition for fV binding sites on platelets has relatively little effect on thrombin time courses. However, for physiological plasma [PSC] ($\sim$ 0.2 - 0.5 nM, panel A), eliminating competition enhances thrombin production, shifting thrombin time courses to the left for plasma [PSC] 0 - 0.3 nM, creating a thrombin burst for [PSC] = 0.4 nM, and increasing [Thrombin] at each time for [PSC] = 0.5 nM and $[\text{TF}]_\text{s}$=1.5 fmol/cm$^2$. Panels B and C show that for all $[\text{TF}]_\text{s}$, removing competition leads to a relatively modest increase in [Thrombin]$_{10}$  (largest for $[\text{TF}]_\text{s}$ $\sim$ 3 fmol/cm$^2$) and a relatively modest decrease in $t_{lag}$ (largest for $[\text{TF}]_\text{s}$ $\sim$ 2 fmol/cm$^2$).

Examining the concentrations of species which occupy fV binding sites on platelets, we see in Fig. \ref{figure:psc_competition_for_fv_binding_sites} D-E that early during coagulation, fVh in PSC (gray curves) occupies far more fV binding sites than do PRO/PROh (blue curves). For plasma [PSC] sufficiently low to allow a thrombin burst, only well after the start of the thrombin burst (panel D) do PRO/PROh catch up. For higher plasma [PSC], the PRO/PROh concentrations slowly increase and then decrease, with PRO/PROh concentrations never exceeding 0.01 nM (panel E). However, panel F shows that removing competition allows PRO/PROh concentration to increase rapidly, eventually producing a thrombin burst. Thus, PSC occupies binding sites that could otherwise go to PRO/PROh production, delaying thrombin production.

\subsection{Pathological clotting conditions}
\label{subsection:extreme_conditions}
Simulations suggest that the inhibitory functionality of PSC may play a critical role in several pathological clotting conditions. Excess PSC may lead to bleeding in individuals with the east Texas fV variant (Section \ref{subsection:excess_psc}). Reduced PSC also may explain the surprisingly mild bleeding phenotype associated with severe fV deficiency and the absence of a bleeding phenotype associated with mild fV deficiency (Section \ref{subsection:factor_v_deficiencies}). Finally, simulations suggest that PSC removal may reduce bleeding in individuals with hemophilia A (Section \ref{subsection:hemophilia_a}).

\subsubsection{Excess PSC significantly inhibits thrombin production}
\label{subsection:excess_psc}
We find that excess PSC dramatically inhibits thrombin production. Fig. \ref{figure:extreme_conditions} A - C show the impact of excess PSC on thrombin time-courses and thrombin metrics across a range of $[\text{TF}]_\text{s}$ values. Panel A shows that a mild excess of plasma PSC (0.5 - 0.75 nM plasma PSC) mildly delays the thrombin time course for $[\text{TF}]_\text{s} = 1.5$ fmol / cm$^2$. More severe excesses of plasma PSC (1.5 - 5 nM) prevent the thrombin burst with higher plasma [PSC] leading to lower [Thrombin] at each time. Panels B - C show that for plasma [PSC]=5 nM, ten times the normal concentration, as expected in the east Texas bleeding disorder, there is no thrombin burst for $[\text{TF}]_\text{s}$ $\sim $ 1.5 - 10 fmol/cm$^2$ and the thrombin metrics ($t_{lag}$ and [Thrombin]$_{10}$) dramatically worsen for relatively high values $[\text{TF}]_\text{s}$ for which a thrombin burst still occurs.

\subsubsection{Thrombin production in factor V deficiencies is restored by plasma PSC deficiencies}
\label{subsection:factor_v_deficiencies}
Fig. \ref{figure:extreme_conditions} D - I show that concurrent PSC deficiencies may be the cause of the surprisingly mild bleeding phenotypes associated with fV deficiencies. They show the impact on thrombin time-courses and thrombin metrics of fV deficiencies with and without concurrent PSC deficiencies across a range of $[\text{TF}]_\text{s}$. To simulate fV deficiencies, we reduce the plasma zymogen [fV] and the copy number of fV and fVh in platelets to 1\% of normal for severe deficiency (0.2 nM fV in the plasma, 15 molecules each of fV and fVh in platelets) or to 10\% of normal for mild deficiency (2 nM fV in plasma, 150 molecules of each fV and fVh in platelets). Simulations use plasma [PSC] corresponding to the baseline (0.5 nM), mild PSC deficiency (0.05 nM), or severe PSC deficiency (0.005 nM). 

Panels D - F show that severe fV deficiency eliminates thrombin bursts for a wide range of $[\text{TF}]_\text{s}$. However, reducing [PSC] shifts thrombin time courses back to the left (panel D), greatly decreases $t_{lag}$ (panel F), and largely restores [Thrombin]$_{10}$ (panel E). Specifically, panel D shows that for $[\text{TF}]_\text{s} = 9$ nM, reducing [PSC] to 0.1 nM restores the thrombin burst, and completely removing PSC further enhances thrombin production. Across $[\text{TF}]_\text{s}$, a concurrent severe PSC deficiency restores a thrombin burst for $[\text{TF}]_\text{s}$ between $\sim$ 2 and 15 fmol/cm$^2$, whereas reducing PSC concentrations to 10\% of normal (0.05 nM) restores thrombin bursts for a smaller but still large range of $[\text{TF}]_\text{s} \sim$ 8 - 15 fmol/cm$^2$ (panel F). 

Panels G - I show that a mild fV deficiency without a concurrent PSC deficiency also leads to significantly less thrombin production, reducing [Thrombin] at each time (panel G), reducing [Thrombin]$_{10}$ for $[\text{TF}]_\text{s}$ $\gtrapprox$ 12 fmol/cm$^2$ (panel H), preventing a thrombin burst for [TF]$_\text{s}$ $\sim$ 1.5 - 6.5 fmol/cm$^2$) and otherwise increasing thrombin lag times (panel I). However, concurrently reducing [PSC] to values as large as 0.2 nM restores a thrombin burst (panel G), and a concurrent PSC deficiency (0.05 nM, 10\% of normal) produces thrombin metrics comparable to normal blood (gray versus black curves in panels H and I). 

\subsubsection{Thrombin generation in hemophilia A is restored by plasma PSC deficiencies for severe injuries}
\label{subsection:hemophilia_a}
\begin{figure}[tp]

        \includegraphics[width = \textwidth]{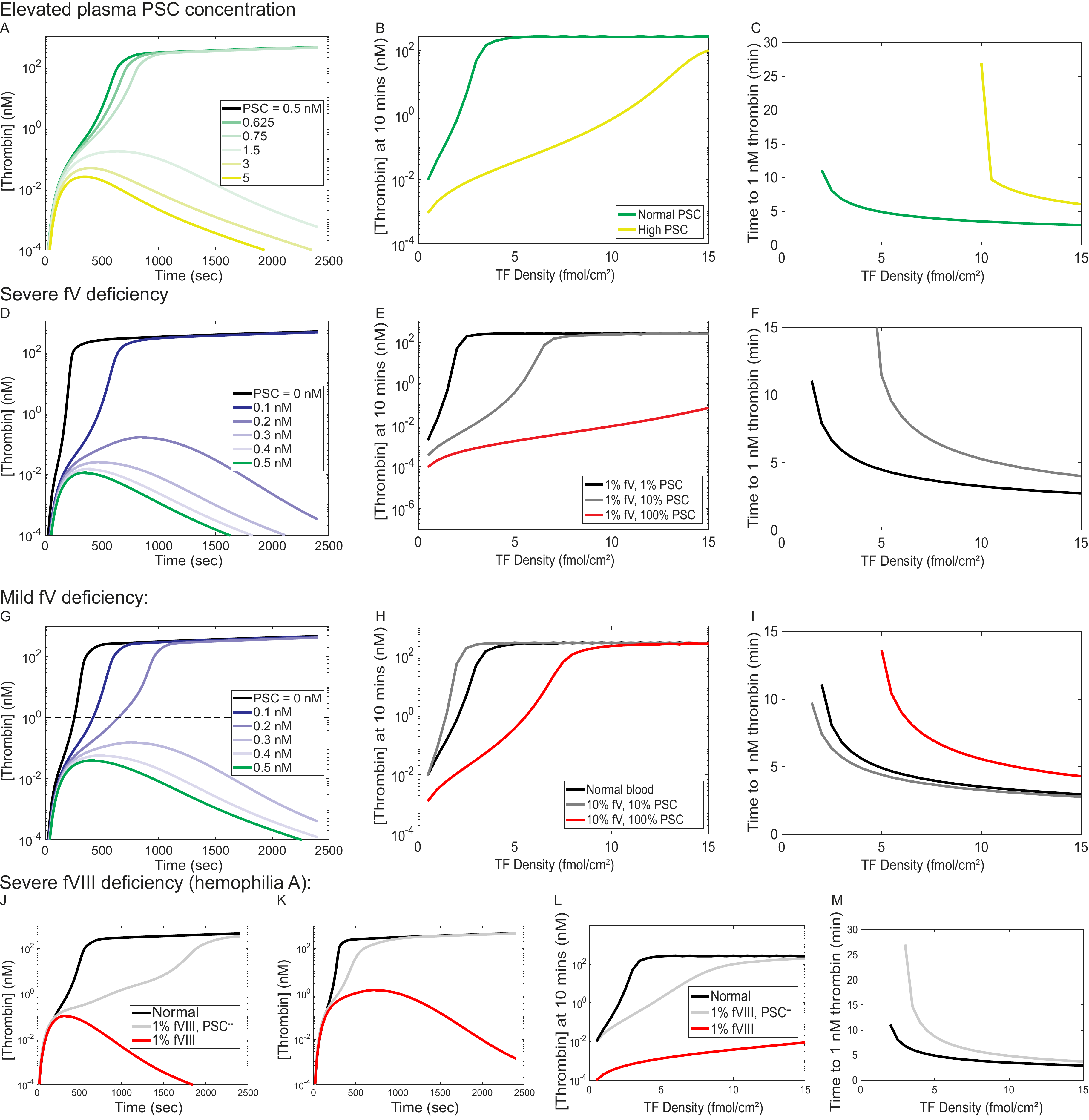}
		  \caption{\footnotesize \textbf{Extreme Conditions.} (A-C) Elevated [PSC]: (A) For $[\text{TF}]_{\text{s}}=$3 fmol/cm$^2$, increasing plasma [PSC] from a high physiological level ($\approx$ 0.5 nM) delays thrombin time courses for [PSC] $\le$ 0.75 nM and prevents a thrombin burst for 15 nM $>$ [PSC] $>$ 5.0 nM. Across a range of $[\text{TF}]_{\text{s}}$ values, [PSC] = 5 nM (as in the east Texas bleeding disorder) B) strongly reduces [Thrombin]$_{10}$ and (C) increases $t_{lag}$ or eliminates the thrombin burst altogether.  (D-F) Severe fV deficiency, 1\% normal fV species:  (D) For $[\text{TF}]_{\text{s}}=$9 fmol/cm$^2$, there is no thrombin burst for [PSC] $\geq$ 0.2 nM, but a burst occurs for [PSC] $\leq$ 0.1 nM. Across a range of $[\text{TF}]_{\text{s}}$ values, severe fV deficiency strongly reduces [Thrombin]$_{10}$ (E) and prevents thrombin from reaching 1 nM ($t_{\text{lag}} > 40$ min), but concurrently reducing PSC to 10\% and more so to 1\% of normal returns [Thrombin]$_{10}$ and $t_{\text{lag}}$ closer to normal. (G-I) Mild fV deficiency, 10\% normal fV species: For $[\text{TF}]_{\text{s}}=$ 3 fmol/cm$^2$,  (G) there is no thrombin burst for [PSC] $\ge$ 0.3 nM.  A burst occurs for [PSC] $\le$ 0.2 nM with burst occurring earlier with further reductions in [PSC]. Across a range of $[\text{TF}]_{\text{s}}$ values, mild fV deficiency (10\% fV and 100\% PSC) reduces [Thrombin]$_{10}$ (H) and increases $t_{\text{lag}}$ (I), preventing thrombin bursts for $[\text{TF}]_{\text{s}}$ near 3 - 6 fmol/cm$^2$. Concurrently reducing PSC to 10\% of normal produces [Thrombin]$_{10}$ and $t_{\text{lag}}$ close to those of normal blood. (J-M) Severe fVIII deficiency, 1\% normal fVIII:  (J) For $[\text{TF}]_{\text{s}}=$3.5 fmol/cm$^2$, severe fVIII deficiency greatly reduces thrombin production. This is partly ameliorated when PSC is also absent. (K) For $[\text{TF}]_{\text{s}}=10$ fmol/cm$^2$, severe fVIII deficiency reduces thrombin production.  Thrombin production is close to normal when PSC is also absent. Across a range of $[\text{TF}]_{\text{s}}$ values, eliminating PSC returns very limited thrombin production for hemophilia A (1\% fVIII) blood closer to that of normal blood both in terms of (L) [Thrombin]$_{10}$ and (M) $t_{\text{lag}}$.
         \normalsize}
          \label{figure:extreme_conditions}

    \end{figure}

We find that lowering plasma [PSC] helps restore thrombin production in severe hemophilia A, simulated by setting the plasma [fVIII] to 1\% of its normal value. Fig. \ref{figure:extreme_conditions} J - M show thrombin time courses and thrombin metrics for severely fVIII-deficient blood, with and without plasma PSC, versus normal blood, for a range of $[\text{TF}]_\text{s}$.  Panel J shows that for 
$[\text{TF}]_\text{s}$ = 3.5 fmol/cm$^2$, normal blood achieves a thrombin burst at around 10 min, but hemophilia A blood with normal [PSC] leads to a peak [Thrombin] that is only 0.1 nM.  Removing PSC in the latter case leads to a gradual growth in [Thrombin], with it reaching almost normal levels after 40 min.  For 
$[\text{TF}]_\text{s}$ = 10 fmol/cm$^2$ (panel K), removal of PSC from severely fVIII-deficient blood can bring thrombin generation close to normal.
Panel L shows that for all tested $[\text{TF}]_\text{s}$ values, removing PSC from the plasma increases [Thrombin]$_{10}$ by a factor of $\sim 100$ or more for severely fVIII-deficient blood, but that [Thrombin] only reaches near normal levels by 10 minutes for $[\text{TF}]_\text{s} \gtrapprox 12$ fmol/cm$^2$. 
Panel M shows that removing PSC leads to [Thrombin] eventually attaining 1 nM for $[\text{TF}]_\text{s}$ $\gtrapprox 3.5$ fmol/cm$^2$.  

\subsubsection{Mechanisms by which PSC inhibits thrombin in selected pathological clotting conditions}
\begin{figure}[H]

        \includegraphics[width = \textwidth]{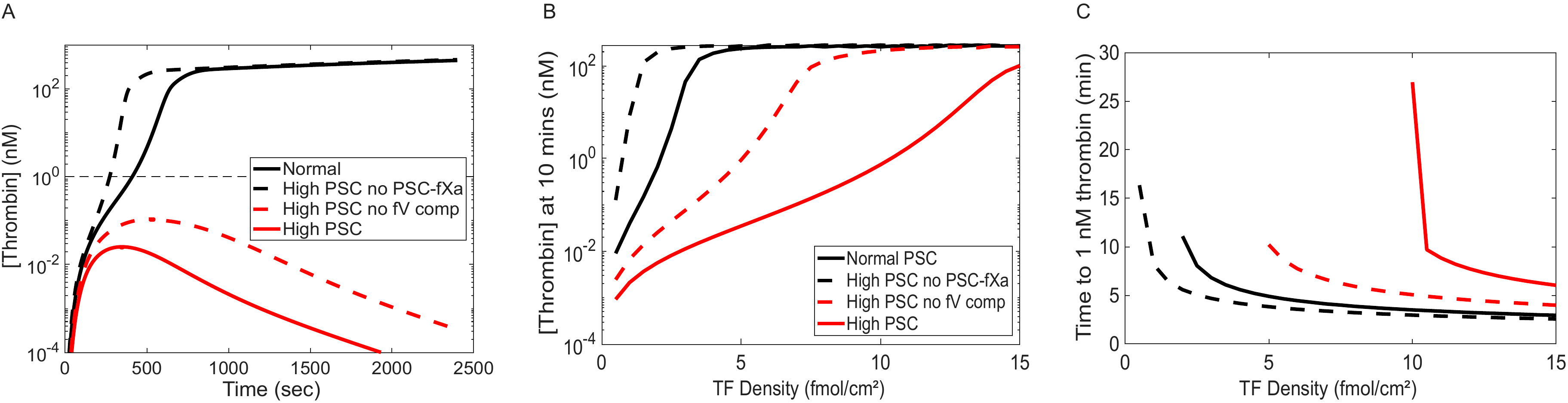}
        \vskip1ex
        \caption{\footnotesize \textbf{Mechanisms of thrombin inhibition by PSC given plasma PSC concentrations 10-fold normal.} 
        Panel (A) respectively compares the time-course of thrombin production given high plasma PSC concentrations (\textcolor{red}{-}) of 5 nM, 10-fold normal; high PSC but where fV species can bind to fV-specific bindings sites on platelets already occupied by PSC (\textcolor{red}{- -}); high PSC but where PSC does not bind fXa species at all (- -); and normal plasma PSC concentration of 0.5 nM (-). Panels (B) and (C), respectively compare [Thrombin]$_{10}$ at 10 minutes and t$_{\text{lag}}$ given the same situations as in (A). Shear rate is 100 1/s for all panels.\normalsize}
        \label{supplemental_figure:mechanisms_high_psc}
        \end{figure}

In the four pathological clotting conditions investigated, PSC inhibits thrombin production mainly by directly binding with fXa (not shown). However, given 10-fold normal plasma [PSC], PSC-fV competition for fV binding sites plays an additional large role (Fig. \ref{supplemental_figure:mechanisms_high_psc}). Fig. \ref{supplemental_figure:mechanisms_high_psc} A shows that for $[\text{TF}]_\text{s}$= 1.5 fmol/cm$^2$, the PSC-fV competition shifts thrombin time courses to the right (solid red curve versus dashed red curve), and PSC-fXa interactions prevent the thrombin burst altogether (dashed black curve). Fig. \ref{supplemental_figure:mechanisms_high_psc} B - C show that with normal PSC-fXa interactions, the PSC-fV competition prevents a thrombin burst for $[\text{TF}]_\text{s}$ between $\sim 5$ and $\sim 11$ fmol/cm$^2$ (panel C, solid red versus dashed red curves), strongly increases the lag time for $[\text{TF}]_\text{s} \gtrapprox 11$ fmol/cm$^2$, and substantially reduces [Thrombin]$_{10}$ for $[\text{TF}]_\text{s}$ up to $\sim 15$ fmol/cm$^2$ (panel B, solid red versus dashed red curves). However, without intact PSC-fXa interactions, the anticoagulant effect of PSC-fVh competition is sufficient only to inhibit thrombin production as if PSC were present in plasma at concentrations smaller than 0.5 nM.

\subsection{Competing effects of free PS}
\label{subsection:free_ps_overall}

\begin{figure}[h!]

			\includegraphics[width = \textwidth]{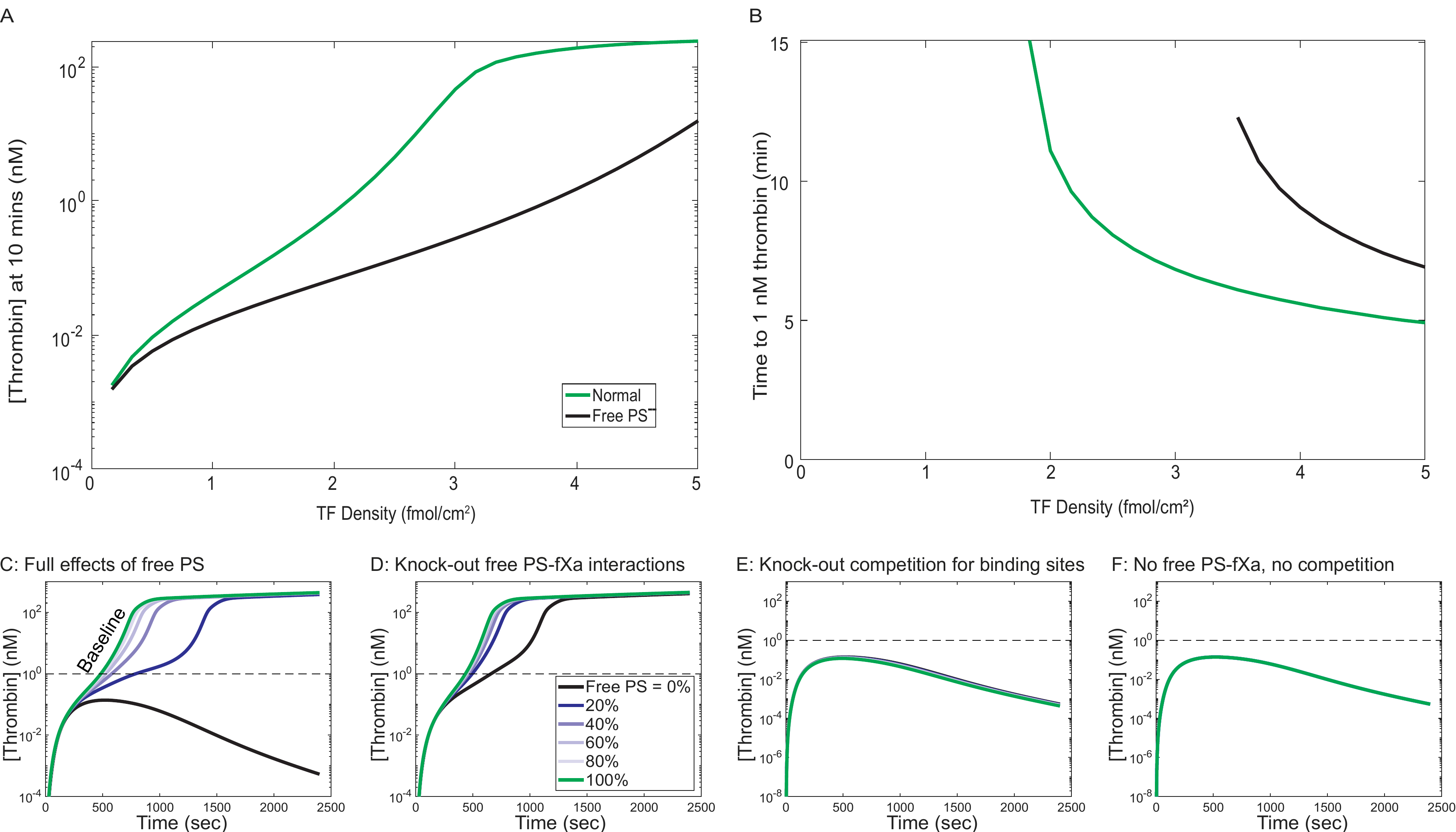}
			
            \vskip1ex
            \caption{\footnotesize \textbf{The overall effect of free protein S.} 
            (A) [Thrombin]$_{10}$ and (B) $t_{\text{lag}}$  in the presence and absence of free PS. For [TF]$_\text{s}$ = 2.5 fmol/cm$^2$, panels (C) - (F) each show time-courses of thrombin for free PS concentrations in the plasma ranging from 0\% to 100\% of normal (160 nM) and with the normal copy number of free PS (5300 PS molecules released per platelet upon activation). (C) Normal blood, (D) no free PS fXa binding, (E) no competition between PSC and free PS for PS-specific binding sites on platelets, and (F) no free PS fXa binding and no competition for binding sites. Shear rate is 100 1/s.
            \normalsize}
			\label{figure:free_ps_overall}
		\end{figure}

Free PS in the plasma has two direct effects in our model. First, it competes with PSC for PS-specific binding sites on platelets, enabling thrombin production by reducing PSC concentration in the injury zone. Second, it directly, but weakly ($\text{K}_{\text{D}}$ = 100 nM), binds and inhibits fXa and fXam, an inhibitory effect. We find that the first of these dominates so that the net effect of free PS is procoagulant. Fig. \ref{figure:free_ps_overall} shows the impact of free PS on thrombin production by comparing time courses and thrombin metrics across $[\text{TF}]_\text{s}$ values in simulations with free PS, without free PS, without free PS binding and inhibiting fXa, or without competition between free PS and PSC for platelet binding sites. The last is achieved by allowing PSC to bind to PS-specific binding sites on platelets that are already occupied by a free PS.

We see that increasing the free PS concentration in the plasma has a moderate procoagulant effect across tissue factor densities. Comparing thrombin metrics for normal blood versus blood without free PS (panels A and B) shows that free PS elevates [Thrombin]$_{10}$ across $[\text{TF}]_\text{s}$ values, prevents a thrombin burst for $[\text{TF}]_\text{s}$ between $\sim 1.75 - 3.5$ fmol/cm$^2$, and decreases $t_{lag}$ for larger $[\text{TF}]_\text{s}$ values. Panel C shows that higher free PS concentration leads to higher [Thrombin] at each time and can lead to a thrombin burst for $[\text{TF}]_\text{s} = 2.5$ fmol/cm$^2$. Panels D - F show the effects of knocking out either or both of the previously mentioned roles of plasma free PS for $[\text{TF}]_\text{s} = 2.5$ fmol/cm$^2$. Without free PS - fXa interactions (panel D), free PS increases [Thrombin] at each time, demonstrating the procoagulant effect of competition between free PS and PSC for PS binding sites on platelets. On the other hand, without the procoagulant effect of free PS competing with PSC for PS-specific binding sites on platelets (panel E), free PS hardly impacts [Thrombin], demonstrating the marginal impact of direct inhibition of fXa by free PS. Panel F shows that without either of these two roles, free PS has essentially no effect on thrombin production, confirming that these two effects are the only ways by which free PS affects [Thrombin] in our model.

\subsubsection{Binding between free PS and fXa has little effect on thrombin inhibition}
\label{subsection:free_ps_fxa_binding_affinity}

\begin{figure}[h!]
	\includegraphics[width = \textwidth]{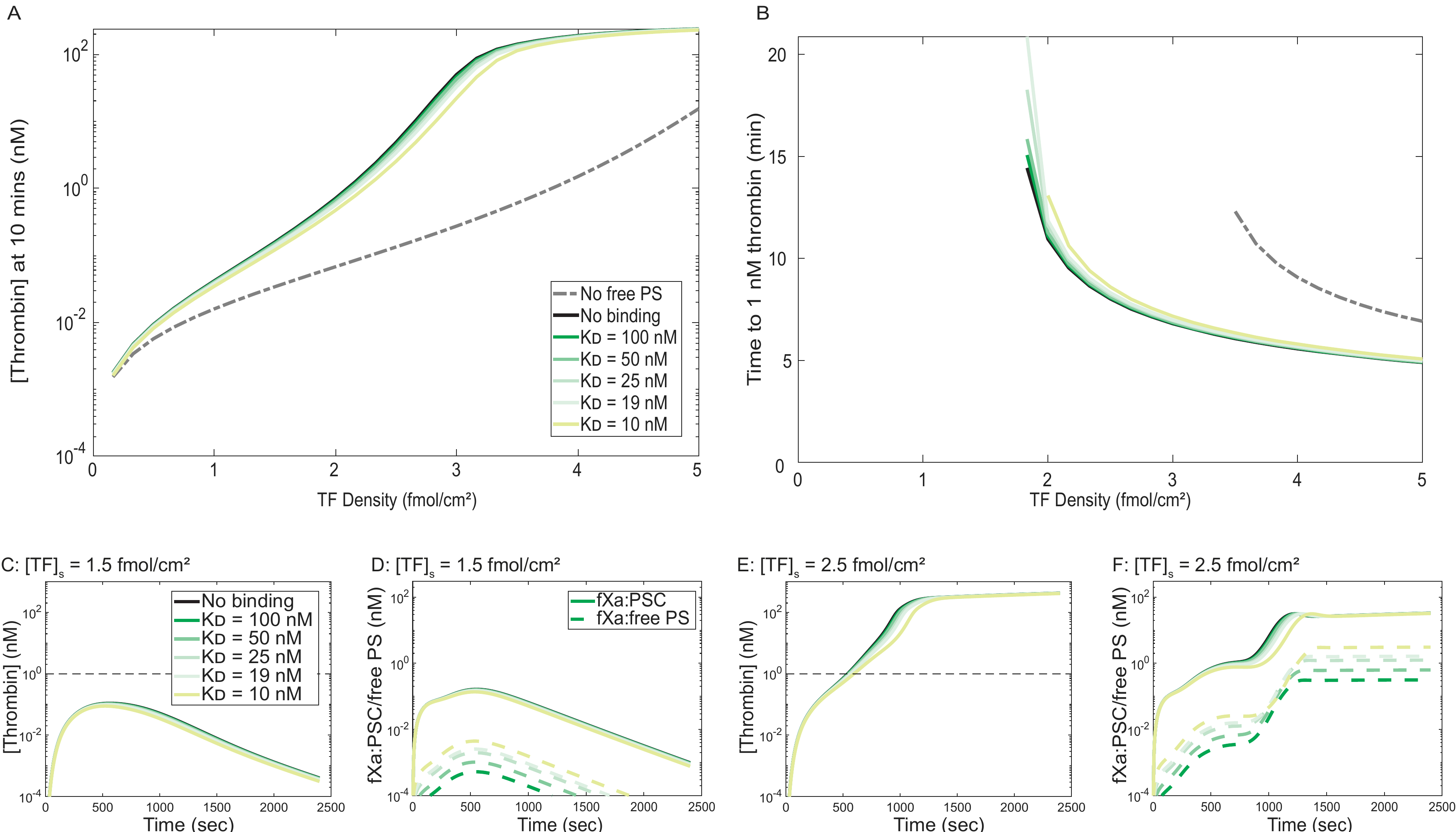}
    \vskip1ex
    \caption{\footnotesize \textbf{Effects of changing Free PS - fXa binding strength.} (A)[Thrombin]$_{10}$ and (B) $t_{lag}$ for a variety of $\text{K}_{\text{D}}$ values for free PS binding to fXa, as well as for the case where free PS is absent. Time courses of (C) thrombin and (D) fXa bound to PSC and fXa bound to free PS for [TF]$_s$ = 1.5 fmol/cm$^2$ and the same $\text{K}_{\text{D}}$ values for PS binding to fXa as in (A-B). Panels (E-F) use [TF]$_s$ = 2.5 fmol/cm$^2$ but are otherwise the same as panels (C-D). All panels use a shear rate of 100 1/s. 
    \normalsize}
	\label{figure:free_ps_fxa_binding_affinity}
\end{figure}

We find that free PS is net procoagulant even if we greatly strengthen the binding affinity between free PS and fXa from baseline ($\text{K}_{\text{D}} = 100$ nM). Fig. \ref{figure:free_ps_fxa_binding_affinity} shows the impact on thrombin time-courses and thrombin metrics across $[\text{TF}]_\text{s}$ values of varying the $\text{K}_{\text{D}}$ for free PS - fXa binding from no binding to 10 nM. Notably, a $\text{K}_{\text{D}}$ in this range (19 nM) was reported in \cite{heeb1994protein}, this is about 5 times lower than the baseline $\text{K}_{\text{D}}$ we use in this paper. 

Simulations further show that free PS - fXa binding hardly impacts thrombin production. Panels A - B show that increasing the $\text{K}_{\text{D}}$ for free PS binding to fXa (A) slightly lowers [Thrombin]$_{10}$, (B) can prevent a thrombin burst only for $[\text{TF}]_\text{s}$ $\sim 2$ fmol/cm$^2$, and otherwise can slightly increase the lag time. Panels C and E show that if a thrombin burst occurs, increasing the $\text{K}_{\text{D}}$ slightly shifts thrombin time-courses to the right ($[\text{TF}]_\text{s}$ = 2.5 fmol/cm$^2$), and if no thrombin burst occurs slightly decreases [Thrombin] at each time. Panels D and F show that even when the $\text{K}_{\text{D}}$ exceeds the 19 nM estimate from \cite{heeb1994protein}, the concentration of fXa bound to free PS is several orders of magnitude smaller than that of fXa bound to PSC. Therefore, free PS likely binds insufficiently large concentrations of fXa to have a strong effect on thrombin production.

\section{Discussion}
\label{section:discussion}
Considering recent evidence that TFPI$\alpha$ circulates in plasma in a complex (PSC) with PS and fVh, we have updated a previously validated model of coagulation to include PSC circulating in the plasma in place of TFPI$\alpha$. We have further extended the model to include free PS in the plasma, as well as free PS and TFPI$\alpha$ released by platelets into plasma near the injury site. Considering recent unpublished evidence, (unpublished, J.A), that PS does not enhance the ability of TFPI$\alpha$ to inhibit the TF-fVIIa complex, we set the rate constants for PSC binding to TF-fVIIa on the subendothelium to be those of TFPI$\alpha$. Our results suggest novel mechanisms by which PSC can inhibit thrombin, and provide quantitative information about literature predictions concerning PS, TFPI$\alpha$, and fVh.

\begin{figure}[h!]
        
		\includegraphics[width = \textwidth]{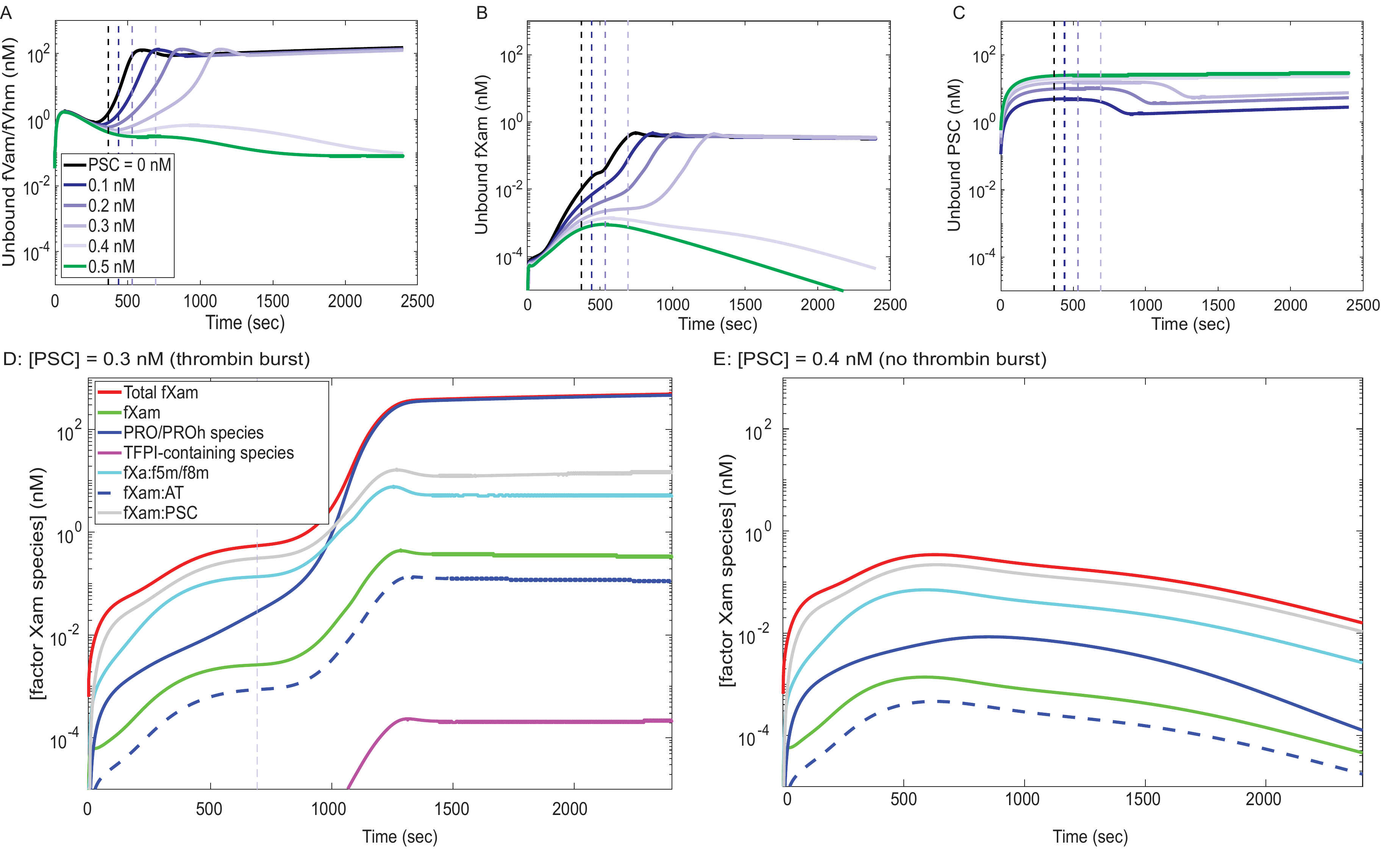}
        \vskip1ex
        \caption{\footnotesize \textbf{Figure PSC competes with fVam and fVhm to bind fXam.} Concentrations of activated and partially activated membrane-bound species (A) fVam and fVhm, (B) unbound fXam, and (C) PSC species (PSC and PSCm) for a variety of plasma PSC concentrations illustrate the competition between fVam/fVhm and PSC for fXam. Dashed vertical lines indicate the times at which [Thrombin] reaches 1 nM given the corresponding PSC level (matching colors). Concentrations of various species which contain fXam (fXam, species which contain PRO/PROh, fXam and PSC, fXam and TFPI but not PSC, fXam bound to antithrombin, or fXam bound to fVm or fVIIIm) illustrate which species dominate the competition for fXam for (D) plasma [PSC] = 0.3 nM which allows a thrombin burst or (E) given [PSC] = 0.4 nM which prevents a thrombin burst. In (A) - (D), the vertical dashed lines indicate the times at which [Thrombin] reaches 1 nM. In (D) and (E), solid blue lines include PRO, PROh, membrane-bound thrombin bound to PRO, membrane-bound thrombin bound to PROh, and fVam:PROh; solid pink lines include TFPI$\alpha$:fXam, and fXam bound to TFPI$\alpha$ bound in turn to fVh or fVhm; solid gray lines include fXam bound to a membrane-bound PSC. Shear rate is 100 1/s for all panels. \normalsize}
		\label{figure:psc_competes_with_fvh_to_bind_fxa}
	\end{figure}

For example, results support the hypothesis (see \cite{dahlback2022hydrophobic}) that PSC's main anticoagulant role is to inhibit fXa when fXa levels are low and procoagulant activity is limited. Specifically, our results show that the addition of PSC at an injury site ceases to be relevant once a thrombin burst has progressed sufficiently (Fig. \ref{figure:psc_only_matters_shortly_after_the_initiation}), likely because fVa and fVh out-compete PSC to bind fXa and because species containing fV out-compete PSC for fV-specific binding sites on platelets, limiting growth of the PSC concentration (Fig. \ref{figure:psc_competes_with_fvh_to_bind_fxa}). In experiments in which PSC was added only when the thrombin concentration reached a target value, we saw that PSC acts very quickly to bind low concentrations of fXa. In particular, PSC quickly binds as much fXa as if PSC were present since the initiation of coagulation, producing similar, very effective, levels of thrombin inhibition. 

In fact, our results indicate that PSC strongly inhibits thrombin production (Fig. \ref{figure::ps_psc_impact}), whereas contrary to our earlier findings, (e.g., \cite{miyazawa2023i}), TFPI$\alpha$ would hardly inhibit thrombin production if it circulated in the plasma in place of PSC (Fig. \ref{supplemental_figure:tfpi}). While TFPI$\alpha$ has advantages over PSC, namely, TFPI$\alpha$ can directly bind fVh by itself or in PROh, whereas PSC cannot, PSC has advantages over TFPI$\alpha$ as well. Namely, the PS and fVh in PSC enhance TFPI$\alpha$'s ability to bind and inhibit fXa (Fig. \ref{figure:psc_competes_with_fvh_to_bind_fxa}) up to 50-fold - 100-fold better than TFPI$\alpha$ can alone \cite{dahlback2018factor, gierula2020anticoagulant} (see \cite{santamaria2017factor}, \cite{ahnstrom2012identification}, \cite{hackeng2006protein}, \cite{huang1993kinetics}). This is largely because the ability of TFPI$\alpha$ to bind and inhibit fXa is relatively weak, (see \cite{santamaria2017factor}, \cite{ahnstrom2012identification}, \cite{hackeng2006protein}, \cite{huang1993kinetics}) compared to the nearly 100-fold stronger dissociation constant of 0.0271 nM from \cite{baugh1998regulation} we used earlier \cite{miyazawa2023i}. Results suggest that without this enhanced ability of PSC to bind and inhibit fXa, PSC's effect on thrombin production (Fig. \ref{figure:psc-e10_binding_affinity}) is similar to that of TFPI$\alpha$ at the same plasma concentration. Results also suggest that different plausible enhancements of the $\text{K}_{\text{D}}$ for PSC-fXa binding versus the $\text{K}_{\text{D}}$ for TFPI$\alpha$-fXa binding ($\sim 5$ nM) could lead to very different levels of thrombin inhibition, with a $\text{K}_{\text{D}} \sim 0.5$ nM leading to considerably more inhibition of thrombin production than would a $\text{K}_{\text{D}}$ of $\sim 5$ nM, and likewise that reducing the $\text{K}_{\text{D}}$ to 0.1 nM versus 0.5 nM, to 0.05 nM versus 0.1 nM, and to 0.01 nM versus 0.05 nM each leads to far stronger thrombin inhibition. To understand the true impact of PSC on thrombin production, our results therefore suggest it is critical to precisely determine the binding affinity between PSC and fXa.

In addition to strongly binding fXa, results show that PSC may moderately inhibit thrombin production by occupying fV-specific binding sites on platelets. The mechanism is that PSC therefore prevents procoagulant species, fVa and fVh, from binding to platelets, limiting thrombin production across a wide range of injury severity (Fig. \ref{figure:psc_competition_for_fv_binding_sites}), by limiting prothrombinase production. To our knowledge, this competition effect has not been investigated before.  

Yet another means by which PSC so effectively inhibits thrombin production lies in its ability to bind to phospholipid surfaces on platelets. Perhaps the most novel of our results, we find that PSC rapidly accumulates on platelets to $\sim 50$ times higher concentration than in the plasma (Fig. \ref{figure:psc_significantly_accumulates}). Without the ability to bind to and thus accumulate on platelet surfaces, PSC has little inhibitory effect on thrombin production for a wide range of TF exposure levels. 

There are two means by which the PS in PSC can augment PSC's ability to bind to platelet surfaces: PSC can directly bind to platelets via its PS, and PS helps ensure that TFPI$\alpha$ and fVh arrive ``pre-packaged'' at the injury site by enhancing TFPI$\alpha$-fVh binding interactions (\cite{dahlback2022hydrophobic, gierula2023c_terminal}. Fig. \ref{figure:psc_significantly_accumulates} F shows that it is critical that TFPI$\alpha$ and fVh come ``pre-packaged'': TFPI$\alpha$ and fVh pre-packaged (in PSC where the PS does nothing but hold the complex together) accumulate to high concentrations, $\sim$10-fold the concentrations of TFPI$\alpha$:fVh if TFPI$\alpha$ and fVh must first encounter each other in the injury zone. As a result the TFPI$\alpha$ in PSC can compete far more effectively with fV species to bind and inhibit fXa, thereby more effectively inhibiting thrombin production.

Surprisingly, but in line with speculation presented in \cite{gierula2023c_terminal}, the ability of PSC to bind to platelets via its PS is relatively \emph{unimportant} for inhibiting thrombin production (Fig. \ref{figure:psc_significantly_accumulates}). Instead, PSC localizes to platelets mainly via its fVh rather than its PS. This is largely because PSC must compete with the far more concentrated free PS in the plasma to bind PS-specific binding sites on platelets.

\begin{figure}[H]

    \includegraphics[width = \textwidth]{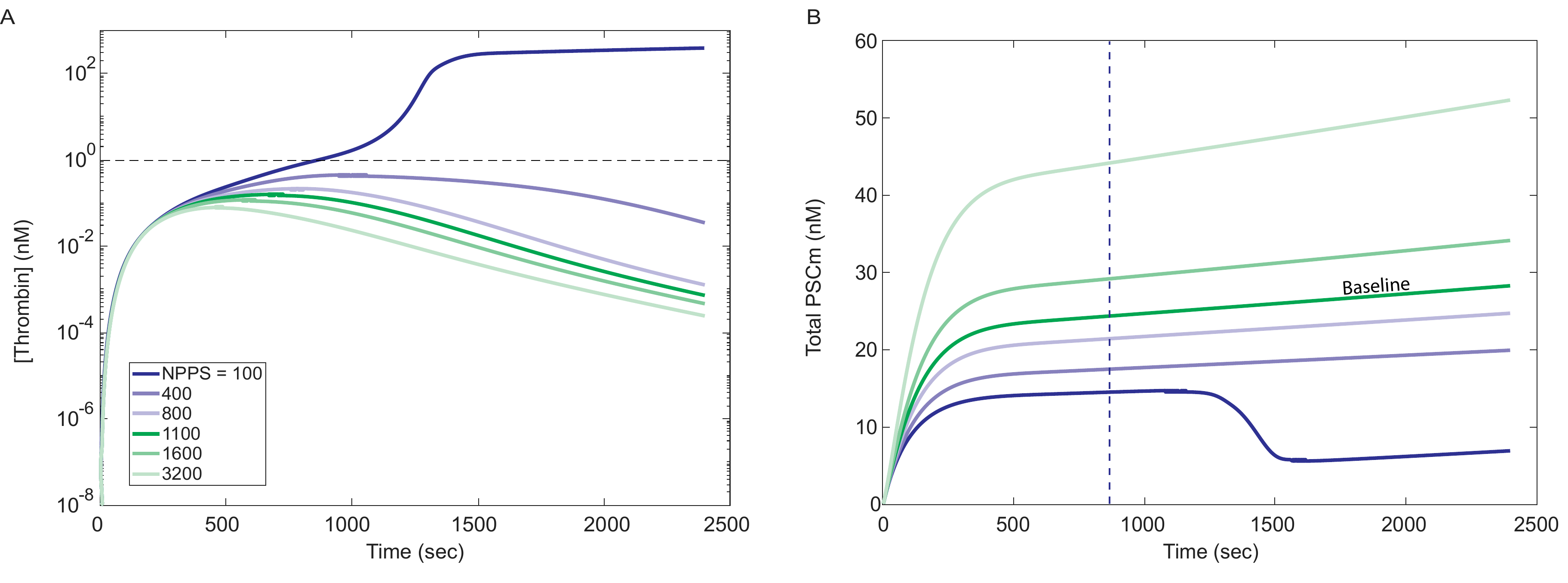}
    \vskip1ex
    \caption{\footnotesize \textbf{PS binding sites on platelets.} Varying the number (NPPS) of binding sites on platelets for PS around its expected physiological value ($400 - 1100$ nM) modestly impacts thrombin time-courses (A) and PSC levels (B) for TF = $1.5$ fmol/cm$^2$. In (B), the vertical dashed line indicates the time at which [Thrombin] reaches 1 nM for the simulation with the corresponding color. Shear rate is 100 1/s in both panels. \normalsize}

    \label{supplemental_figure:ps_binding_sites_on_platelets}
    \end{figure}

In fact, results suggest that competition between free PS and PSC for PS-specific binding sites on platelets increases thrombin production for a wide range of TF exposure (Fig. \ref{figure:free_ps_overall}). In the absence of free PS (figure not shown), far more PS-specific binding sites on platelets are available for PSC, so PSC is much better able to attach to platelets via its PS (even better than via its fVh), and concentrations of membrane bound PSC grows dramatically. Increasing the number of PS-specific binding sites has a similar although less dramatic effect (Fig. \ref{supplemental_figure:ps_binding_sites_on_platelets}). In either of those two cases, PSC inhibits thrombin production to an even greater degree, and the PS portion of PSC becomes important for localizing PSC to phospholipid surfaces on platelet surfaces. However, any procoagulant effect of free PS has yet to be observed in lab experiments. Consequently, this finding concerning the procoagulant effect of competition between free PS and PSC needs experimental validation.

Outside of competing with PSC for PS-specific binding sites on platelets, free PS can affect thrombin production only by directly binding and inhibiting fXa. While there is some controversy over the value of $\text{K}_{\text{D}}$ for free PS binding to fXa, with \cite{heeb1994protein} reporting $\text{K}_{\text{D}} = 19$ nM, but e.g., \cite{ndonwi2008ps_needs_fxa} reporting that PS does not bind fXa, our results suggest that for $\text{K}_{\text{D}}$ ranging from no binding to 10 nM, free PS binding to fXa has a relatively small impact on thrombin production across a wide range of TF exposures. This is largely because free PS binds very little fXa (Fig. \ref{figure:free_ps_fxa_binding_affinity}). Further, for all tested $\text{K}_{\text{D}}$, the effect of competition between free PS and PSC to bind to PS-specific binding sites on platelets is greater than the direct inhibitory effect of free PS on fXa, thus giving free PS a net procoagulant effect. So, unless free PS binding to fXa proves much stronger than expected, free PS is unlikely to significantly inhibit thrombin production.

\subsection{Future directions}
\label{subsection:discussion:future_directions}
While results indicate that only a certain number of PS molecules may bind to a platelet at once \citep{stavenuiter2013platelet, mitchell1987cleavage}, it has yet to be shown that PS-specific binding sites exist. If it turns out that PS can bind to platelets in the same spot as fV or fX species, for example, then in addition to competing with each other to bind to platelets, PS and PSC would compete with fV and fX species as well. It is possible that the importance of PSC binding to platelets via its fVh, via its PS, or of free PS binding to platelets may change. How exactly they would change is unclear. We plan to address the importance of species-specific versus species-nonspecific binding sites on platelets in a forthcoming study concerning the importance of various platelet characteristics on blood coagulation.

\begin{figure}[h!]

		\includegraphics[width = \textwidth]{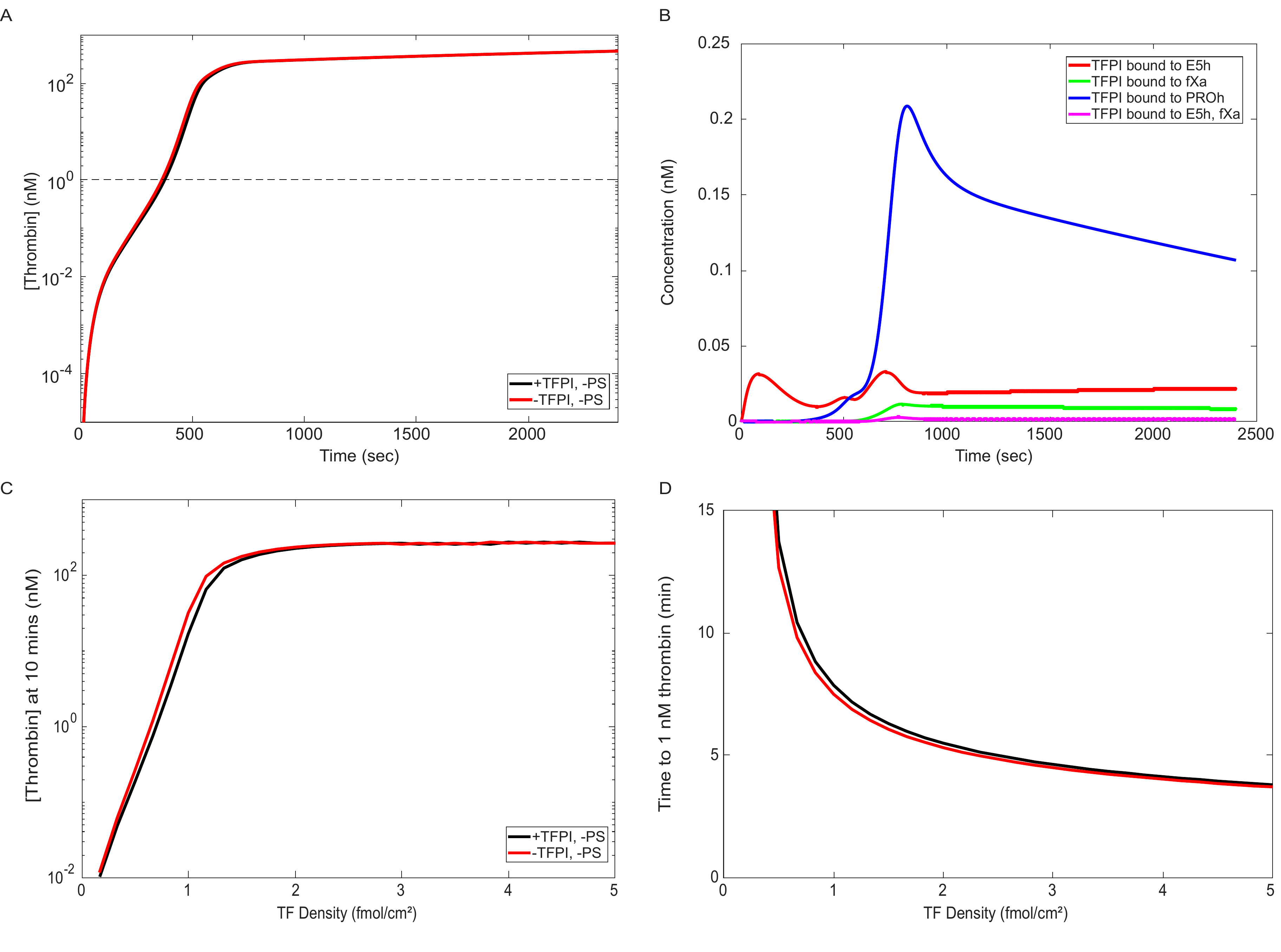}

        \vskip1ex
        \caption{\footnotesize \textbf{The overall effect of TFPI without Protein S.} Increasing the concentration of TFPI$\alpha$ in plasma and platelets from 0 to physiological levels ($\sim$0.5 nM TFPI$\alpha$ in the plasma, and 1200 TFPI$\alpha$ molecules in each platelet) has little effect on time-course of thrombin (panel A) and has little effect on inhibition of species containing factor Xa (panel B) at a tissue factor density of 1.5 fmol/cm$^2$. Introducing TFPI (black curves) has little effect on thrombin levels at 10 minutes across tissue factor levels (panel C). Introducing PS (black curves) also has little effect on the time until 1 nM thrombin is reached (lag time, panel D), at which point a rapid burst in thrombin production usually occurs. Shear rate is 100 1/s in all panels.\normalsize}
		\label{supplemental_figure:tfpi}
    \end{figure}

In addition to providing binding sites for coagulation proteins, platelets release a significant amount of free PS, in our model, and of TFPI$\alpha$ and fV species as well. However, we refrain from considering the formation of a PSC by platelet-released species due to incomplete information (see \cite{dahlback2022hydrophobic} and \cite{gierula2023c_terminal} for recent work). We find that blood washes away most platelet-released PS before it can bind to platelets or to fXa (not shown). An inconsequential concentration of free PS released by platelets therefore binds to platelet membranes and/or to fXa. Platelet-released TFPI$\alpha$ has a similarly small impact on thrombin production (Fig. \ref{supplemental_figure:tfpi}). Platelet-released fVh, on the other hand, strongly accelerates thrombin production (not shown). We plan to address the roles of platelet-released TFPI$\alpha$, PS, and fV species (including the make-up of platelet-released fV species) in more detail, as well as the potential of formation of PSC from platelet-released species, in more detail in a the previously mentioned forthcoming study. 

While we consider the roles of PS in enhancing inhibitory functions of TFPI$\alpha$ and in directly inhibiting fXa, we do not consider the role of free PS in enhancing activated protein C (APC). We plan to study the effects of PS-APC interactions in forthcoming work.

\subsection{Insight on several clotting diseases}
\label{subsection:discussion:bleeding_disorders}
Our model provides insight on several clotting diseases (Fig \ref{figure:extreme_conditions}). Results support the hypothesis that elevated fVh generated in the east Texas bleeding disorder is largely incorporated into PSC, leading to the excess bleeding associated with the disorder. Our results also support the hypothesis that the unexpectedly mild bleeding associated with fV deficiencies is due to a concurrent deficiency in PSC.  In our model, mild to severe PSC deficiencies largely rescue the lack of thrombin production for a severe fV deficiency, and a mild PSC deficiency completely restores the limited thrombin production associated with mild fVh deficiency, predicting the mild and normal bleeding phenotypes reported for patients with severe and mild fV deficiencies, respectively. Variations in plasma PSC concentrations among these individuals could therefore explain the variations in reported severity.

Further, PSC could be a target for reducing bleeding in bleeding disorders. For instance, blocking most of the inhibitory functionality of PSC likely would restore normal bleeding for both individuals with east Texas bleeding disorder and individuals with fV deficiency but no PSC deficiency. Further, results suggest that blocking the inhibitory functionality of PSC in individuals with hemophilia A could partially or even largely restore thrombin production, depending on injury severity. This is supported by recent experimental findings where inhibiting PS in the plasma strongly increased thrombin generation in hemophilia A (and hemophilia B) blood \cite{eladnani2025protein_s_drug_target_hem_a, wilson2024protein_s_antibody}. The drugs concizumab and marstacimab, which target the Kunitz 2 domain of TFPI \cite{pittman2022marstacimab, hansen2014concizumab, hilden2012concizumab}, presumably limiting the ability of TFPI to bind and inhibit fXa, would thus likely also limit the ability of PSC to bind and inhibit fXa and thus also its ability to inhibit thrombin production. Indeed, to block the inhibitory functionality of PSC, our model suggests that outside of directly blocking PSC formation, a good strategy could be to reduce the ability of PSC to bind fXa or to reduce the ability of PSC to bind to platelets and thereby accumulate at the injury site. Our model has shown that either of these changes largely eliminates the enhanced inhibition due to PSC in normal blood.

\section{Author Contributions} 
ALF, DMM, KBN, KL, and SFS acquired funding. AGG, ALF, JA, JTBC, and KL designed the model. AGG and ALF implemented the model. AGG performed simulations. AGG, ALF, JA, and JTBC analyzed and interpreted simulations.
AGG and ALF wrote the original paper draft. AGG, ALF, DMM, KBN, KL, JA, JTBC, and SFS edited paper drafts.

\section{Acknowledgments} This work was supported, in part, by National Institutes of Health grant R01HL151984 to SFS, DMM, KBN, ALF, and KL. This project was supported by the Health Resources and Services Administration (HRSA) of the U.S. Department of Health and Human Services (HHS) under grant number H30MC24049 Hemophilia Treatment Centers (SPRANS) to KBN. This information or content and conclusions are those of the author and should not be construed as the official position or policy of, nor should any endorsements be inferred by HRSA, HHS or the U.S. Government. JTBC was supported by a British Heart Foundation programme grant (RG/18/3/33405).  Portions of this work were conducted while ALF was visiting the Department of Immunology and Inflammation of Imperial College, London.  

\section{Declaration of Interests} JA has research funding from AstraZeneca Ltd that is unrelated to this study.  Other authors declare no competing interests.

\newpage
\bibliography{references}

\newpage
\supplementarysection
\section{Supplemental Material}

\subsection{Reaction diagram before the extension adding PS species}

\newpage
\subsection{TFPI by itself has little impact on thrombin}

\newpage
\subsection{Dip in PSCm vs time}

	\begin{figure}[H]

    \includegraphics[width = \textwidth]{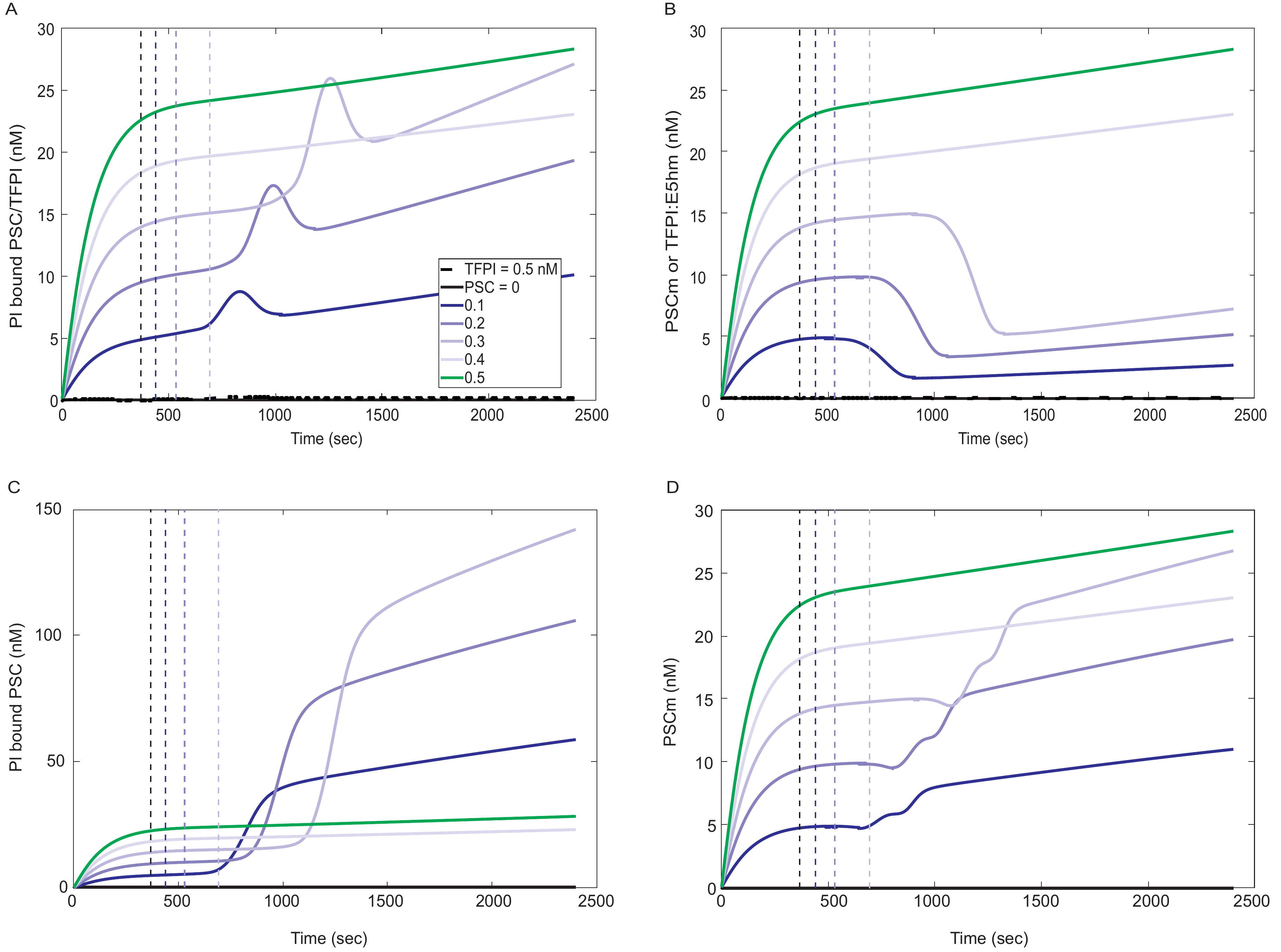}
  
    \caption{\footnotesize \textbf{Dip in PSC vs time.} When a thrombin burst occurs, the concentrations of all membrane-bound species containing PSC (namely membrane-bound PSC and PSC-fXa) undergo a transient, broad spike (panel A), whereas membrane-bound PSC (PSC bound to platelets via its fVh, its PS, or both) drops after the thrombin burst has begun (panel B). Allowing PSC to bind to PS-specific platelet-binding sites already occupied by PRO and vice-versa replaces the transient spike in total membrane-bound species containing PSC with a large, non-transient increase (panel C) and removes the drop in membrane-bound PSC (panel D).  In all panels, the vertical dashed lines indicate the times at which [Thrombin] reaches 1 nM, $[\text{TF}]_{\text{s}}$ = 1.5 fmol/cm$^2$ and shear rate is 100 1/s.\normalsize }
    \label{supplemental_figure:dip_in_psc_vs_time}
	
\end{figure}

\newpage
\subsection{Competition for fXam between PSC and fVam/fVhm}

\subsection{PS binding sites on platelets}

\subsection{Mechanisms of thrombin inhibition by PSC given plasma PSC concentrations 10-fold normal}

\subsection{New variables and notation used in the full equations}
Concentration of a species A is indicated with square brackets: [A]. z$_{\text{i}}$ and e$_{\text{i}}$ refer to zymogen i and enzyme i, respectively, and a complex of species A, B, C, ... is indicated by A:B:C:.... For example, the protein S complex is represented by ps:TFPI:e$_{\text{5h}}$. Parentheses indicate order of binding when otherwise unclear. For example, ((ps:TFPI:e$_{\text{5h}}$):e$_{\text{10}}$):e$_{\text{7}}^{\text{m}}$ indicates that PSC binds to fXa via its TFPI$\alpha$, which then binds to the TF:fVIIa complex. Note that fXa (e$_{\text{10}}$) only binds to PSC (ps:TFPI:e$_{\text{5h}}$) via the TFPI$\alpha$ in PSC, and that TF:fVIIa (e$_{\text{7}}^{\text{m}}$) does not bind to PSC; it instead binds to fXa already bound to PSC. Special names for species are:
            \begin{itemize}[noitemsep,topsep=0pt]
                \item Tissue factor = TF
                \item TF:fVIIa complex on the subendothelium = e$_{\text{7}}^{\text{m}}$
                \item Platelet-bound ‘tenase’ VIIIa:IXa = TEN
                \item Prothrombinase, with fVa:fXa, both bound to platelets = PRO 
                \item Prothrombinase with a fVh instead of fVa = PROh 
                \item Thrombomodulin = TM
                \item Unactivated and activated protein C are respectivcely denoted PC and APC. 
                \item TFPI:PROhv5 indicates that TFPI$\alpha$ binds to PROh via the fVh in PROh
                \item TFPI:PROhv10 indicates that TFPI$\alpha$ binds to the fXa in PROh
                \item TFPI$\beta$ = TFPIb
                \item Antithrombin = AT
                \item Heparin = H
                \item Antithrombin-heparin complex = ATH
                \item The complex where fXa and fVh are bound to TFPI$\alpha$ but not to each other = e$_{\text{10}}$:TFPI:e$_{\text{5h}}$ and the corresponding membrane-bound forms
                \item Free PS (not bound to TFPI) = ``ps\_fr'
                \item The fraction of platelet-released factor V that is fVh = h5
            \end{itemize}
            
Species are fluid-phase, i.e, in the plasma, unless otherwise indicated. Membrane-bound versions of species are indicated with a superscript ``m'', and endothelial zone versions of species are indicated with a superscript ``ec''.
            
Binding and unbinding rate constants between a species $A$ and a species $B$ are given by $\text{k}_{\text{A:B}}^{+}$ and $\text{k}_{\text{A:B}}^{-}$, respectively. If either $A$ or $B$ is a complex and the binding between A and B would otherwise be unclear, parentheses indicate the order of binding, as described above. Catalysis between two species A and B is indicated by $\text{k}_{ \text{A:B}}^{\text{cat}}$. Which species catalyzes which is assumed to be clear from context: for example, $\text{k}_{ \text{z$_{5}^{\text{m}}$:e$_{10}^{\text{m}}$} }^{\text{cat}}$ indicates that membrane-bound fXa catalyzes membrane-bound fV, since the ``e'' and ``z'' indicate fXa is the enzyme in the reaction whereas fV is the zymogen (substrate).
            
Binding and unbinding rate constants of a species A to a platelet or to the subendothelium are indicated by $\text{k}_{\text{A}}^{\text{on}}$ and $\text{k}_{\text{A}}^{\text{off}}$, respectively. Binding to a platelet is proportional to the concentration $p_{\text{A}}^{\text{avail}}$ of binding sites available to $A$ on activated platelets. $p_{\text{A}}^{\text{avail}}$ is the total concentration $p_{\text{A}}$ of binding sites on activated platelets for A less the total concentration of species in which A is already bound to a platelet. $p_{\text{A}}$ in turn is the number $N_{\text{A}}$ of binding sites available to A per platelet multiplied by the concentration of activated platelets. Activated platelets are either on the subendothelium or otherwise bound to other platelets, indicated respectively by $\text{PL}^{\text{as}}$ and $\text{PL}^{\text{av}}$. Hence the total concentration of activated platelets is $[\text{PL}^{\text{as}}]$ + $[\text{PL}^{\text{av}}]$. Unactivated platelets are indicated by PL. 
           
The concentration of TF on the subendothelium available to bind plasma-phase species is given by $[\text{TF}^{\text{avail}}]$. Coverage of the subendothelium by platelets is accounted for by keeping track of the total concentration  $P_{max} - [\text{PL}^{\text{as}}]$ of binding sites for platelets on the subendothelium. $P_{\text{max}}$ is the total number of binding sites on the subendothelium, and subtracting off $[\text{PL}^{\text{as}}]$ accounts for coverage of binding sites for platelets.

Flow of species A into and out of the reaction zone is accounted for via a flow parameter $\text{k}_{\text{flow}}$ and by the difference between the upstream concentration of A, $[\text{A}]^\text{up}$, and the concentration of A in the reaction zone. Flow of species into and out of the endothelial zone is analogous, accounted for via a diffusion parameter $\text{k}_{\text{diff}}$ and the difference between the concentration of the species in the reaction zone versus the concentration in the endothelial zone.
        $\text{k}_{\text{flow}}$ and $\text{k}_{\text{diff}}$ are defined as follows:
            \begin{align*}
                \text{k}_{\text{flow}} &= \frac{3}{4}\left( \frac{\gamma^2\text{D}}{4\text{L}^2}\right)^{1/3}\\
                \text{k}_{\text{diff}} &= \frac{2\text{D}}{\text{W}_{\text{ec}}\left( \frac{\text{W}}{2} + \text{W}_{\text{ec}} \right)}
            \end{align*}
            where
            \begin{align*}
                \text{W}_{\text{ec}} &= \sqrt{4/3}\left( \frac{2\text{LD}}{\gamma}\right)^{1/3}
            \end{align*}
            is the effective width of the endothelial zone, L and W are the length and width of the injury, $\gamma$ is the diffusion coefficient, and $\gamma$ is the shear rate. (See \cite{fogelson2006coagulation, kuharsky2001surface} for further discussion). The rate constant for flow out of the reaction zone towards the endothelial zone is $\text{k}_{\text{diffRZ}} = (\text{W}_{\text{ec}}/\text{W})\text{k}_{\text{diff}}$,  computed by rescaling $\text{k}_{\text{diff}}$ by the ratio of volumes of the endothelial zone to the reaction zone. Note further that $\frac{d}{dt} \left( [\text{PL}^{\text{as}}] \right) /\text{(P$_\text{{max}}$ - [PL$^{\text{as}}$])}$ describes coverage of subendothelial species by platelets. 
        
Other constants that do not precisely follow the naming convention are:
        \begin{itemize}[noitemsep,topsep=0pt]
            \item The rate constant for unactivated platelets binding to platelet-binding sites on the subendothelium and immediately activating = $\text{k}^{\text{+}}$
            \item The rate constant for activated platelets in the fluid binding to platelet-binding sites on the subendothelium = $\text{k}^{\text{+1}}$
            \item The rate constant for activation of platelets by thrombin = $\text{k}_\text{act}$
            \item The rate constant for activation of platelets by other activated platelets, $\text{k}_{\text{plact}}$
        \end{itemize}

A list of all new species included in the model follows:
        \begin{itemize}[noitemsep,topsep=0pt]    
			\item ps:TFPI:e$_{\text{5h}}$ and its various membrane-bound forms:
			\begin{itemize}[noitemsep,topsep=0pt]
				\item	ps$^{\text{m}}$:TFPI:e$_{\text{5h}}$
				\item	ps:TFPI:e$_{\text{5h}}^{\text{m}}$
				\item	ps$^{\text{m}}$:TFPI:e$_{\text{5h}}^{\text{m}}$
			\end{itemize}
			
			\item (ps:TFPI:e$_{\text{5h}}$):e$_{\text{10}}$ and its various membrane-bound forms
			\begin{itemize}[noitemsep,topsep=0pt]
				\item (ps$^{\text{m}}$:TFPI:e$_{\text{5h}}$):e$_{\text{10}}$
				\item	(ps:TFPI:e$_{\text{5h}}^{\text{m}}$):e$_{\text{10}}$
				\item	(ps:TFPI:e$_{\text{5h}}$):e$_{\text{10}}^{\text{m}}$
				\item	(ps$^{\text{m}}$:TFPI:e$_{\text{5h}}^{\text{m}}$):e$_{\text{10}}$
				\item	(ps$^{\text{m}}$:TFPI:e$_{\text{5h}}$):e$_{\text{10}}^{\text{m}}$
				\item	(ps:TFPI:e$_{\text{5h}}^{\text{m}}$):e$_{\text{10}}^{\text{m}}$
				\item	(ps$^{\text{m}}$:TFPI:e$_{\text{5h}}^{\text{m}}$):e$_{\text{10}}^{\text{m}}$
			\end{itemize}

			\item ((ps:TFPI:e$_{\text{5h}}$):e$_{\text{10}}$):e$_{\text{7}}^{\text{m}}$
			
			\item ps\_fr (protein S by itself) and its membrane bound form
			\begin{itemize}[noitemsep,topsep=0pt]
				\item ps$^{\text{m}}$\_fr:
			\end{itemize}
			
			\item ps\_fr:e$_{\text{10}}$ and its various membrane bound forms
			\begin{itemize}[noitemsep,topsep=0pt]
				\item ps$^{\text{m}}$\_fr:e$_{\text{10}}$
				\item ps\_fr:e$_{\text{10}}^{\text{m}}$
				\item ps$^{\text{m}}$\_fr:e$_{\text{10}}^{\text{m}}$
			\end{itemize}

		\end{itemize}

\newpage

\subsection{Full Equations}
\label{supplemental_section:full_equations}

 \begin{flalign*} 
 \text{[TF$^{\text{avail}}$]} &= \left[ \text{TF} \right] - \left[ \text{z$_7^{\text{m}}$} \right] - \left[ \text{e$_7^{\text{m}}$} \right] - \left[ \text{z$_7^{\text{m}}$:e$_{10}$} \right] - \left[ \text{z$_7^{\text{m}}$:e$_{\text{2}}$} \right] - \left[ \text{z$_{10}$:e$_7^{\text{m}}$} \right]\\
 &\hskip2.5ex     - \left[ \text{z$_{9}$:e$_7^{\text{m}}$} \right] - \left[ \text{TFPI:e$_{10}$:e$_{7}^{\text{m}}$} \right] - \left[ \text{z$_7^{\text{m}}$:e$_{9}$} \right] - \left[ \text{((ps:TFPI:e$_{\text{5h}}$):e$_{10}$):e$_7^{\text{m}}$} \right] &
 \end{flalign*}

 \begin{flalign*} 
 \left[ \text{p}_{ \text{10} }^{\text{avail}} \right] &= ( \left[ \text{p$_{\text{10}}$} \right] - \left[ \text{z$_{10}^{\text{m}}$} \right] - \left[ \text{e$_{10}^{\text{m}}$} \right] - \left[ \text{z$_{10}^{\text{m}}$:TEN} \right] - \left[ \text{z$_5^{\text{m}}$:e$_{10}^{\text{m}}$} \right] - \left[ \text{z$_8^{\text{m}}$:e$_{10}^{\text{m}}$} \right]\\
 &\hskip2.5ex     - \left[ \text{PRO} \right] - \left[ \text{z$_2^{\text{m}}$:PRO} \right] - \left[ \text{z$_{10}^{\text{m}}$:TEN$^{*}$} \right] ) \\
 &\hskip2.5ex     - \left[ \text{PROh} \right] - \left[ \text{z$_2^{\text{m}}$:PROh} \right] - \left[ \text{TFPI:e$_{10}^{\text{m}}$} \right] \\
 &\hskip2.5ex     - \left[ \text{TFPI:PROhv10} \right] - \left[ \text{TFPI:PROhv5} \right] \\
 &\hskip2.5ex     - \left[ \text{e$_{10}^{\text{m}}$:TFPI:e$_{5\text{h}}^{\text{m}}$} \right] - \left[ \text{e$_{10}^{\text{m}}$:TFPI:e$_{\text{5h}}$} \right] \\
 &\hskip2.5ex     - \left[ \text{e$_{10}^{\text{m}}$:AT} \right] - \left[ \text{e$_{10}^{\text{m}}$:ATH} \right]\\
 &\hskip2.5ex
 - \left[ \text{(ps:TFPI:e$_{\text{5h}}$):e$_{10}^{\text{m}}$} \right] - \left[ \text{(ps:TFPI:e$_{5\text{h}}^{\text{m}}$):e$_{10}^{\text{m}}$} \right] - \left[ \text{(ps$^{\text{m}}$:TFPI:e$_{\text{5h}}$):e$_{10}^{\text{m}}$} \right] - \left[ \text{(ps$^{\text{m}}$:TFPI:e$_{5\text{h}}^{\text{m}}$):e$_{10}^{\text{m}}$} \right] \\
 &\hskip2.5ex - \left[ \text{ps\_fr:e$_{10}^{\text{m}}$} \right] - \left[ \text{ps$^{\text{m}}$\_fr:e$_{10}^{\text{m}}$} \right] &
 \end{flalign*} 

 \begin{flalign*} 
 \left[ \text{p}_{ \text{5} }^{\text{avail}}  \right] &= \left[ \text{p$_{\text{5}}$} \right] - \left[ \text{z$_5^{\text{m}}$} \right] - \left[ \text{e$_5^{\text{m}}$} \right] - \left[ \text{z$_5^{\text{m}}$:e$_{10}^{\text{m}}$} \right] - \left[ \text{z$_5^{\text{m}}$:e$_{2}^{\text{m,*}}$} \right] - \left[ \text{APC:e$_5^{\text{m}}$} \right]\\
 &\hskip2.5ex     - \left[ \text{PRO} \right] - \left[ \text{z$_2^{\text{m}}$:PRO} \right] - \left[ \text{e$_{5\text{h}}^{\text{m}}$} \right] - \left[ \text{PROh} \right] - \left[ \text{z$_2^{\text{m}}$:PROh} \right] - \left[ \text{e$_{5\text{h}}^{\text{m}}$:e$_2^{\text{m}}$} \right] \\
 &\hskip2.5ex     - \left[ \text{TFPI:e$_{5\text{h}}^{\text{m}}$} \right] - \left[ \text{APC:e$_{5\text{h}}^{\text{m}}$} \right] - \left[ \text{TFPI:PROhv10} \right] - \left[ \text{TFPI:PROhv5} \right] \\
 &\hskip2.5ex     - \left[ \text{e$_{10}^{\text{m}}$:TFPI:e$_{5\text{h}}^{\text{m}}$} \right] - \left[ \text{e$_{10}$:TFPI:e$_{5\text{h}}^{\text{m}}$} \right] - \left[ \text{PROh:e$_2^{\text{m}}$} \right] 
\\
 &\hskip2.5ex - \left[ \text{ps:TFPI:e$_{5\text{h}}^{\text{m}}$} \right] - \left[ \text{(ps:TFPI:e$_{5\text{h}}^{\text{m}}$):e$_{10}$} \right] - \left[ \text{(ps:TFPI:e$_{5\text{h}}^{\text{m}}$):e$_{10}^{\text{m}}$} \right]  - \left[ \text{ps$^{\text{m}}$:TFPI:e$_{5\text{h}}^{\text{m}}$} \right]  \\
 &\hskip2.5ex - \left[ \text{(ps$^{\text{m}}$:TFPI:e$_{5\text{h}}^{\text{m}}$):e$_{10}$} \right] - \left[ \text{(ps$^{\text{m}}$:TFPI:e$_{5\text{h}}^{\text{m}}$):e$_{10}^{\text{m}}$} \right]&
 \end{flalign*} 

 \begin{flalign*} 
 \text{p}_{ \text{8} }^{\text{avail}}  &= \left[ \text{p$_{\text{8}}$} \right] - \left[ \text{z$_8^{\text{m}}$} \right] - \left[ \text{e$_8^{\text{m}}$} \right] - \left[ \text{TEN} \right] - \left[ \text{z$_8^{\text{m}}$:e$_{10}^{\text{m}}$} \right] - \left[ \text{z$_8^{\text{m}}$:e$_2^{\text{m}}$} \right] - \left[ \text{z$_{10}^{\text{m}}$:TEN} \right] - \left[ \text{APC:e$_8^{\text{m}}$} \right]\\
 &\hskip2.5ex     - \left[ \text{TEN$^{*}$} \right] - \left[ \text{z$_{10}^{\text{m}}$:TEN$^{*}$} \right]&
 \end{flalign*} 

 \begin{flalign*} 
 \text{p}_{ \text{9} }^{\text{avail}}  &= \left[ \text{p$_{\text{9}}$} \right] - \left[ \text{z$_9^{\text{m}}$} \right] - \left[ \text{e$_9^{\text{m}}$} \right] - \left[ \text{TEN} \right] - \left[ \text{z$_{10}^{\text{m}}$:TEN} \right]\\
 &\hskip2.5ex  - \left[ \text{z$_9^{\text{m}}$:e$_{11\text{h}}^{\text{m}}$} \right] - \left[ \text{z$_9^{\text{m}}$:e$_{11}^{\text{m,*}}$} \right] \\
 &\hskip2.5ex  - \left[ \text{e$_9^{\text{m}}$:AT} \right] \\
 &\hskip2.5ex  - \left[ \text{e$_9^{\text{m}}$:ATH} \right]&
 \end{flalign*} 

 \begin{flalign*} 
 \text{p}_{ \text{2} }^{\text{avail}}  &= \left[ \text{p$_{\text{2}}$} \right] - \left[ \text{z$_2^{\text{m}}$} \right] - \left[ \text{z$_2^{\text{m}}$:PRO} \right]  - \left[ \text{z$_2^{\text{m}}$:PROh} \right]&
 \end{flalign*} 

 \begin{flalign*} 
 \text{p}_{ \text{2,*} }^{\text{avail}} &= \left[ \text{p$_{\text{2,*}}$} \right] - \left[ \text{e$_2^{\text{m}}$} \right] - \left[ \text{z$_5^{\text{m}}$:e$_2^{\text{m}}$} \right] - \left[ \text{z$_8^{\text{m}}$:e$_{2}^{\text{m,*}}$} \right]\\
 &\hskip2.5ex  - \left[ \text{z$_{11}^{\text{m}}$:e$_{2}^{\text{m,*}}$} \right] - \left[ \text{e$_{11\text{h}}^{\text{m,*}}$:e$_{2}^{\text{m,*}}$} \right] \\
 &\hskip2.5ex  - \left[ \text{e$_{5\text{h}}^{\text{m}}$:e$_{2}^{\text{m,*}}$} \right] \\
 &\hskip2.5ex  - \left[ \text{e$_2^{\text{m}}$:AT} \right] - \left[ \text{e$_2^{\text{m}}$:ATH} \right]&
 \end{flalign*} 

 \begin{flalign*} 
 \left[ \text{p}_{ \text{9,*} }^{\text{avail}} \right] &= \left[ \text{p$_{\text{9,*}}$} \right] - \left[ \text{e$_{9}^{\text{m,*}}$} \right] - ( \left[ \text{TEN$^{*}$} \right] + \left[ \text{z$_{10}^{\text{m}}$:TEN$^{*}$} \right] ) - \left[ \text{e$_{9}^{\text{m,*}}$:AT} \right]  - \left[ \text{e$_{9}^{\text{m,*}}$:ATH} \right]&
 \end{flalign*} 

 \begin{flalign*} 
 \left[ \text{p}_{ \text{11} }^{\text{avail}} \right] &= \left[ \text{p$_{\text{11}}$} \right] - \left[ \text{z$_{11}^{\text{m}}$} \right] - \left[ \text{e$_{11\text{h}}^{\text{m}}$} \right] - \left[ \text{z$_9^{\text{m}}$:e$_{11\text{h}}^{\text{m}}$} \right]\\
 &\hskip2.5ex  - 2\left[ \text{z$_{11}^{\text{m}}$:e$_{11\text{h}}^{\text{m}}$} \right] - \left[ \text{z$_{11}^{\text{m}}$:e$_{11}^{\text{m,*}}$} \right] - \left[ \text{z$_{11}^{\text{m}}$:e$_2^{\text{m}}$} \right] \\
 &\hskip2.5ex  - \left[ \text{e$_{11\text{h}}^{\text{m,*}}$:e$_{11\text{h}}^{\text{m}}$} \right] \\
 &\hskip2.5ex  - \left[ \text{e$_{11\text{h}}^{\text{m}}$:AT} \right] \\
 &\hskip2.5ex  - \left[ \text{e$_{11\text{h}}^{\text{m}}$:ATH} \right]&
 \end{flalign*} 

 \begin{flalign*} 
 \left[ \text{p}_{ \text{11,*} }^{\text{avail}} \right] &= \left[ \text{p$_{\text{11,*}}$} \right] - \left[ \text{e$_{11\text{h}}^{\text{m,*}}$} \right] - \left[ \text{e$_{11}^{\text{m,*}}$} \right] - \left[ \text{z$_9^{\text{m}}$:e$_{11}^{\text{m,*}}$} \right]\\
 &\hskip2.5ex - \left[ \text{z$_{11}^{\text{m}}$:e$_{11}^{\text{m,*}}$} \right] - \left[ \text{e$_{11\text{h}}^{\text{m,*}}$:e$_{11\text{h}}^{\text{m}}$} \right] - 2  \left[ \text{e$_{11\text{h}}^{\text{m,*}}$:e$_{11}^{\text{m,*}}$} \right] \\
 &\hskip2.5ex - \left[ \text{e$_{11\text{h}}^{\text{m,*}}$:e$_2^{\text{m}}$} \right] \\
 &\hskip2.5ex - \left[ \text{e$_{11}^{\text{m,*}}$:AT} \right] \\
 &\hskip2.5ex - \left[ \text{e$_{11}^{\text{m,*}}$:ATH} \right]&
 \end{flalign*}

 \begin{flalign*} 
 \text{p}_{ \text{ps} }^{\text{avail}} &= \left[ \text{p$_{\text{ps}}$} \right] - \left[ \text{ps$^{\text{m}}$:TFPI:e$_{\text{5h}}$} \right] - \left[ \text{(ps$^{\text{m}}$:TFPI:e$_{\text{5h}}$):e$_{10}$} \right] - \left[ \text{(ps$^{\text{m}}$:TFPI:e$_{\text{5h}}$):e$_{10}^{\text{m}}$} \right]   - \left[ \text{ps$^{\text{m}}$:TFPI:e$_{5\text{h}}^{\text{m}}$} \right] - \left[ \text{(ps$^{\text{m}}$:TFPI:e$_{5\text{h}}^{\text{m}}$):e$_{10}$} \right]  \\
 &\hskip2.5ex - \left[ \text{(ps$^{\text{m}}$:TFPI:e$_{5\text{h}}^{\text{m}}$):e$_{10}^{\text{m}}$} \right]  -  \left[ \text{ps$^{\text{m}}$\_fr} \right] - \left[ \text{ps$^{\text{m}}$\_fr:e$_{10}$} \right] &
 \end{flalign*}

\begin{flalign*} 
 \frac{d}{dt}\left[ \text{  z$_{7}$  } \right]&= - \text{k}_{ \text{7} }^{\text{on}}  \left[ \text{z$_{7}$} \right]  \left[ \text{TF$^{\text{avail}}$} \right] + \text{k}_{ \text{7} }^{\text{off}}  \left[ \text{z$_7^{\text{m}}$} \right]\\
 &\hskip2.5ex - \text{k}_{ \text{z$_{7}$:e$_{\text{2}}$} }^{\text{+}}  \left[ \text{z$_{7}$} \right]  \left[ \text{e$_{\text{2}}$} \right] + \left[ \text{z$_{7}$:e$_{\text{2}}$} \right]  \text{k}_{ \text{z$_{7}$:e$_{\text{2}}$} }^{\text{-}} \\
 &\hskip2.5ex - \text{k}_{ \text{z$_{7}$:e$_{10}$} }^{\text{+}}  \left[ \text{z$_{7}$} \right]  \left[ \text{e$_{10}$} \right] + \text{k}_{ \text{z$_{7}$:e$_{10}$} }^{\text{-}}  \left[ \text{z$_{7}$:e$_{10}$} \right] \\
 &\hskip2.5ex + \text{k}_{\text{flow}}  ( \text{ [z$_{7}$]$^{\text{up}}$}  - \left[ \text{z$_{7}$} \right] ) \\
 &\hskip2.5ex - \text{k}_{ \text{z$_{7}$:e$_{9}$} }^{\text{+}}  \left[ \text{z$_{7}$} \right]  \left[ \text{e$_{9}$} \right] + \text{k}_{ \text{z$_{7}$:e$_{9}$} }^{\text{-}}  \left[ \text{z$_{7}$:e$_{9}$} \right]&
 \end{flalign*} 

 \begin{flalign*} 
 \frac{d}{dt}\left[ \text{  e$_{7}$  } \right]&= - \text{k}_{ \text{e$_{7}$} }^{\text{on}}  \left[ \text{e$_{7}$} \right]  \left[ \text{TF$^{\text{avail}}$} \right] + \text{k}_{ \text{e$_{7}$} }^{\text{off}}  \left[ \text{e$_7^{\text{m}}$} \right]\\
 &\hskip2.5ex + \text{k}_{ \text{z$_{7}$:e$_{\text{2}}$} }^{\text{cat}}  \left[ \text{z$_{7}$:e$_{\text{2}}$} \right] + \text{k}_{ \text{z$_{7}$:e$_{10}$} }^{\text{cat}}  \left[ \text{z$_{7}$:e$_{10}$} \right] + \text{k}_{\text{flow}}  ( \text{ [e$_{7}$]$^{\text{up}}$}  - \left[ \text{e$_{7}$} \right] ) \\
 &\hskip2.5ex + \text{k}_{ \text{z$_{7}$:e$_{9}$} }^{\text{cat}}  \left[ \text{z$_{7}$:e$_{9}$} \right]&
 \end{flalign*} 

 \begin{flalign*} 
 \frac{d}{dt}\left[ \text{  z$_7^{\text{m}}$  } \right]&= \text{k}_{ \text{7} }^{\text{on}}  \left[ \text{z$_{7}$} \right]  \left[ \text{TF$^{\text{avail}}$} \right] - \text{k}_{ \text{7} }^{\text{off}}  \left[ \text{z$_7^{\text{m}}$} \right]\\
 &\hskip2.5ex - \text{k}_{ \text{z$_7^{\text{m}}$:e$_{10}$} }^{\text{+}}  \left[ \text{z$_7^{\text{m}}$} \right]  \left[ \text{e$_{10}$} \right] - \text{k}_{ \text{z$_7^{\text{m}}$:e$_{\text{2}}$} }^{\text{+}}  \left[ \text{z$_7^{\text{m}}$} \right]  \left[ \text{e$_{\text{2}}$} \right] + \text{k}_{ \text{z$_7^{\text{m}}$:e$_{10}$} }^{\text{-}}  \left[ \text{z$_7^{\text{m}}$:e$_{10}$} \right] \\
 &\hskip2.5ex + \text{k}_{ \text{z$_7^{\text{m}}$:e$_{\text{2}}$} }^{\text{-}}  \left[ \text{z$_7^{\text{m}}$:e$_{\text{2}}$} \right] \\
 &\hskip2.5ex - \text{k}_{ \text{z$_7^{\text{m}}$:e$_{9}$} }^{\text{+}}  \left[ \text{z$_7^{\text{m}}$} \right]  \left[ \text{e$_{9}$} \right] + \text{k}_{ \text{z$_7^{\text{m}}$:e$_{9}$} }^{\text{-}}  \left[ \text{z$_7^{\text{m}}$:e$_{9}$} \right] \\
 &\hskip2.5ex - \left[ \text{z$_7^{\text{m}}$} \right]  \frac{d}{dt} \left( [\text{PL}^{\text{as}}] \right) /\text{(P$_\text{{max}}$ - [PL$^{\text{as}}$])} &
 \end{flalign*} 

 \begin{flalign*} 
 \frac{d}{dt}\left[ \text{  e$_7^{\text{m}}$  } \right]&= \text{k}_{ \text{7} }^{\text{on}}  \left[ \text{e$_{7}$} \right]  \left[ \text{TF$^{\text{avail}}$} \right] \\
 &\hskip2.5ex - \text{k}_{ \text{7} }^{\text{off}}  \left[ \text{e$_7^{\text{m}}$} \right] - \text{k}_{ \text{(TFPI:e$_{10}$):e$_7^{\text{m}}$} }^{\text{+}}  \left[ \text{e$_7^{\text{m}}$} \right]  \left[ \text{TFPI:e$_{10}$} \right] + \text{k$_{\text{(TFPI:e$_{10}$):e$_7^{\text{m}}$}}^{-}$}  \left[ \text{TFPI:e$_{10}$:e$_{7}^{\text{m}}$} \right] \\
 &\hskip2.5ex + \text{k}_{ \text{z$_7^{\text{m}}$:e$_{10}$} }^{\text{cat}}  \left[ \text{z$_7^{\text{m}}$:e$_{10}$} \right] + \text{k}_{ \text{z$_7^{\text{m}}$:e$_{\text{2}}$} }^{\text{cat}}  \left[ \text{z$_7^{\text{m}}$:e$_{\text{2}}$} \right] \\
 &\hskip2.5ex + ( \text{k}_{ \text{z$_{10}$:e$_7^{\text{m}}$} }^{\text{cat}} + \text{k}_{ \text{z$_{10}$:e$_7^{\text{m}}$} }^{\text{-}} )  \left[ \text{z$_{10}$:e$_7^{\text{m}}$} \right] \\
 &\hskip2.5ex - \text{k}_{ \text{z$_{10}$:e$_7^{\text{m}}$} }^{\text{+}}  \left[ \text{e$_7^{\text{m}}$} \right]  \left[ \text{z$_{10}$} \right] - \text{k}_{ \text{z$_{9}$:e$_7^{\text{m}}$} }^{\text{+}}  \left[ \text{e$_7^{\text{m}}$} \right]  \left[ \text{z$_{9}$} \right] \\
 &\hskip2.5ex + ( \text{k}_{ \text{z$_{9}$:e$_7^{\text{m}}$} }^{\text{cat}} + \text{k}_{ \text{z$_{9}$:e$_7^{\text{m}}$} }^{\text{-}} )  \left[ \text{z$_{9}$:e$_7^{\text{m}}$} \right] \\
 &\hskip2.5ex + \text{k}_{ \text{z$_7^{\text{m}}$:e$_{9}$} }^{\text{cat}}  \left[ \text{z$_7^{\text{m}}$:e$_{9}$} \right] \\
 &\hskip2.5ex - \left[ \text{e$_7^{\text{m}}$} \right]  \frac{d}{dt} \left( [\text{PL}^{\text{as}}] \right) /\text{(P$_\text{{max}}$ - [PL$^{\text{as}}$])} \\
 &\hskip2.5ex - \text{k}_{ \text{((ps:TFPI:e$_{\text{5h}}$):e$_{10}$):e$_7^{\text{m}}$} }^{+}  \left[ \text{(ps:TFPI:e$_{\text{5h}}$):e$_{10}$} \right]  \left[ \text{e$_7^{\text{m}}$} \right] + \text{k}_{ \text{((ps:TFPI:e$_{\text{5h}}$):e$_{10}$):e$_7^{\text{m}}$} }^{-}  \left[ \text{((ps:TFPI:e$_{\text{5h}}$):e$_{10}$):e$_7^{\text{m}}$} \right]&
 \end{flalign*} 

 \begin{flalign*} 
 \frac{d}{dt}\left[ \text{  z$_{10}$  } \right]&= - \text{k}_{ \text{10} }^{\text{on}}  \left[ \text{z$_{10}$} \right]  \left[ \text{p}_{ \text{10} }^{\text{avail}} \right] + \text{k}_{ \text{10} }^{\text{off}}  \left[ \text{z$_{10}^{\text{m}}$} \right]\\
 &\hskip2.5ex - \text{k}_{ \text{z$_{10}$:e$_7^{\text{m}}$} }^{\text{+}}  \left[ \text{z$_{10}$} \right]  \left[ \text{e$_7^{\text{m}}$} \right] + \text{k}_{ \text{z$_{10}$:e$_7^{\text{m}}$} }^{\text{-}}  \left[ \text{z$_{10}$:e$_7^{\text{m}}$} \right] \\
 &\hskip2.5ex + \text{k}_{\text{flow}}  ( \text{ [z$_{10}$]$^{\text{up}}$}  - \left[ \text{z$_{10}$} \right] )  &
 \end{flalign*} 

 \begin{flalign*} 
 \frac{d}{dt}\left[ \text{  e$_{10}$  } \right]&= - \text{k}_{ \text{10} }^{\text{on}}  \left[ \text{e$_{10}$} \right]  \left[ \text{p}_{ \text{10} }^{\text{avail}} \right] + \text{k}_{ \text{10} }^{\text{off}}  \left[ \text{e$_{10}^{\text{m}}$} \right]\\
 &\hskip2.5ex + \text{k}_{ \text{z$_{10}$:e$_7^{\text{m}}$} }^{\text{cat}}  \left[ \text{z$_{10}$:e$_7^{\text{m}}$} \right] \\
 &\hskip2.5ex + ( \text{k}_{ \text{z$_{7}$:e$_{10}$} }^{\text{cat}} + \text{k}_{ \text{z$_{7}$:e$_{10}$} }^{\text{-}} )  \left[ \text{z$_{7}$:e$_{10}$} \right] - \text{k}_{ \text{z$_{7}$:e$_{10}$} }^{\text{+}}  \left[ \text{e$_{10}$} \right]  \left[ \text{z$_{7}$} \right] \\
 &\hskip2.5ex + ( \text{k}_{ \text{z$_7^{\text{m}}$:e$_{10}$} }^{\text{cat}} + \text{k}_{ \text{z$_7^{\text{m}}$:e$_{10}$} }^{\text{-}} )  \left[ \text{z$_7^{\text{m}}$:e$_{10}$} \right] - \text{k}_{ \text{z$_7^{\text{m}}$:e$_{10}$} }^{\text{+}}  \left[ \text{e$_{10}$} \right]  \left[ \text{z$_7^{\text{m}}$} \right] \\
 &\hskip2.5ex - \text{k}_{ \text{TFPI:e$_{10}$} }^{\text{+}}  \left[ \text{e$_{10}$} \right]  \left[ \text{TFPI} \right] + \text{k$_{\text{TFPI:e$_{10}$}}^{-}$}  \left[ \text{TFPI:e$_{10}$} \right] \\
 &\hskip2.5ex + \text{k}_{\text{flow}}  ( \text{ [e$_{10}$]$^{\text{up}}$}  - \left[ \text{e$_{10}$} \right] ) \\
 &\hskip2.5ex - \text{k}_{ \text{e$_{10}$:AT} }  \left[ \text{AT} \right]  \left[ \text{e$_{10}$} \right] \\
 &\hskip2.5ex - \text{k}_{ \text{(ps:TFPI:e$_{\text{5h}}$):e$_{10}$} }^{+}  \left[ \text{ps:TFPI:e$_{\text{5h}}$} \right]  \left[ \text{e$_{10}$} \right] + \text{k}_{ \text{(ps:TFPI:e$_{\text{5h}}$):e$_{10}$} }^{-}  \left[ \text{(ps:TFPI:e$_{\text{5h}}$):e$_{10}$} \right] \\
 &\hskip2.5ex - \text{k}_{ \text{(ps$^{\text{m}}$:TFPI:e$_{\text{5h}}$):e$_{10}$} }^{+}  \left[ \text{ps$^{\text{m}}$:TFPI:e$_{\text{5h}}$} \right]  \left[ \text{e$_{10}$} \right] + \text{k}_{ \text{(ps$^{\text{m}}$:TFPI:e$_{\text{5h}}$):e$_{10}$} }^{-}  \left[ \text{(ps$^{\text{m}}$:TFPI:e$_{\text{5h}}$):e$_{10}$} \right] \\
 &\hskip2.5ex - \text{k}_{ \text{(ps:TFPI:e$_{5\text{h}}^{\text{m}}$):e$_{10}$} }^{+}  \left[ \text{ps:TFPI:e$_{5\text{h}}^{\text{m}}$} \right]  \left[ \text{e$_{10}$} \right] + \text{k}_{ \text{(ps:TFPI:e$_{5\text{h}}^{\text{m}}$):e$_{10}$} }^{-}  \left[ \text{(ps:TFPI:e$_{5\text{h}}^{\text{m}}$):e$_{10}$} \right] \\
 &\hskip2.5ex - \text{k}_{ \text{(ps$^{\text{m}}$:TFPI:e$_{5\text{h}}^{\text{m}}$):e$_{10}$} }^{+}  \left[ \text{ps$^{\text{m}}$:TFPI:e$_{5\text{h}}^{\text{m}}$} \right]  \left[ \text{e$_{10}$} \right] + \text{k}_{ \text{(ps$^{\text{m}}$:TFPI:e$_{5\text{h}}^{\text{m}}$):e$_{10}$} }^{-}  \left[ \text{(ps$^{\text{m}}$:TFPI:e$_{5\text{h}}^{\text{m}}$):e$_{10}$} \right] \\
 &\hskip2.5ex - \text{k}_\text{diffRZ}  ( \left[ \text{e$_{10}$} \right] - \left[ \text{e$_{10}$$^{\text{ec}}$} \right] ) \\
 &\hskip2.5ex - \text{k}_{ \text{(TFPI:e$_{\text{5h}}$):e$_{10}$} }^{\text{+}}  \left[ \text{TFPI:e$_{\text{5h}}$} \right]  \left[ \text{e$_{10}$} \right] + \text{k}_{ \text{(TFPI:e$_{\text{5h}}$):e$_{10}$} }^{\text{-}}  \left[ \text{e$_{10}$:TFPI:e$_{\text{5h}}$} \right] \\
 &\hskip2.5ex - \text{k}_{ \text{TFPI:e$_{5\text{h}}^{\text{m}}$:e$_{10}$} }^{\text{+}}  \left[ \text{TFPI:e$_{5\text{h}}^{\text{m}}$} \right]  \left[ \text{e$_{10}$} \right] \\
 &\hskip2.5ex + \text{k}_{ \text{TFPI:e$_{5\text{h}}^{\text{m}}$:e$_{10}$} }^{\text{-}}  \left[ \text{e$_{10}$:TFPI:e$_{5\text{h}}^{\text{m}}$} \right] \\
 &\hskip2.5ex - \text{k}_{ \text{e$_{10}$:ATH} }  \left[ \text{ATH} \right]  \left[ \text{e$_{10}$} \right] \\
 &\hskip2.5ex - \text{k}_{ \text{ps\_fr:e$_{10}$} }^{+}  \left[ \text{ps\_fr} \right]  \left[ \text{e$_{10}$} \right] + \text{k}_{ \text{ps\_fr:e$_{10}$} }^{-}  \left[ \text{ps\_fr:e$_{10}$} \right] \\
 &\hskip2.5ex - \text{k}_{ \text{ps$^{\text{m}}$\_fr:e$_{10}$} }^{+}  \left[ \text{ps$^{\text{m}}$\_fr} \right]  \left[ \text{e$_{10}$} \right] + \text{k}_{ \text{ps$^{\text{m}}$\_fr:e$_{10}$} }^{-}  \left[ \text{ps$^{\text{m}}$\_fr:e$_{10}$} \right]&
 \end{flalign*} 

 \begin{flalign*} 
 \frac{d}{dt}\left[ \text{  z$_{10}^{\text{m}}$  } \right]&= \text{k}_{ \text{10} }^{\text{on}}  \left[ \text{z$_{10}$} \right]  \left[ \text{p}_{ \text{10} }^{\text{avail}} \right] - \text{k}_{ \text{10} }^{\text{off}}  \left[ \text{z$_{10}^{\text{m}}$} \right]\\
 &\hskip2.5ex - \text{k}_{ \text{z$_{10}^{\text{m}}$:TEN} }^{\text{+}}  \left[ \text{z$_{10}^{\text{m}}$} \right]  \left[ \text{TEN} \right] + \text{k}_{ \text{z$_{10}^{\text{m}}$:TEN} }^{\text{-}}  \left[ \text{z$_{10}^{\text{m}}$:TEN} \right] \\
 &\hskip2.5ex - \text{k}_{ \text{z$_{10}^{\text{m}}$:TEN} }^{\text{+}}  \left[ \text{z$_{10}^{\text{m}}$} \right]  \left[ \text{TEN$^{*}$} \right] + \text{k}_{ \text{z$_{10}^{\text{m}}$:TEN} }^{\text{-}}  \left[ \text{z$_{10}^{\text{m}}$:TEN$^{*}$} \right]&
 \end{flalign*} 

 \begin{flalign*} 
 \frac{d}{dt}\left[ \text{  e$_{10}^{\text{m}}$  } \right]&= \text{k}_{ \text{10} }^{\text{on}}  \left[ \text{e$_{10}$} \right]  \left[ \text{p}_{ \text{10} }^{\text{avail}} \right] - \text{k}_{ \text{10} }^{\text{off}}  \left[ \text{e$_{10}^{\text{m}}$} \right]\\
 &\hskip2.5ex + \text{k}_{ \text{z$_{10}^{\text{m}}$:TEN} }^{\text{cat}}  \left[ \text{z$_{10}^{\text{m}}$:TEN} \right] + ( \text{k}_{ \text{z$_5^{\text{m}}$:e$_{10}^{\text{m}}$} }^{\text{cat}} + \text{k}_{ \text{z$_5^{\text{m}}$:e$_{10}^{\text{m}}$} }^{\text{-}} )  \left[ \text{z$_5^{\text{m}}$:e$_{10}^{\text{m}}$} \right] \\
 &\hskip2.5ex - \text{k}_{ \text{z$_5^{\text{m}}$:e$_{10}^{\text{m}}$} }^{\text{+}}  \left[ \text{e$_{10}^{\text{m}}$} \right]  \left[ \text{z$_5^{\text{m}}$} \right] + ( \text{k}_{ \text{z$_8^{\text{m}}$:e$_{10}^{\text{m}}$} }^{\text{cat}} + \text{k}_{ \text{z$_8^{\text{m}}$:e$_{10}^{\text{m}}$} }^{\text{-}} )  \left[ \text{z$_8^{\text{m}}$:e$_{10}^{\text{m}}$} \right] \\
 &\hskip2.5ex - \text{k}_{ \text{z$_8^{\text{m}}$:e$_{10}^{\text{m}}$} }^{\text{+}}  \left[ \text{e$_{10}^{\text{m}}$} \right]  \left[ \text{z$_8^{\text{m}}$} \right] \\
 &\hskip2.5ex + \text{k}_{ \text{e$_5^{\text{m}}$:e$_{10}^{\text{m}}$} }^{\text{-}}  \left[ \text{PRO} \right] - \text{k}_{ \text{e$_5^{\text{m}}$:e$_{10}^{\text{m}}$} }^{\text{+}}  \left[ \text{e$_{10}^{\text{m}}$} \right]  \left[ \text{e$_5^{\text{m}}$} \right] \\
 &\hskip2.5ex + \text{k}_{ \text{z$_{10}^{\text{m}}$:TEN} }^{\text{cat}}  \left[ \text{z$_{10}^{\text{m}}$:TEN$^{*}$} \right] \\
 &\hskip2.5ex - \text{k}_{ \text{(ps:TFPI:e$_{\text{5h}}$):e$_{10}^{\text{m}}$} }^{+}  \left[ \text{ps:TFPI:e$_{\text{5h}}$} \right]  \left[ \text{e$_{10}^{\text{m}}$} \right] + \text{k}_{ \text{(ps:TFPI:e$_{\text{5h}}$):e$_{10}^{\text{m}}$} }^{-}  \left[ \text{(ps:TFPI:e$_{\text{5h}}$):e$_{10}^{\text{m}}$} \right] \\
 &\hskip2.5ex - \text{k}_{ \text{ps$^{\text{m}}$:TFPI:e$_{\text{5h}}$:e$_{10}^{\text{m}}$} }^{+}  \left[ \text{ps$^{\text{m}}$:TFPI:e$_{\text{5h}}$} \right]  \left[ \text{e$_{10}^{\text{m}}$} \right] + \text{k}_{ \text{ps$^{\text{m}}$:TFPI:e$_{\text{5h}}$:e$_{10}^{\text{m}}$} }^{-}  \left[ \text{(ps$^{\text{m}}$:TFPI:e$_{\text{5h}}$):e$_{10}^{\text{m}}$} \right] \\
 &\hskip2.5ex - \text{k}_{ \text{(ps:TFPI:e$_{5\text{h}}^{\text{m}}$):e$_{10}^{\text{m}}$} }^{+}  \left[ \text{ps:TFPI:e$_{5\text{h}}^{\text{m}}$} \right]  \left[ \text{e$_{10}^{\text{m}}$} \right] + \text{k}_{ \text{(ps:TFPI:e$_{5\text{h}}^{\text{m}}$):e$_{10}^{\text{m}}$} }^{-}  \left[ \text{(ps:TFPI:e$_{5\text{h}}^{\text{m}}$):e$_{10}^{\text{m}}$} \right] \\
 &\hskip2.5ex - \text{k}_{ \text{(ps$^{\text{m}}$:TFPI:e$_{5\text{h}}^{\text{m}}$):e$_{10}^{\text{m}}$} }^{+}  \left[ \text{ps$^{\text{m}}$:TFPI:e$_{5\text{h}}^{\text{m}}$} \right]  \left[ \text{e$_{10}^{\text{m}}$} \right] + \text{k}_{ \text{(ps$^{\text{m}}$:TFPI:e$_{5\text{h}}^{\text{m}}$):e$_{10}^{\text{m}}$} }^{-}  \left[ \text{(ps$^{\text{m}}$:TFPI:e$_{5\text{h}}^{\text{m}}$):e$_{10}^{\text{m}}$} \right] \\
 &\hskip2.5ex - \text{k}_{ \text{e$_{5\text{h}}^{\text{m}}$:e$_{10}^{\text{m}}$} }^{\text{+}}  \left[ \text{e$_{10}^{\text{m}}$} \right]  \left[ \text{e$_{5\text{h}}^{\text{m}}$} \right] + \text{k}_{ \text{e$_{5\text{h}}^{\text{m}}$:e$_{10}^{\text{m}}$} }^{\text{-}}  \left[ \text{PROh} \right] \\
 &\hskip2.5ex - \text{k}_{ \text{TFPI:e$_{10}^{\text{m}}$} }^{\text{+}}  \left[ \text{TFPI} \right]  \left[ \text{e$_{10}^{\text{m}}$} \right] + \text{k}_{ \text{TFPI:e$_{10}^{\text{m}}$} }^{\text{-}}  \left[ \text{TFPI:e$_{10}^{\text{m}}$} \right] \\
 &\hskip2.5ex - \text{k}_{ \text{TFPI:e$_{5\text{h}}^{\text{m}}$:e$_{10}^{\text{m}}$} }^{\text{+}}  \left[ \text{TFPI:e$_{5\text{h}}^{\text{m}}$} \right]  \left[ \text{e$_{10}^{\text{m}}$} \right] \\
 &\hskip2.5ex + \text{k}_{ \text{TFPI:e$_{5\text{h}}^{\text{m}}$:e$_{10}^{\text{m}}$} }^{\text{-}}  \left[ \text{e$_{10}^{\text{m}}$:TFPI:e$_{5\text{h}}^{\text{m}}$} \right] \\
 &\hskip2.5ex - \text{k}_{ \text{e$_{10}^{\text{m}}$:AT} }  \left[ \text{AT} \right]  \left[ \text{e$_{10}^{\text{m}}$} \right] \\
 &\hskip2.5ex - \text{k}_{ \text{e$_{10}^{\text{m}}$:ATH} }  \left[ \text{ATH} \right]  \left[ \text{e$_{10}^{\text{m}}$} \right] \\
 &\hskip2.5ex - \text{k}_{ \text{ps\_fr:e$_{10}^{\text{m}}$} }^{+}  \left[ \text{ps\_fr} \right]  \left[ \text{e$_{10}^{\text{m}}$} \right] + \text{k}_{ \text{ps\_fr:e$_{10}^{\text{m}}$} }^{-}  \left[ \text{ps\_fr:e$_{10}^{\text{m}}$} \right] \\
 &\hskip2.5ex - \text{k}_{ \text{ps$^{\text{m}}$\_fr:e$_{10}^{\text{m}}$} }^{+}  \left[ \text{ps$^{\text{m}}$\_fr} \right]  \left[ \text{e$_{10}^{\text{m}}$} \right] + \text{k}_{ \text{ps$^{\text{m}}$\_fr:e$_{10}^{\text{m}}$} }^{-}  \left[ \text{ps$^{\text{m}}$\_fr:e$_{10}^{\text{m}}$} \right]&
 \end{flalign*} 

 \begin{flalign*} 
 \frac{d}{dt}\left[ \text{  z$_{5}$  } \right]&= - \text{k}_{ \text{5} }^{\text{on}}  \left[ \text{z$_{5}$} \right]  \left[ \text{p}_{ \text{5} }^{\text{avail}}  \right] + \text{k}_{ \text{5} }^{\text{off}}  \left[ \text{z$_5^{\text{m}}$} \right]\\
 &\hskip2.5ex - \text{k}_{ \text{z$_{5}$:e$_{\text{2}}$} }^{\text{+}}  \left[ \text{z$_{5}$} \right]  \left[ \text{e$_{\text{2}}$} \right] + \text{k}_{ \text{z$_{5}$:e$_{\text{2}}$} }^{\text{-}}  \left[ \text{z$_{5}$:e$_{\text{2}}$} \right] \\
 &\hskip2.5ex + \text{k}_{\text{flow}}  ( \text{ [z$_{5}$]$^{\text{up}}$}  - \left[ \text{z$_{5}$} \right] ) \\
 &\hskip2.5ex + (1 - \text{h5}) \cdot \text{n5} \left[ \frac{d}{dt} \left( [\text{PL}^{\text{as}}] + [\text{PL}^{\text{av}}] \right) \right]  &
 \end{flalign*} 

 \begin{flalign*} 
 \frac{d}{dt}\left[ \text{  e$_{\text{5}}$  } \right]&= - \text{k}_{ \text{5} }^{\text{on}}  \left[ \text{e$_{\text{5}}$} \right]  \left[ \text{p}_{ \text{5} }^{\text{avail}}  \right] + \text{k}_{ \text{5} }^{\text{off}}  \left[ \text{e$_5^{\text{m}}$} \right]\\
 &\hskip2.5ex + \text{k}_{ \text{z$_{5}$:e$_{\text{2}}$} }^{\text{cat}}  \left[ \text{z$_{5}$:e$_{\text{2}}$} \right] \\
 &\hskip2.5ex + \text{k}_{\text{flow}}  ( \text{ [e$_{\text{5}}$]$^{\text{up}}$}  - \left[ \text{e$_{\text{5}}$} \right] ) \\
 &\hskip2.5ex + \text{k}_{ \text{e$_{\text{5}}$:APC} }^{\text{-}}  \left[ \text{APC:e$_{\text{5}}$} \right] - \text{k}_{ \text{e$_{\text{5}}$:APC} }^{\text{+}}  \left[ \text{APC} \right]  \left[ \text{e$_{\text{5}}$} \right] \\
 &\hskip2.5ex + \text{k}_{ \text{e$_{\text{5h}}$:e$_{\text{2}}$} }^{\text{cat}}  \left[ \text{e$_{\text{5h}}$:e$_{\text{2}}$} \right]&
 \end{flalign*} 

 \begin{flalign*} 
 \frac{d}{dt}\left[ \text{  z$_5^{\text{m}}$  } \right]&= \text{k}_{ \text{5} }^{\text{on}}  \left[ \text{z$_{5}$} \right]  \left[ \text{p}_{ \text{5} }^{\text{avail}}  \right] - \text{k}_{ \text{5} }^{\text{off}}  \left[ \text{z$_5^{\text{m}}$} \right]\\
 &\hskip2.5ex - \text{k}_{ \text{z$_5^{\text{m}}$:e$_{10}^{\text{m}}$} }^{\text{+}}  \left[ \text{z$_5^{\text{m}}$} \right]  \left[ \text{e$_{10}^{\text{m}}$} \right] + \text{k}_{ \text{z$_5^{\text{m}}$:e$_{10}^{\text{m}}$} }^{\text{-}}  \left[ \text{z$_5^{\text{m}}$:e$_{10}^{\text{m}}$} \right] \\
 &\hskip2.5ex - \text{k}_{ \text{z$_5^{\text{m}}$:e$_2^{\text{m}}$} }^{\text{+}}  \left[ \text{z$_5^{\text{m}}$} \right]  \left[ \text{e$_2^{\text{m}}$} \right] + \text{k}_{ \text{z$_5^{\text{m}}$:e$_2^{\text{m}}$} }^{\text{-}}  \left[ \text{z$_5^{\text{m}}$:e$_2^{\text{m}}$} \right]&
 \end{flalign*} 

 \begin{flalign*} 
 \frac{d}{dt}\left[ \text{  e$_5^{\text{m}}$  } \right]&= \text{k}_{ \text{5} }^{\text{on}}  \left[ \text{e$_{\text{5}}$} \right]  \left[ \text{p}_{ \text{5} }^{\text{avail}}  \right] - \text{k}_{ \text{5} }^{\text{off}}  \left[ \text{e$_5^{\text{m}}$} \right]\\
 &\hskip2.5ex + \text{k}_{ \text{z$_5^{\text{m}}$:e$_2^{\text{m}}$} }^{\text{cat}}  \left[ \text{z$_5^{\text{m}}$:e$_2^{\text{m}}$} \right] \\
 &\hskip2.5ex + \text{k}_{ \text{e$_5^{\text{m}}$:APC} }^{\text{-}}  \left[ \text{APC:e$_5^{\text{m}}$} \right] \\
 &\hskip2.5ex - \text{k}_{ \text{e$_5^{\text{m}}$:APC} }^{\text{+}}  \left[ \text{APC} \right]  \left[ \text{e$_5^{\text{m}}$} \right] \\
 &\hskip2.5ex - \text{k}_{ \text{e$_5^{\text{m}}$:e$_{10}^{\text{m}}$} }^{\text{+}}  \left[ \text{e$_5^{\text{m}}$} \right]  \left[ \text{e$_{10}^{\text{m}}$} \right] + \text{k}_{ \text{e$_5^{\text{m}}$:e$_{10}^{\text{m}}$} }^{\text{-}}  \left[ \text{PRO} \right] \\
 &\hskip2.5ex + \text{k}_{ \text{e$_{5\text{h}}^{\text{m}}$:e$_2^{\text{m}}$} }^{\text{cat}}  \left[ \text{e$_{5\text{h}}^{\text{m}}$:e$_2^{\text{m}}$} \right]&
 \end{flalign*} 

 \begin{flalign*} 
 \frac{d}{dt}\left[ \text{  z$_{8}$  } \right]&= - \text{k}_{ \text{8} }^{\text{on}}  \left[ \text{z$_{8}$} \right]  \left[ \text{p}_{ \text{8} }^{\text{avail}} \right] + \text{k}_{ \text{8} }^{\text{off}}  \left[ \text{z$_8^{\text{m}}$} \right]\\
 &\hskip2.5ex + \text{k}_{\text{flow}}  ( \text{ [z$_{8}$]$^{\text{up}}$}  - \left[ \text{z$_{8}$} \right] ) \\
 &\hskip2.5ex - \text{k}_{ \text{z$_{8}$:e$_{\text{2}}$} }^{\text{+}}  \left[ \text{z$_{8}$} \right]  \left[ \text{e$_{\text{2}}$} \right] + \text{k}_{ \text{z$_{8}$:e$_{\text{2}}$} }^{\text{-}}  \left[ \text{z$_{8}$:e$_{\text{2}}$} \right]&
 \end{flalign*} 

 \begin{flalign*} 
 \frac{d}{dt}\left[ \text{  e$_{8}$  } \right]&= - \text{k}_{ \text{8} }^{\text{on}}  \left[ \text{e$_{8}$} \right]  \left[ \text{p}_{ \text{8} }^{\text{avail}} \right] + \text{k}_{ \text{8} }^{\text{off}}  \left[ \text{e$_8^{\text{m}}$} \right]\\
 &\hskip2.5ex + \text{k}_{\text{flow}}  ( \text{ [e$_{8}$]$^{\text{up}}$}  - \left[ \text{e$_{8}$} \right] ) \\
 &\hskip2.5ex + \text{k}_{ \text{z$_{8}$:e$_{\text{2}}$} }^{\text{cat}}  \left[ \text{z$_{8}$:e$_{\text{2}}$} \right] - 0.005  \left[ \text{e$_{8}$} \right] \\
 &\hskip2.5ex + \text{k}_{ \text{e$_{8}$:APC} }^{\text{-}}  \left[ \text{APC:e$_{8}$} \right] - \text{k}_{ \text{e$_{8}$:APC} }^{\text{+}}  \left[ \text{APC} \right]  \left[ \text{e$_{8}$} \right]&
 \end{flalign*} 

 \begin{flalign*} 
 \frac{d}{dt}\left[ \text{  z$_8^{\text{m}}$  } \right]&= \text{k}_{ \text{8} }^{\text{on}}  \left[ \text{z$_{8}$} \right]  \left[ \text{p}_{ \text{8} }^{\text{avail}} \right] - \text{k}_{ \text{8} }^{\text{off}}  \left[ \text{z$_8^{\text{m}}$} \right]\\
 &\hskip2.5ex - \text{k}_{ \text{z$_8^{\text{m}}$:e$_{10}^{\text{m}}$} }^{\text{+}}  \left[ \text{z$_8^{\text{m}}$} \right]  \left[ \text{e$_{10}^{\text{m}}$} \right] + \text{k}_{ \text{z$_8^{\text{m}}$:e$_{10}^{\text{m}}$} }^{\text{-}}  \left[ \text{z$_8^{\text{m}}$:e$_{10}^{\text{m}}$} \right] - \text{k}_{ \text{z$_8^{\text{m}}$:e$_2^{\text{m}}$} }^{\text{+}}  \left[ \text{z$_8^{\text{m}}$} \right]  \left[ \text{e$_2^{\text{m}}$} \right] \\
 &\hskip2.5ex + \text{k}_{ \text{z$_8^{\text{m}}$:e$_2^{\text{m}}$} }^{\text{-}}  \left[ \text{z$_8^{\text{m}}$:e$_2^{\text{m}}$} \right]&
 \end{flalign*} 

 \begin{flalign*} 
 \frac{d}{dt}\left[ \text{  e$_8^{\text{m}}$  } \right]&= \text{k}_{ \text{8} }^{\text{on}}  \left[ \text{e$_{8}$} \right]  \left[ \text{p}_{ \text{8} }^{\text{avail}} \right] - \text{k}_{ \text{8} }^{\text{off}}  \left[ \text{e$_8^{\text{m}}$} \right]\\
 &\hskip2.5ex + \text{k}_{ \text{z$_8^{\text{m}}$:e$_{10}^{\text{m}}$} }^{\text{cat}}  \left[ \text{z$_8^{\text{m}}$:e$_{10}^{\text{m}}$} \right] + \text{k}_{ \text{z$_8^{\text{m}}$:e$_2^{\text{m}}$} }^{\text{cat}}  \left[ \text{z$_8^{\text{m}}$:e$_2^{\text{m}}$} \right] - \text{k}_{ \text{e$_8^{\text{m}}$:APC} }^{\text{+}}  \left[ \text{APC} \right]  \left[ \text{e$_8^{\text{m}}$} \right] \\
 &\hskip2.5ex + \text{k}_{ \text{e$_8^{\text{m}}$:APC} }^{\text{-}}  \left[ \text{APC:e$_8^{\text{m}}$} \right] \\
 &\hskip2.5ex - \text{k}_{ \text{e$_8^{\text{m}}$:e$_9^{\text{m}}$} }^{\text{+}}  \left[ \text{e$_9^{\text{m}}$} \right]  \left[ \text{e$_8^{\text{m}}$} \right] + \text{k}_{ \text{e$_8^{\text{m}}$:e$_9^{\text{m}}$} }^{\text{-}}  \left[ \text{TEN} \right] - 0.005  \left[ \text{e$_8^{\text{m}}$} \right] \\
 &\hskip2.5ex - \text{k}_{ \text{e$_8^{\text{m}}$:e$_9^{\text{m}}$} }^{\text{+}}  \left[ \text{e$_8^{\text{m}}$} \right]  \left[ \text{e$_{9}^{\text{m,*}}$} \right] + \text{k}_{ \text{e$_8^{\text{m}}$:e$_9^{\text{m}}$} }^{\text{-}}  \left[ \text{TEN$^{*}$} \right]&
 \end{flalign*} 

 \begin{flalign*} 
 \frac{d}{dt}\left[ \text{  z$_{9}$  } \right]&= \text{k}_{ \text{\text{flow}}}  ( \text{[z$_{9}$]$^{\text{up}}$} - \left[\text{z$_{9}$}\right] ) - \text{k}_{9} ^{\text{on}} \left[ \text{z$_{9}$} \right]  \left[ \text{p}_{ \text{9} }^{\text{avail}} \right]   + \text{k}_{ \text{9} }^{\text{off}}  \left[ \text{z$_9^{\text{m}}$} \right]\\
 &\hskip2.5ex - \text{k}_{ \text{z$_{9}$:e$_7^{\text{m}}$} }^{\text{+}}  \left[ \text{z$_{9}$} \right]  \left[ \text{e$_7^{\text{m}}$} \right] + \text{k}_{ \text{z$_{9}$:e$_7^{\text{m}}$} }^{\text{-}}  \left[ \text{z$_{9}$:e$_7^{\text{m}}$} \right] \\
 &\hskip2.5ex - \text{k}_{ \text{z$_{9}$:e$_{11\text{h}}$} }^{\text{+}}  \left[ \text{z$_{9}$} \right]  \left[ \text{e$_{11\text{h}}$} \right] + \text{k}_{ \text{z$_{9}$:e$_{11\text{h}}$} }^{\text{-}}  \left[ \text{z$_{9}$:e$_{11\text{h}}$} \right] \\
 &\hskip2.5ex - \text{k}_{ \text{z$_{9}$:e$_{11}$} }^{\text{+}}  \left[ \text{z$_{9}$} \right]  \left[ \text{e$_{11}$} \right] + \text{k}_{ \text{z$_{9}$:e$_{11}$} }^{\text{-}}  \left[ \text{z$_{9}$:e$_{11}$} \right]&
 \end{flalign*} 

 \begin{flalign*} 
 \frac{d}{dt}\left[ \text{  e$_{9}$  } \right]&= \text{k}_{\text{flow}}  ( \text{ [e$_{9}$]$^{\text{up}}$}  - \left[ \text{e$_{9}$} \right] ) \\
 &\hskip2.5ex - \text{k}_{ \text{9} }^{\text{on}}  \left[ \text{p}_{ \text{9} }^{\text{avail}} \right]  \left[ \text{e$_{9}$} \right] + \text{k}_{ \text{9} }^{\text{off}}  \left[ \text{e$_9^{\text{m}}$} \right] \\
 &\hskip2.5ex + \text{k}_{ \text{z$_{9}$:e$_7^{\text{m}}$} }^{\text{cat}}  \left[ \text{z$_{9}$:e$_7^{\text{m}}$} \right] \\
 &\hskip2.5ex - \text{k}_{ \text{ \text{e$_{9}$:AT} }}  \left[ \text{AT} \right]  \left[ \text{e$_{9}$} \right] - \text{k}_{\text{z$_{7}$:e$_{9}$}}^{\text{+}}  \left[ \text{z$_{7}$} \right]  \left[ \text{e$_{9}$} \right] \\
 &\hskip2.5ex + ( \text{k}_{ \text{z$_{7}$:e$_{9}$} }^{\text{cat}} + \text{k}_{ \text{z$_{7}$:e$_{9}$} }^{\text{-}} )  \left[ \text{z$_{7}$:e$_{9}$} \right] \\
 &\hskip2.5ex + ( \text{k}_{ \text{z$_7^{\text{m}}$:e$_{9}$} }^{\text{cat}} + \text{k}_{ \text{z$_7^{\text{m}}$:e$_{9}$} }^{\text{-}} )  \left[ \text{z$_7^{\text{m}}$:e$_{9}$} \right] - \text{k}_{ \text{z$_7^{\text{m}}$:e$_{9}$} }^{\text{+}}  \left[ \text{z$_7^{\text{m}}$} \right]  \left[ \text{e$_{9}$} \right] \\
 &\hskip2.5ex - \text{k}_{ \text{9} }^{\text{on}}  \left[ \text{p}_{ \text{9,*} }^{\text{avail}} \right]  \left[ \text{e$_{9}$} \right] + \text{k}_{ \text{9} }^{\text{off}}  \left[ \text{e$_{9}^{\text{m,*}}$} \right] \\
 &\hskip2.5ex - \text{k}_\text{diff}  ( \left[ \text{e$_{9}$} \right] - \left[ \text{e$_{9}$$^{\text{ec}}$} \right] ) \\
 &\hskip2.5ex + \text{k}_{ \text{z$_{9}$:e$_{11\text{h}}$} }^{\text{cat}}  \left[ \text{z$_{9}$:e$_{11\text{h}}$} \right] + \text{k}_{ \text{z$_{9}$:e$_{11}$} }^{\text{cat}}  \left[ \text{z$_{9}$:e$_{11}$} \right] \\
 &\hskip2.5ex - \text{k}_{ \text{e$_{9}$:ATH} }  \left[ \text{ATH} \right]  \left[ \text{e$_{9}$} \right] &
 \end{flalign*} 

 \begin{flalign*} 
 \frac{d}{dt}\left[ \text{  z$_9^{\text{m}}$  } \right]&= \text{k}_{ \text{9} }^{\text{on}}  \left[ \text{p}_{ \text{9} }^{\text{avail}} \right]  \left[ \text{z$_{9}$} \right] - \text{k}_{ \text{9} }^{\text{off}}  \left[ \text{z$_9^{\text{m}}$} \right]\\
 &\hskip2.5ex - \text{k}_{ \text{z$_9^{\text{m}}$:e$_{11\text{h}}^{\text{m}}$} }^{\text{+}}  \left[ \text{z$_9^{\text{m}}$} \right]  \left[ \text{e$_{11\text{h}}^{\text{m}}$} \right] + \text{k}_{ \text{z$_9^{\text{m}}$:e$_{11\text{h}}^{\text{m}}$} }^{\text{-}}  \left[ \text{z$_9^{\text{m}}$:e$_{11\text{h}}^{\text{m}}$} \right] \\
 &\hskip2.5ex - \text{k}_{ \text{z$_9^{\text{m}}$:e$_{11}^{\text{m,*}}$} }^{\text{+}}  \left[ \text{z$_9^{\text{m}}$} \right]  \left[ \text{e$_{11}^{\text{m,*}}$} \right] + \text{k}_{ \text{z$_9^{\text{m}}$:e$_{11}^{\text{m,*}}$} }^{\text{-}}  \left[ \text{z$_9^{\text{m}}$:e$_{11}^{\text{m,*}}$} \right]&
 \end{flalign*} 

 \begin{flalign*} 
 \frac{d}{dt}\left[ \text{  e$_9^{\text{m}}$  } \right]&= \text{k}_{ \text{9} }^{\text{on}}  \left[ \text{p}_{ \text{9} }^{\text{avail}} \right]  \left[ \text{e$_{9}$} \right] - \text{k}_{ \text{9} }^{\text{off}}  \left[ \text{e$_9^{\text{m}}$} \right]\\
 &\hskip2.5ex - \text{k}_{ \text{e$_8^{\text{m}}$:e$_9^{\text{m}}$} }^{\text{+}}  \left[ \text{e$_8^{\text{m}}$} \right]  \left[ \text{e$_9^{\text{m}}$} \right] + \text{k}_{ \text{e$_8^{\text{m}}$:e$_9^{\text{m}}$} }^{\text{-}}  \left[ \text{TEN} \right] \\
 &\hskip2.5ex + \text{k}_{ \text{z$_9^{\text{m}}$:e$_{11\text{h}}^{\text{m}}$} }^{\text{cat}}  \left[ \text{z$_9^{\text{m}}$:e$_{11\text{h}}^{\text{m}}$} \right] \\
 &\hskip2.5ex + \text{k}_{ \text{z$_9^{\text{m}}$:e$_{11}^{\text{m,*}}$} }^{\text{cat}}  \left[ \text{z$_9^{\text{m}}$:e$_{11}^{\text{m,*}}$} \right] \\
 &\hskip2.5ex - \text{k}_{ \text{e$_9^{\text{m}}$:AT} }  \left[ \text{AT} \right]  \left[ \text{e$_9^{\text{m}}$} \right] \\
 &\hskip2.5ex - \text{k}_{ \text{e$_9^{\text{m}}$:ATH} }  \left[ \text{ATH} \right]  \left[ \text{e$_9^{\text{m}}$} \right] &
 \end{flalign*} 

 \begin{flalign*} 
 \frac{d}{dt}\left[ \text{  z$_{2}$  } \right]&= - \text{k}_{ \text{2} }^{\text{on}}  \left[ \text{p}_{ \text{2} }^{\text{avail}} \right]  \left[ \text{z$_{2}$} \right] + \text{k}_{ \text{2} }^{\text{off}}  \left[ \text{z$_2^{\text{m}}$} \right]\\
 &\hskip2.5ex + \text{k}_{\text{flow}}  ( \text{ [z$_{2}$]$^{\text{up}}$}  - \left[ \text{z$_{2}$} \right] )  &
 \end{flalign*} 

 \begin{flalign*} 
 \frac{d}{dt}\left[ \text{  e$_{\text{2}}$  } \right]&= \text{k}_{\text{flow}}  ( \text{ [e$_{\text{2}}$]$^{\text{up}}$}  - \left[ \text{e$_{\text{2}}$} \right] ) \\
 &\hskip2.5ex - \text{k}_{ \text{2,*} }^{\text{on}}  \left[ \text{p}_{ \text{2,*} }^{\text{avail}} \right]  \left[ \text{e$_{\text{2}}$} \right] + \text{k}_{ \text{2,*} }^{\text{off}}  \left[ \text{e$_2^{\text{m}}$} \right] \\
 &\hskip2.5ex + \text{k}_{ \text{z$_2^{\text{m}}$:PRO} }^{\text{cat}}  \left[ \text{z$_2^{\text{m}}$:PRO} \right] \\
 &\hskip2.5ex - \text{k}_{ \text{e$_{\text{2}}$:AT} }  \left[ \text{AT} \right]  \left[ \text{e$_{\text{2}}$} \right] \\
 &\hskip2.5ex - \text{k}_{ \text{z$_{5}$:e$_{\text{2}}$} }^{\text{+}}  \left[ \text{z$_{5}$} \right]  \left[ \text{e$_{\text{2}}$} \right] + ( \text{k}_{ \text{z$_{5}$:e$_{\text{2}}$} }^{\text{cat}} + \text{k}_{ \text{z$_{5}$:e$_{\text{2}}$} }^{\text{-}} )  \left[ \text{z$_{5}$:e$_{\text{2}}$} \right] \\
 &\hskip2.5ex - \text{k}_{ \text{z$_{8}$:e$_{\text{2}}$} }^{\text{+}}  \left[ \text{z$_{8}$} \right]  \left[ \text{e$_{\text{2}}$} \right] + ( \text{k}_{ \text{z$_{8}$:e$_{\text{2}}$} }^{\text{cat}} + \text{k}_{ \text{z$_{8}$:e$_{\text{2}}$} }^{\text{-}} )  \left[ \text{z$_{8}$:e$_{\text{2}}$} \right] \\
 &\hskip2.5ex - \text{k}_{ \text{z$_{7}$:e$_{\text{2}}$} }^{\text{+}}  \left[ \text{z$_{7}$} \right]  \left[ \text{e$_{\text{2}}$} \right] + ( \text{k}_{ \text{z$_{7}$:e$_{\text{2}}$} }^{\text{cat}} + \text{k}_{ \text{z$_{7}$:e$_{\text{2}}$} }^{\text{-}} )  \left[ \text{z$_{7}$:e$_{\text{2}}$} \right] \\
 &\hskip2.5ex - \text{k}_{ \text{z$_7^{\text{m}}$:e$_{\text{2}}$} }^{\text{+}}  \left[ \text{z$_7^{\text{m}}$} \right]  \left[ \text{e$_{\text{2}}$} \right] + ( \text{k}_{ \text{z$_7^{\text{m}}$:e$_{\text{2}}$} }^{\text{cat}} + \text{k}_{ \text{z$_7^{\text{m}}$:e$_{\text{2}}$} }^{\text{-}} )  \left[ \text{z$_7^{\text{m}}$:e$_{\text{2}}$} \right] \\
 &\hskip2.5ex - \text{k}_\text{diffRZ}  ( \left[ \text{e$_{\text{2}}$} \right] - \left[ \text{e$_{\text{2}}$$^{\text{ec}}$} \right] ) \\
 &\hskip2.5ex - \text{k}_{ \text{z$_{11}$:e$_{\text{2}}$} }^{\text{+}}  \left[ \text{z$_{11}$} \right]  \left[ \text{e$_{\text{2}}$} \right] + ( \text{k}_{ \text{z$_{11}$:e$_{\text{2}}$} }^{\text{-}} + \text{k}_{ \text{z$_{11}$:e$_{\text{2}}$} }^{\text{cat}} )  \left[ \text{z$_{11}$:e$_{\text{2}}$} \right] \\
 &\hskip2.5ex - \text{k}_{ \text{e$_{11\text{h}}$:e$_{\text{2}}$} }^{\text{+}}  \left[ \text{e$_{11\text{h}}$} \right]  \left[ \text{e$_{\text{2}}$} \right] + ( \text{k}_{ \text{e$_{11\text{h}}$:e$_{\text{2}}$} }^{\text{-}} + \text{k}_{ \text{e$_{11\text{h}}$:e$_{\text{2}}$} }^{\text{cat}} )  \left[ \text{e$_{11\text{h}}$:e$_{\text{2}}$} \right] \\
 &\hskip2.5ex + \text{k}_{ \text{z$_2^{\text{m}}$:PROh} }^{\text{cat}}  \left[ \text{z$_2^{\text{m}}$:PROh} \right] \\
 &\hskip2.5ex - \text{k}_{ \text{e$_{\text{5h}}$:e$_{\text{2}}$} }^{\text{+}}  \left[ \text{e$_{\text{2}}$} \right]  \left[ \text{e$_{\text{5h}}$} \right] + \text{k}_{ \text{e$_{\text{5h}}$:e$_{\text{2}}$} }^{\text{-}}  \left[ \text{e$_{\text{5h}}$:e$_{\text{2}}$} \right] + \text{k}_{ \text{e$_{\text{5h}}$:e$_{\text{2}}$} }^{\text{cat}}  \left[ \text{e$_{\text{5h}}$:e$_{\text{2}}$} \right] \\
 &\hskip2.5ex - \text{k}_{ \text{e$_{\text{2}}$:ATH} }  \left[ \text{ATH} \right]  \left[ \text{e$_{\text{2}}$} \right]&
 \end{flalign*} 

 \begin{flalign*} 
 \frac{d}{dt}\left[ \text{  z$_2^{\text{m}}$  } \right]&= \text{k}_{ \text{2} }^{\text{on}}  \left[ \text{p}_{ \text{2} }^{\text{avail}} \right]  \left[ \text{z$_{2}$} \right] - \text{k}_{ \text{2} }^{\text{off}}  \left[ \text{z$_2^{\text{m}}$} \right]\\
 &\hskip2.5ex - \text{k}_{ \text{z$_2^{\text{m}}$:PRO} }^{\text{+}}  \left[ \text{PRO} \right]  \left[ \text{z$_2^{\text{m}}$} \right] + \text{k}_{ \text{z$_2^{\text{m}}$:PRO} }^{\text{-}}  \left[ \text{z$_2^{\text{m}}$:PRO} \right] \\
 &\hskip2.5ex - \text{k}_{ \text{z$_2^{\text{m}}$:PROh} }^{\text{+}}  \left[ \text{PROh} \right]  \left[ \text{z$_2^{\text{m}}$} \right] + \text{k}_{ \text{z$_2^{\text{m}}$:PROh} }^{\text{-}}  \left[ \text{z$_2^{\text{m}}$:PROh} \right]&
 \end{flalign*} 

 \begin{flalign*} 
 \frac{d}{dt}\left[ \text{  e$_2^{\text{m}}$  } \right]&= \text{k}_{ \text{2,*} }^{\text{on}}  \left[ \text{p}_{ \text{2,*} }^{\text{avail}} \right]  \left[ \text{e$_{\text{2}}$} \right] - \text{k}_{ \text{2,*} }^{\text{off}}  \left[ \text{e$_2^{\text{m}}$} \right]\\
 &\hskip2.5ex + ( \text{k}_{ \text{z$_5^{\text{m}}$:e$_2^{\text{m}}$} }^{\text{cat}} + \text{k}_{ \text{z$_5^{\text{m}}$:e$_2^{\text{m}}$} }^{\text{-}} )  \left[ \text{z$_5^{\text{m}}$:e$_2^{\text{m}}$} \right] \\
 &\hskip2.5ex - \text{k}_{ \text{z$_5^{\text{m}}$:e$_2^{\text{m}}$} }^{\text{+}}  \left[ \text{z$_5^{\text{m}}$} \right]  \left[ \text{e$_2^{\text{m}}$} \right] \\
 &\hskip2.5ex + ( \text{k}_{ \text{z$_8^{\text{m}}$:e$_2^{\text{m}}$} }^{\text{cat}} + \text{k}_{ \text{z$_8^{\text{m}}$:e$_2^{\text{m}}$} }^{\text{-}} )  \left[ \text{z$_8^{\text{m}}$:e$_2^{\text{m}}$} \right] - \text{k}_{ \text{z$_8^{\text{m}}$:e$_2^{\text{m}}$} }^{\text{+}}  \left[ \text{z$_8^{\text{m}}$} \right]  \left[ \text{e$_2^{\text{m}}$} \right] \\
 &\hskip2.5ex - \text{k}_{ \text{z$_{11}^{\text{m}}$:e$_2^{\text{m}}$} }^{\text{+}}  \left[ \text{z$_{11}^{\text{m}}$} \right]  \left[ \text{e$_2^{\text{m}}$} \right] \\
 &\hskip2.5ex + ( \text{k}_{ \text{z$_{11}^{\text{m}}$:e$_2^{\text{m}}$} }^{\text{-}} + \text{k}_{ \text{z$_{11}^{\text{m}}$:e$_2^{\text{m}}$} }^{\text{cat}} )  \left[ \text{z$_{11}^{\text{m}}$:e$_2^{\text{m}}$} \right] \\
 &\hskip2.5ex - \text{k}_{ \text{e$_{11\text{h}}^{\text{m,*}}$:e$_2^{\text{m}}$} }^{\text{+}}  \left[ \text{e$_{11\text{h}}^{\text{m,*}}$} \right]  \left[ \text{e$_2^{\text{m}}$} \right] \\
 &\hskip2.5ex + ( \text{k}_{ \text{e$_{11\text{h}}^{\text{m,*}}$:e$_2^{\text{m}}$} }^{\text{-}} + \text{k}_{ \text{e$_{11\text{h}}^{\text{m,*}}$:e$_2^{\text{m}}$} }^{\text{cat}} )  \left[ \text{e$_{11\text{h}}^{\text{m,*}}$:e$_2^{\text{m}}$} \right] \\
 &\hskip2.5ex - \text{k}_{ \text{e$_{5\text{h}}^{\text{m}}$:e$_2^{\text{m}}$} }^{\text{+}}  \left[ \text{e$_2^{\text{m}}$} \right]  \left[ \text{e$_{5\text{h}}^{\text{m}}$} \right] + \text{k}_{ \text{e$_{5\text{h}}^{\text{m}}$:e$_2^{\text{m}}$} }^{\text{-}}  \left[ \text{e$_{5\text{h}}^{\text{m}}$:e$_2^{\text{m}}$} \right] \\
 &\hskip2.5ex + \text{k}_{ \text{e$_{5\text{h}}^{\text{m}}$:e$_2^{\text{m}}$} }^{\text{cat}}  \left[ \text{e$_{5\text{h}}^{\text{m}}$:e$_2^{\text{m}}$} \right] \\
 &\hskip2.5ex - \text{k}_{ \text{PROh:e$_2^{\text{m}}$} }^{\text{+}}  \left[ \text{PROh} \right]  \left[ \text{e$_2^{\text{m}}$} \right] + \text{k}_{ \text{PROh:e$_2^{\text{m}}$} }^{\text{-}}  \left[ \text{PROh:e$_2^{\text{m}}$} \right] \\
 &\hskip2.5ex + \text{k}_{ \text{PROh:e$_2^{\text{m}}$} }^{\text{cat}}  \left[ \text{PROh:e$_2^{\text{m}}$} \right] \\
 &\hskip2.5ex - \text{k}_{ \text{e$_2^{\text{m}}$:AT} }  \left[ \text{AT} \right]  \left[ \text{e$_2^{\text{m}}$} \right] \\
 &\hskip2.5ex - \text{k}_{ \text{e$_2^{\text{m}}$:ATH} }  \left[ \text{ATH} \right]  \left[ \text{e$_2^{\text{m}}$} \right]&
 \end{flalign*} 

 \begin{flalign*} 
 \frac{d}{dt}\left[ \text{  TEN  } \right]&= - \text{k}_{ \text{e$_8^{\text{m}}$:e$_9^{\text{m}}$} }^{\text{-}}  \left[ \text{TEN} \right] + \text{k}_{ \text{e$_8^{\text{m}}$:e$_9^{\text{m}}$} }^{\text{+}}  \left[ \text{e$_8^{\text{m}}$} \right]  \left[ \text{e$_9^{\text{m}}$} \right]\\
 &\hskip2.5ex + ( \text{k}_{ \text{z$_{10}^{\text{m}}$:TEN} }^{\text{cat}} + \text{k}_{ \text{z$_{10}^{\text{m}}$:TEN} }^{\text{-}} )  \left[ \text{z$_{10}^{\text{m}}$:TEN} \right] - \text{k}_{ \text{z$_{10}^{\text{m}}$:TEN} }^{\text{+}}  \left[ \text{TEN} \right]  \left[ \text{z$_{10}^{\text{m}}$} \right]&
 \end{flalign*} 

 \begin{flalign*} 
 \frac{d}{dt}\left[ \text{  PRO  } \right]&= - \text{k}_{ \text{e$_5^{\text{m}}$:e$_{10}^{\text{m}}$} }^{\text{-}}  \left[ \text{PRO} \right] + \text{k}_{ \text{e$_5^{\text{m}}$:e$_{10}^{\text{m}}$} }^{\text{+}}  \left[ \text{e$_{10}^{\text{m}}$} \right]  \left[ \text{e$_5^{\text{m}}$} \right]\\
 &\hskip2.5ex + ( \text{k}_{ \text{z$_2^{\text{m}}$:PRO} }^{\text{cat}} + \text{k}_{ \text{z$_2^{\text{m}}$:PRO} }^{\text{-}} )  \left[ \text{z$_2^{\text{m}}$:PRO} \right] - \text{k}_{ \text{z$_2^{\text{m}}$:PRO} }^{\text{+}}  \left[ \text{PRO} \right]  \left[ \text{z$_2^{\text{m}}$} \right] \\
 &\hskip2.5ex + \text{k}_{ \text{PROh:e$_2^{\text{m}}$} }^{\text{cat}}  \left[ \text{PROh:e$_2^{\text{m}}$} \right]&
 \end{flalign*} 

 \begin{flalign*} 
 \frac{d}{dt}\left[ \text{  PL$^{\text{as}}$  } \right]&= \text{k}^{\text{+}}  \text{(P$_\text{{max}}$ - [PL$^{\text{as}}$])}  \left[ \text{PL} \right] - \text{k}^{\text{-}}  \left[ \text{PL$^{\text{as}}$} \right]\\
 &\hskip2.5ex + \text{k}^{\text{+1}}  \left[ \text{PL$^{\text{av}}$} \right]  \text{(P$_\text{{max}}$ - [PL$^{\text{as}}$])} &
 \end{flalign*} 

\begin{flalign*} 
 \frac{d}{dt}\left[ \text{  PL  } \right]&= - \text{k}^{\text{+}}  \left[ \text{PL} \right]  \text{(P$_\text{{max}}$ - [PL$^{\text{as}}$])}  - \text{k}_{\text{plact}}  ( \left[ \text{PL$^{\text{av}}$} \right] + \left[ \text{PL$^{\text{as}}$} \right] ) \left[ \text{PL} \right] &
\end{flalign*}

\begin{flalign*} 
 \frac{d}{dt}\left[ \text{  PL$^{\text{av}}$  } \right]&= \text{k}^{\text{-}}  \left[ \text{PL$^{\text{as}}$} \right]
+ \text{k}_{\text{plact}}  \left[ \text{PL} \right]  ( \left[ \text{PL$^{\text{av}}$} \right] + \left[ \text{PL$^{\text{as}}$} \right]) \\
 &\hskip2.5ex + \text{k}_{\text{act}}  \left[ \text{PL} \right]  \left[ \text{e$_{\text{2}}$} \right]/ ( \left[ \text{e$_{\text{2}}$} \right] + 0.001 ) \\
 &\hskip2.5ex - \text{k}^{\text{+1}}  \left[ \text{PL$^{\text{av}}$} \right]  \text{(P$_\text{{max}}$ - [PL$^{\text{as}}$])} &
 \end{flalign*} 

 \begin{flalign*} 
 \frac{d}{dt}\left[ \text{  TFPI  } \right]&= - \text{k}_{ \text{TFPI:e$_{10}$} }^{\text{+}}  \left[ \text{e$_{10}$} \right]  \left[ \text{TFPI} \right]\\
 &\hskip2.5ex + \text{k$_{\text{TFPI:e$_{10}$}}^{-}$}  \left[ \text{TFPI:e$_{10}$} \right] + \text{k}_{\text{flow}}  ( \text{ [TFPI]$^{\text{up}}$}  - \left[ \text{TFPI} \right] ) \\
 &\hskip2.5ex - \text{k}_{ \text{TFPI:e$_{5\text{h}}^{\text{m}}$} }^{\text{+}}  \left[ \text{e$_{5\text{h}}^{\text{m}}$} \right]  \left[ \text{TFPI} \right] + \text{k}_{ \text{TFPI:e$_{5\text{h}}^{\text{m}}$} }^{\text{-}}  \left[ \text{TFPI:e$_{5\text{h}}^{\text{m}}$} \right] \\
 &\hskip2.5ex - \text{k}_{ \text{TFPI:e$_{\text{5h}}$} }^{\text{+}}  \left[ \text{e$_{\text{5h}}$} \right]  \left[ \text{TFPI} \right] + \text{k}_{ \text{TFPI:e$_{\text{5h}}$} }^{\text{-}}  \left[ \text{TFPI:e$_{\text{5h}}$} \right] \\
 &\hskip2.5ex - \text{k}_{ \text{TFPI:e$_{10}^{\text{m}}$} }^{\text{+}}  \left[ \text{TFPI} \right]  \left[ \text{e$_{10}^{\text{m}}$} \right] + \text{k}_{ \text{TFPI:e$_{10}^{\text{m}}$} }^{\text{-}}  \left[ \text{TFPI:e$_{10}^{\text{m}}$} \right] \\
 &\hskip2.5ex - \text{k}_{ \text{TFPI:PROhv10} }^{\text{+}}  \left[ \text{PROh} \right]  \left[ \text{TFPI} \right] + \text{k}_{ \text{TFPI:PROhv10} }^{\text{-}}  \left[ \text{TFPI:PROhv10} \right] \\
 &\hskip2.5ex - \text{k}_{ \text{TFPI:PROhv5} }^{\text{+}}  \left[ \text{PROh} \right]  \left[ \text{TFPI} \right] + \text{k}_{ \text{TFPI:PROhv5} }^{\text{-}}  \left[ \text{TFPI:PROhv5} \right] \\
 &\hskip2.5ex +\text{n}_\text{TFPI}  \left[ \frac{d}{dt} \left( [\text{PL}^{\text{as}}] + [\text{PL}^{\text{av}}] \right) \right]  &
 \end{flalign*} 

 \begin{flalign*} 
 \frac{d}{dt}\left[ \text{  TFPI:e$_{10}$  } \right]&= \text{k}_{ \text{TFPI:e$_{10}$} }^{\text{+}}  \left[ \text{e$_{10}$} \right]  \left[ \text{TFPI} \right] - \text{k$_{\text{TFPI:e$_{10}$}}^{-}$}  \left[ \text{TFPI:e$_{10}$} \right]\\
 &\hskip2.5ex + \text{k$_{\text{(TFPI:e$_{10}$):e$_7^{\text{m}}$}}^{-}$}  \left[ \text{TFPI:e$_{10}$:e$_{7}^{\text{m}}$} \right] \\
 &\hskip2.5ex - \text{k$_\text{(TFPI:e$_{10}$):e$_7^{\text{m}}$}^{+}$}   \left[ \text{e$_7^{\text{m}}$} \right]  \left[ \text{TFPI:e$_{10}$} \right] + \text{k}_{\text{flow}}  ( \text{ [TFPI:e$_{10}$]$^{\text{up}}$}  - \left[ \text{TFPI:e$_{10}$} \right] ) \\
 &\hskip2.5ex - \text{k}_{ \text{(TFPI:e$_{10}$):e$_{\text{5h}}$} }^{\text{+}}  \left[ \text{TFPI:e$_{10}$} \right]  \left[ \text{e$_{\text{5h}}$} \right] + \text{k}_{ \text{(TFPI:e$_{10}$):e$_{\text{5h}}$} }^{\text{-}}  \left[ \text{e$_{10}$:TFPI:e$_{\text{5h}}$} \right] \\
 &\hskip2.5ex - \text{k}_{ \text{10} }^{\text{on}}  \left[ \text{TFPI:e$_{10}$} \right]  \left[ \text{p}_{ \text{10} }^{\text{avail}} \right] + \text{k}_{ \text{10} }^{\text{off}}  \left[ \text{TFPI:e$_{10}^{\text{m}}$} \right]&
 \end{flalign*} 

 \begin{flalign*} 
 \frac{d}{dt}\left[ \text{  TFPI:e$_{10}$:e$_{7}^{\text{m}}$  } \right]&= - \text{k$_{\text{(TFPI:e$_{10}$):e$_7^{\text{m}}$}}^{-}$}  \left[ \text{TFPI:e$_{10}$:e$_{7}^{\text{m}}$} \right] + \text{k}_{ \text{(TFPI:e$_{10}$):e$_7^{\text{m}}$} }^{\text{+}}  \left[ \text{e$_7^{\text{m}}$} \right]  \left[ \text{TFPI:e$_{10}$} \right]\\
 &\hskip2.5ex - \left[ \text{TFPI:e$_{10}$:e$_{7}^{\text{m}}$} \right]  \frac{d}{dt} \left( [\text{PL}^{\text{as}}] \right) /\text{(P$_\text{{max}}$ - [PL$^{\text{as}}$])} &
 \end{flalign*} 

 \begin{flalign*} 
 \frac{d}{dt}\left[ \text{  APC  } \right]&=  ( \text{k}_{ \text{e$_5^{\text{m}}$:APC} }^{\text{cat}} + \text{k}_{ \text{e$_5^{\text{m}}$:APC} }^{\text{-}} )  \left[ \text{APC:e$_5^{\text{m}}$} \right]\\
 &\hskip2.5ex - \text{k}_{ \text{e$_5^{\text{m}}$:APC} }^{\text{+}}  \left[ \text{APC} \right]  \left[ \text{e$_5^{\text{m}}$} \right] \\
 &\hskip2.5ex + ( \text{k}_{ \text{e$_8^{\text{m}}$:APC} }^{\text{cat}} + \text{k}_{ \text{e$_8^{\text{m}}$:APC} }^{\text{-}} )  \left[ \text{APC:e$_8^{\text{m}}$} \right] - \text{k}_{ \text{e$_8^{\text{m}}$:APC} }^{\text{+}}  \left[ \text{APC} \right]  \left[ \text{e$_8^{\text{m}}$} \right] \\
 &\hskip2.5ex + \text{k}_{\text{flow}}  ( \text{ [APC]$^{\text{up}}$}  - \left[ \text{APC} \right] ) \\
 &\hskip2.5ex - \text{k}_\text{diffRZ}  ( \left[ \text{APC} \right] - \left[ \text{APC$^{\text{ec}}$} \right] ) \\
 &\hskip2.5ex + ( \text{k}_{ \text{e$_{\text{5}}$:APC} }^{\text{cat}} + \text{k}_{ \text{e$_{\text{5}}$:APC} }^{\text{-}} )  \left[ \text{APC:e$_{\text{5}}$} \right] - \text{k}_{ \text{e$_{\text{5}}$:APC} }^{\text{+}}  \left[ \text{APC} \right]  \left[ \text{e$_{\text{5}}$} \right] \\
 &\hskip2.5ex + ( \text{k}_{ \text{e$_{8}$:APC} }^{\text{cat}} + \text{k}_{ \text{e$_{8}$:APC} }^{\text{-}} )  \left[ \text{APC:e$_{8}$} \right] - \text{k}_{ \text{e$_{8}$:APC} }^{\text{+}}  \left[ \text{APC} \right]  \left[ \text{e$_{8}$} \right] \\
 &\hskip2.5ex - \text{k}_{ \text{e$_{5\text{h}}^{\text{m}}$:APC} }^{\text{+}}  \left[ \text{e$_{5\text{h}}^{\text{m}}$} \right]  \left[ \text{APC} \right] + \text{k}_{ \text{e$_{5\text{h}}^{\text{m}}$:APC} }^{\text{-}}  \left[ \text{APC:e$_{5\text{h}}^{\text{m}}$} \right] \\
 &\hskip2.5ex + \text{k}_{ \text{e$_{5\text{h}}^{\text{m}}$:APC} }^{\text{cat}}  \left[ \text{APC:e$_{5\text{h}}^{\text{m}}$} \right] - \text{k}_{ \text{e$_{\text{5h}}$:APC} }^{\text{+}}  \left[ \text{APC} \right]  \left[ \text{e$_{\text{5h}}$} \right] \\
 &\hskip2.5ex + \text{k}_{ \text{e$_{\text{5h}}$:APC} }^{\text{-}}  \left[ \text{APC:e$_{\text{5h}}$} \right] + \text{k}_{ \text{e$_{\text{5h}}$:APC} }^{\text{cat}}  \left[ \text{APC:e$_{\text{5h}}$} \right]&
 \end{flalign*} 

 \begin{flalign*} 
 \frac{d}{dt}\left[ \text{  z$_{7}$:e$_{\text{2}}$  } \right]&= \text{k}_{\text{flow}}  ( \text{ [z$_{7}$:e$_{\text{2}}$]$^{\text{up}}$}  - \left[ \text{z$_{7}$:e$_{\text{2}}$} \right] ) \\
 &\hskip2.5ex + \text{k}_{ \text{z$_{7}$:e$_{\text{2}}$} }^{\text{+}}  \left[ \text{e$_{\text{2}}$} \right]  \left[ \text{z$_{7}$} \right] - ( \text{k}_{ \text{z$_{7}$:e$_{\text{2}}$} }^{\text{cat}} + \text{k}_{ \text{z$_{7}$:e$_{\text{2}}$} }^{\text{-}} )  \left[ \text{z$_{7}$:e$_{\text{2}}$} \right]&
 \end{flalign*} 

 \begin{flalign*} 
 \frac{d}{dt}\left[ \text{  z$_{7}$:e$_{10}$  } \right]&= \text{k}_{ \text{z$_{7}$:e$_{10}$} }^{\text{+}}  \left[ \text{e$_{10}$} \right]  \left[ \text{z$_{7}$} \right] - ( \text{k}_{ \text{z$_{7}$:e$_{10}$} }^{\text{cat}} + \text{k}_{ \text{z$_{7}$:e$_{10}$} }^{\text{-}} )  \left[ \text{z$_{7}$:e$_{10}$} \right]\\
 &\hskip2.5ex + \text{k}_{\text{flow}}  ( \text{ [z$_{7}$:e$_{10}$]$^{\text{up}}$}  - \left[ \text{z$_{7}$:e$_{10}$} \right] )  &
 \end{flalign*} 

 \begin{flalign*} 
 \frac{d}{dt}\left[ \text{  z$_7^{\text{m}}$:e$_{10}$  } \right]&= \text{k}_{ \text{z$_7^{\text{m}}$:e$_{10}$} }^{\text{+}}  \left[ \text{e$_{10}$} \right]  \left[ \text{z$_7^{\text{m}}$} \right] - ( \text{k}_{ \text{z$_7^{\text{m}}$:e$_{10}$} }^{\text{cat}} + \text{k}_{ \text{z$_7^{\text{m}}$:e$_{10}$} }^{\text{-}} )  \left[ \text{z$_7^{\text{m}}$:e$_{10}$} \right]\\
 &\hskip2.5ex - \left[ \text{z$_7^{\text{m}}$:e$_{10}$} \right]  \frac{d}{dt} \left( [\text{PL}^{\text{as}}] \right) /\text{(P$_\text{{max}}$ - [PL$^{\text{as}}$])} &
 \end{flalign*} 

 \begin{flalign*} 
 \frac{d}{dt}\left[ \text{  z$_7^{\text{m}}$:e$_{\text{2}}$  } \right]&= \text{k}_{ \text{z$_7^{\text{m}}$:e$_{\text{2}}$} }^{\text{+}}  \left[ \text{e$_{\text{2}}$} \right]  \left[ \text{z$_7^{\text{m}}$} \right] - ( \text{k}_{ \text{z$_7^{\text{m}}$:e$_{\text{2}}$} }^{\text{cat}} + \text{k}_{ \text{z$_7^{\text{m}}$:e$_{\text{2}}$} }^{\text{-}} )  \left[ \text{z$_7^{\text{m}}$:e$_{\text{2}}$} \right]\\
 &\hskip2.5ex - \left[ \text{z$_7^{\text{m}}$:e$_{\text{2}}$} \right]  \frac{d}{dt} \left( [\text{PL}^{\text{as}}] \right) /\text{(P$_\text{{max}}$ - [PL$^{\text{as}}$])} &
 \end{flalign*} 

 \begin{flalign*} 
 \frac{d}{dt}\left[ \text{  z$_{10}$:e$_7^{\text{m}}$  } \right]&= \text{k}_{ \text{z$_{10}$:e$_7^{\text{m}}$} }^{\text{+}}  \left[ \text{e$_7^{\text{m}}$} \right]  \left[ \text{z$_{10}$} \right] - ( \text{k}_{ \text{z$_{10}$:e$_7^{\text{m}}$} }^{\text{cat}} + \text{k}_{ \text{z$_{10}$:e$_7^{\text{m}}$} }^{\text{-}} )  \left[ \text{z$_{10}$:e$_7^{\text{m}}$} \right]\\
 &\hskip2.5ex - \left[ \text{z$_{10}$:e$_7^{\text{m}}$} \right]  \frac{d}{dt} \left( [\text{PL}^{\text{as}}] \right) /\text{(P$_\text{{max}}$ - [PL$^{\text{as}}$])} &
 \end{flalign*} 

 \begin{flalign*} 
 \frac{d}{dt}\left[ \text{  z$_{10}^{\text{m}}$:TEN  } \right]&= \text{k}_{ \text{z$_{10}^{\text{m}}$:TEN} }^{\text{+}}  \left[ \text{TEN} \right]  \left[ \text{z$_{10}^{\text{m}}$} \right] - ( \text{k}_{ \text{z$_{10}^{\text{m}}$:TEN} }^{\text{cat}} + \text{k}_{ \text{z$_{10}^{\text{m}}$:TEN} }^{\text{-}} )  \left[ \text{z$_{10}^{\text{m}}$:TEN} \right]&
 \end{flalign*} 

 \begin{flalign*} 
 \frac{d}{dt}\left[ \text{  z$_{5}$:e$_{\text{2}}$  } \right]&= \text{k}_{ \text{z$_{5}$:e$_{\text{2}}$} }^{\text{+}}  \left[ \text{e$_{\text{2}}$} \right]  \left[ \text{z$_{5}$} \right] - ( \text{k}_{ \text{z$_{5}$:e$_{\text{2}}$} }^{\text{cat}} + \text{k}_{ \text{z$_{5}$:e$_{\text{2}}$} }^{\text{-}} )  \left[ \text{z$_{5}$:e$_{\text{2}}$} \right]\\
 &\hskip2.5ex + \text{k}_{\text{flow}}  ( \text{ [z$_{5}$:e$_{\text{2}}$]$^{\text{up}}$}  - \left[ \text{z$_{5}$:e$_{\text{2}}$} \right] )  &
 \end{flalign*} 

 \begin{flalign*} 
 \frac{d}{dt}\left[ \text{  z$_5^{\text{m}}$:e$_{10}^{\text{m}}$  } \right]&= \text{k}_{ \text{z$_5^{\text{m}}$:e$_{10}^{\text{m}}$} }^{\text{+}}  \left[ \text{e$_{10}^{\text{m}}$} \right]  \left[ \text{z$_5^{\text{m}}$} \right] - ( \text{k}_{ \text{z$_5^{\text{m}}$:e$_{10}^{\text{m}}$} }^{\text{cat}} + \text{k}_{ \text{z$_5^{\text{m}}$:e$_{10}^{\text{m}}$} }^{\text{-}} )  \left[ \text{z$_5^{\text{m}}$:e$_{10}^{\text{m}}$} \right]&
 \end{flalign*} 

 \begin{flalign*} 
 \frac{d}{dt}\left[ \text{    z$_5^{\text{m}}$:e$_2^{\text{m}}$  } \right]&= \text{k}_{ \text{z$_5^{\text{m}}$:e$_2^{\text{m}}$} }^{\text{+}}  \left[ \text{e$_2^{\text{m}}$} \right]  \left[ \text{z$_5^{\text{m}}$} \right] - ( \text{k}_{ \text{z$_5^{\text{m}}$:e$_2^{\text{m}}$} }^{\text{cat}} + \text{k}_{ \text{z$_5^{\text{m}}$:e$_2^{\text{m}}$} }^{\text{-}} )  \left[ \text{z$_5^{\text{m}}$:e$_2^{\text{m}}$} \right]&
 \end{flalign*} 

 \begin{flalign*} 
 \frac{d}{dt}\left[ \text{  z$_8^{\text{m}}$:e$_{10}^{\text{m}}$  } \right]&= \text{k}_{ \text{z$_8^{\text{m}}$:e$_{10}^{\text{m}}$} }^{\text{+}}  \left[ \text{e$_{10}^{\text{m}}$} \right]  \left[ \text{z$_8^{\text{m}}$} \right] - ( \text{k}_{ \text{z$_8^{\text{m}}$:e$_{10}^{\text{m}}$} }^{\text{cat}} + \text{k}_{ \text{z$_8^{\text{m}}$:e$_{10}^{\text{m}}$} }^{\text{-}} )  \left[ \text{z$_8^{\text{m}}$:e$_{10}^{\text{m}}$} \right]&
 \end{flalign*} 

 \begin{flalign*} 
 \frac{d}{dt}\left[ \text{  z$_8^{\text{m}}$:e$_2^{\text{m}}$  } \right]&= \text{k}_{ \text{z$_8^{\text{m}}$:e$_2^{\text{m}}$} }^{\text{+}}  \left[ \text{e$_2^{\text{m}}$} \right]  \left[ \text{z$_{8}^{\text{m,*}}$} \right] - ( \text{k}_{ \text{z$_8^{\text{m}}$:e$_2^{\text{m}}$} }^{\text{cat}} + \text{k}_{ \text{z$_8^{\text{m}}$:e$_2^{\text{m}}$} }^{\text{-}} )  \left[ \text{z$_8^{\text{m}}$:e$_2^{\text{m}}$} \right]&
 \end{flalign*} 

 \begin{flalign*} 
 \frac{d}{dt}\left[ \text{  z$_{8}$:e$_{\text{2}}$  } \right]&= \text{k}_{ \text{z$_{8}$:e$_{\text{2}}$} }^{\text{+}}  \left[ \text{e$_{\text{2}}$} \right]  \left[ \text{z$_{8}$} \right] - ( \text{k}_{ \text{z$_{8}$:e$_{\text{2}}$} }^{\text{cat}} + \text{k}_{ \text{z$_{8}$:e$_{\text{2}}$} }^{\text{-}} )  \left[ \text{z$_{8}$:e$_{\text{2}}$} \right]\\
 &\hskip2.5ex + \text{k}_{\text{flow}}  ( \text{ [z$_{8}$:e$_{\text{2}}$]$^{\text{up}}$}  - \left[ \text{z$_{8}$:e$_{\text{2}}$} \right] )  &
 \end{flalign*} 

 \begin{flalign*} 
 \frac{d}{dt}\left[ \text{  APC:e$_8^{\text{m}}$  } \right]&= \text{k}_{ \text{e$_8^{\text{m}}$:APC} }^{\text{+}}  \left[ \text{APC} \right]  \left[ \text{e$_8^{\text{m}}$} \right] - ( \text{k}_{ \text{e$_8^{\text{m}}$:APC} }^{\text{cat}} + \text{k}_{ \text{e$_8^{\text{m}}$:APC} }^{\text{-}} )  \left[ \text{APC:e$_8^{\text{m}}$} \right]&
 \end{flalign*} 

 \begin{flalign*} 
 \frac{d}{dt}\left[ \text{  z$_{9}$:e$_7^{\text{m}}$  } \right]&= \text{k}_{ \text{z$_{9}$:e$_7^{\text{m}}$} }^{\text{+}}  \left[ \text{e$_7^{\text{m}}$} \right]  \left[ \text{z$_{9}$} \right] - ( \text{k}_{ \text{z$_{9}$:e$_7^{\text{m}}$} }^{\text{cat}} + \text{k}_{ \text{z$_{9}$:e$_7^{\text{m}}$} }^{\text{-}} )  \left[ \text{z$_{9}$:e$_7^{\text{m}}$} \right]\\
 &\hskip2.5ex - \left[ \text{z$_{9}$:e$_7^{\text{m}}$} \right]  \frac{d}{dt} \left( [\text{PL}^{\text{as}}] \right) /\text{(P$_\text{{max}}$ - [PL$^{\text{as}}$])} &
 \end{flalign*} 

 \begin{flalign*} 
 \frac{d}{dt}\left[ \text{  z$_2^{\text{m}}$:PRO  } \right]&= \text{k}_{ \text{z$_2^{\text{m}}$:PRO} }^{\text{+}}  \left[ \text{z$_2^{\text{m}}$} \right]  \left[ \text{PRO} \right] - ( \text{k}_{ \text{z$_2^{\text{m}}$:PRO} }^{\text{cat}} + \text{k}_{ \text{z$_2^{\text{m}}$:PRO} }^{\text{-}} )  \left[ \text{z$_2^{\text{m}}$:PRO} \right]&
 \end{flalign*} 

 \begin{flalign*} 
 \frac{d}{dt}\left[ \text{  APC:e$_5^{\text{m}}$  } \right]&= \text{k}_{ \text{e$_5^{\text{m}}$:APC} }^{\text{+}}  \left[ \text{APC} \right]  \left[ \text{e$_5^{\text{m}}$} \right] - ( \text{k}_{ \text{e$_5^{\text{m}}$:APC} }^{\text{cat}} + \text{k}_{ \text{e$_5^{\text{m}}$:APC} }^{\text{-}} )  \left[ \text{APC:e$_5^{\text{m}}$} \right]&
 \end{flalign*} 

 \begin{flalign*} 
 \frac{d}{dt}\left[ \text{  TF  } \right]&= - \left[ \text{TF} \right]  \frac{d}{dt} \left( [\text{PL}^{\text{as}}] \right) /\text{(P$_\text{{max}}$ - [PL$^{\text{as}}$])} &
 \end{flalign*} 

 \begin{flalign*} 
 \frac{d}{dt}\left[ \text{  z$_{7}$:e$_{9}$  } \right]&= \text{k}_{ \text{z$_{7}$:e$_{9}$} }^{\text{+}}  \left[ \text{e$_{9}$} \right]  \left[ \text{z$_{7}$} \right] - ( \text{k}_{ \text{z$_{7}$:e$_{9}$} }^{\text{cat}} + \text{k}_{ \text{z$_{7}$:e$_{9}$} }^{\text{-}} )  \left[ \text{z$_{7}$:e$_{9}$} \right]&
 \end{flalign*} 

 \begin{flalign*} 
 \frac{d}{dt}\left[ \text{  z$_7^{\text{m}}$:e$_{9}$  } \right]&= \text{k}_{ \text{z$_7^{\text{m}}$:e$_{9}$} }^{\text{+}}  \left[ \text{e$_{9}$} \right]  \left[ \text{z$_7^{\text{m}}$} \right] - ( \text{k}_{ \text{z$_7^{\text{m}}$:e$_{9}$} }^{\text{cat}} + \text{k}_{ \text{z$_7^{\text{m}}$:e$_{9}$} }^{\text{-}} )  \left[ \text{z$_7^{\text{m}}$:e$_{9}$} \right]\\
 &\hskip2.5ex - \left[ \text{z$_7^{\text{m}}$:e$_{9}$} \right]  \frac{d}{dt} \left( [\text{PL}^{\text{as}}] \right) /\text{(P$_\text{{max}}$ - [PL$^{\text{as}}$])} &
 \end{flalign*} 

 \begin{flalign*} 
 \frac{d}{dt}\left[ \text{  e$_{9}^{\text{m,*}}$  } \right]&= \text{k}_{ \text{9} }^{\text{on}}  ( \left[ \text{p$_{9}^{*}$} \right] - \left[ \text{e$_{9}^{\text{m,*}}$} \right] - ( \left[ \text{TEN$^{*}$} \right] + \left[ \text{z$_{10}^{\text{m}}$:TEN$^{*}$} \right] )   \left[ \text{e$_{9}$} \right]\\
 &\hskip2.5ex - \text{k}_{ \text{9} }^{\text{off}}  \left[ \text{e$_{9}^{\text{m,*}}$} \right] + \text{k}_{ \text{e$_8^{\text{m}}$:e$_9^{\text{m}}$} }^{\text{-}}  \left[ \text{TEN$^{*}$} \right] - \text{k}_{ \text{e$_8^{\text{m}}$:e$_9^{\text{m}}$} }^{\text{+}}  \left[ \text{e$_8^{\text{m}}$} \right]  \left[ \text{e$_{9}^{\text{m,*}}$} \right] \\
 &\hskip2.5ex - \text{k}_{ \text{e$_9^{\text{m}}$:AT} }  \left[ \text{AT} \right]  \left[ \text{e$_{9}^{\text{m,*}}$} \right] \\
 &\hskip2.5ex - \text{k}_{ \text{e$_9^{\text{m}}$:ATH} }  \left[ \text{ATH} \right]  \left[ \text{e$_{9}^{\text{m,*}}$} \right]&
 \end{flalign*} 

 \begin{flalign*} 
 \frac{d}{dt}\left[ \text{  TEN$^{*}$  } \right]&= - \text{k}_{ \text{e$_8^{\text{m}}$:e$_9^{\text{m}}$} }^{\text{-}}  \left[ \text{TEN$^{*}$} \right] + \text{k}_{ \text{e$_8^{\text{m}}$:e$_9^{\text{m}}$} }^{\text{+}}  \left[ \text{e$_8^{\text{m}}$} \right]  \left[ \text{e$_{9}^{\text{m,*}}$} \right]\\
 &\hskip2.5ex + ( \text{k}_{ \text{z$_{10}^{\text{m}}$:TEN} }^{\text{cat}} + \text{k}_{ \text{z$_{10}^{\text{m}}$:TEN} }^{\text{-}} )  \left[ \text{z$_{10}^{\text{m}}$:TEN$^{*}$} \right] - \text{k}_{ \text{z$_{10}^{\text{m}}$:TEN} }^{\text{+}}  \left[ \text{TEN$^{*}$} \right]  \left[ \text{z$_{10}^{\text{m}}$} \right]&
 \end{flalign*} 

 \begin{flalign*} 
 \frac{d}{dt}\left[ \text{  z$_{10}^{\text{m}}$:TEN$^{*}$  } \right]&= \text{k}_{ \text{z$_{10}^{\text{m}}$:TEN} }^{\text{+}}  \left[ \text{TEN$^{*}$} \right]  \left[ \text{z$_{10}^{\text{m}}$} \right]\\
 &\hskip2.5ex - ( \text{k}_{ \text{z$_{10}^{\text{m}}$:TEN} }^{\text{cat}} + \text{k}_{ \text{z$_{10}^{\text{m}}$:TEN} }^{\text{-}} )  \left[ \text{z$_{10}^{\text{m}}$:TEN$^{*}$} \right]&
 \end{flalign*} 

 \begin{flalign*} 
 \frac{d}{dt}\left[ \text{  e$_{\text{2}}$$^{\text{ec}}$  } \right]&= \text{k}_\text{diff}  ( \left[ \text{e$_{\text{2}}$} \right] - \left[ \text{e$_{\text{2}}$$^{\text{ec}}$} \right] ) + \text{k}_{\text{flow}}  ( \text{ [e$_{\text{2}}$$^{\text{ec,up}}$}  - \left[ \text{e$_{\text{2}}$$^{\text{ec}}$} \right] ) \\
 &\hskip2.5ex - \text{k}_{ \text{TM} }^{\text{+}}  \left[ \text{e$_{\text{2}}$$^{\text{ec}}$} \right]  ( 1 - \left[ \text{TM:e$_{\text{2}}$$^{\text{ec}}$} \right] - \left[ \text{TM:e$_{\text{2}}$$^{\text{ec}}$:APC} \right] ) \\
 &\hskip2.5ex + \text{k}_{\text{TM}}^{\text{-}}  \left[ \text{TM:e$_{\text{2}}$$^{\text{ec}}$} \right] - \text{k}_{ \text{e$_{\text{2}}$:AT} }  \left[ \text{e$_{\text{2}}$$^{\text{ec}}$} \right]  \left[ \text{AT$^{\text{ec}}$} \right]&
 \end{flalign*} 

 \begin{flalign*} 
 \frac{d}{dt}\left[ \text{  APC$^{\text{ec}}$  } \right]&= \text{k}_\text{diff}  ( \left[ \text{APC} \right] - \left[ \text{APC$^{\text{ec}}$} \right] ) + \text{k}_{\text{flow}}  ( \text{ [APC]$^{\text{ec,up}}$}  - \left[ \text{APC$^{\text{ec}}$} \right] ) \\
 &\hskip2.5ex + \text{k}_{ \text{PCTM:e$_{\text{2}}$} }^{\text{cat}}  \left[ \text{TM:e$_{\text{2}}$$^{\text{ec}}$:APC} \right]&
 \end{flalign*} 

 \begin{flalign*} 
 \frac{d}{dt}\left[ \text{  TM:e$_{\text{2}}$$^{\text{ec}}$  } \right]&= \text{k}_{ \text{TM} }^{\text{+}}  \left[ \text{e$_{\text{2}}$$^{\text{ec}}$} \right]  ( 1 - \left[ \text{TM:e$_{\text{2}}$$^{\text{ec}}$} \right] - \left[ \text{TM:e$_{\text{2}}$$^{\text{ec}}$:APC} \right] ) \\
 &\hskip2.5ex - \text{k}_{\text{TM}}^{\text{-}}  \left[ \text{TM:e$_{\text{2}}$$^{\text{ec}}$} \right] \\
 &\hskip2.5ex - \text{k}_{ \text{PCTM:e$_{\text{2}}$} }^{\text{+}}  \left[ \text{TM:e$_{\text{2}}$$^{\text{ec}}$} \right] + ( \text{k}_{ \text{PCTM:e$_{\text{2}}$} }^{\text{-}} + \text{k}_{ \text{PCTM:e$_{\text{2}}$} }^{\text{cat}} )  \left[ \text{TM:e$_{\text{2}}$$^{\text{ec}}$:APC} \right]&
 \end{flalign*} 

 \begin{flalign*} 
 \frac{d}{dt}\left[ \text{  TM:e$_{\text{2}}$$^{\text{ec}}$:APC  } \right]&= \text{k}_{ \text{PCTM:e$_{\text{2}}$} }^{\text{+}}  \left[ \text{TM:e$_{\text{2}}$$^{\text{ec}}$} \right] - ( \text{k}_{ \text{PCTM:e$_{\text{2}}$} }^{\text{-}} + \text{k}_{ \text{PCTM:e$_{\text{2}}$} }^{\text{cat}} )  \left[ \text{TM:e$_{\text{2}}$$^{\text{ec}}$:APC} \right]&
 \end{flalign*} 

 \begin{flalign*} 
 \frac{d}{dt}\left[ \text{  e$_{9}$$^{\text{ec}}$  } \right]&= \text{k}_\text{diff}  ( \left[ \text{e$_{9}$} \right] - \left[ \text{e$_{9}$$^{\text{ec}}$} \right] ) + \text{k}_{\text{flow}}  ( \text{ [e$_{9}$$^{\text{ec}}$]$^{\text{up}}$}  - \left[ \text{e$_{9}$$^{\text{ec}}$} \right] ) \\
 &\hskip2.5ex - \text{k}_{ \text{e$_{9}$:AT} }  \left[ \text{e$_{9}$$^{\text{ec}}$} \right]  \left[ \text{AT$^{\text{ec}}$} \right]&
 \end{flalign*} 

 \begin{flalign*} 
 \frac{d}{dt}\left[ \text{  e$_{10}$$^{\text{ec}}$  } \right]&= \text{k}_\text{diff}  ( \left[ \text{e$_{10}$} \right] - \left[ \text{e$_{10}$$^{\text{ec}}$} \right] ) + \text{k}_{\text{flow}}  ( \text{ [e$_{10}$$^{\text{ec}}$]$^{\text{up}}$}  - \left[ \text{e$_{10}$$^{\text{ec}}$} \right] ) \\
 &\hskip2.5ex - \text{k}_{ \text{e$_{10}$:AT} }  \left[ \text{AT$^{\text{ec}}$} \right]  \left[ \text{e$_{10}$$^{\text{ec}}$} \right] \\
 &\hskip2.5ex - \text{k}_{ \text{TFPIb$^{\text{m}}$} }^{\text{+}}  \left[ \text{e$_{10}$$^{\text{ec}}$} \right]  \left[ \text{TFPIb} \right] + \text{k}_{ \text{TFPIb$^{\text{m}}$} }^{\text{-}}  \left[ \text{TFPIb$^{\text{m}}$} \right]&
 \end{flalign*} 

 \begin{flalign*} 
 \frac{d}{dt}\left[ \text{  APC:e$_{\text{5}}$  } \right]&= - ( \text{k}_{ \text{e$_{\text{5}}$:APC} }^{\text{cat}} + \text{k}_{ \text{e$_{\text{5}}$:APC} }^{\text{-}} )  \left[ \text{APC:e$_{\text{5}}$} \right] + \text{k}_{ \text{e$_{\text{5}}$:APC} }^{\text{+}}  \left[ \text{APC} \right]  \left[ \text{e$_{\text{5}}$} \right]&
 \end{flalign*} 

 \begin{flalign*} 
 \frac{d}{dt}\left[ \text{  APC:e$_{8}$  } \right]&= - ( \text{k}_{ \text{e$_{8}$:APC} }^{\text{cat}} + \text{k}_{ \text{e$_{8}$:APC} }^{\text{-}} )  \left[ \text{APC:e$_{8}$} \right] + \text{k}_{ \text{e$_{8}$:APC} }^{\text{+}}  \left[ \text{APC} \right]  \left[ \text{e$_{8}$} \right]&
 \end{flalign*} 

 \begin{flalign*} 
 \frac{d}{dt}\left[ \text{  z$_{11}$  } \right]&=  \text{k}_{\text{flow}}  ( \text{ [z$_{11}$]$^{\text{up}}$}  - \left[ \text{z$_{11}$} \right] ) \\
 &\hskip2.5ex - \text{k}_{ \text{z$_{11}$} }^{\text{on}}  \left[ \text{z$_{11}$} \right]  \left[ \text{p}_{ \text{11} }^{\text{avail}} \right] + \text{k}_{ \text{z$_{11}$} }^{\text{off}}  \left[ \text{z$_{11}^{\text{m}}$} \right] \\
 &\hskip2.5ex - \text{k}_{ \text{z$_{11}$:e$_{11\text{h}}$} }^{\text{+}}  \left[ \text{z$_{11}$} \right]  \left[ \text{e$_{11\text{h}}$} \right] + \text{k}_{ \text{z$_{11}$:e$_{11\text{h}}$} }^{\text{-}}  \left[ \text{z$_{11}$:e$_{11\text{h}}$} \right] \\
 &\hskip2.5ex - \text{k}_{ \text{z$_{11}$:e$_{11}$} }^{\text{+}}  \left[ \text{z$_{11}$} \right]  \left[ \text{e$_{11}$} \right] + \text{k}_{ \text{z$_{11}$:e$_{11}$} }^{\text{-}}  \left[ \text{z$_{11}$:e$_{11}$} \right] \\
 &\hskip2.5ex - \text{k}_{ \text{z$_{11}$:e$_{\text{2}}$} }^{\text{+}}  \left[ \text{z$_{11}$} \right]  \left[ \text{e$_{\text{2}}$} \right] + \text{k}_{ \text{z$_{11}$:e$_{\text{2}}$} }^{\text{-}}  \left[ \text{z$_{11}$:e$_{\text{2}}$} \right]&
 \end{flalign*} 

 \begin{flalign*} 
 \frac{d}{dt}\left[ \text{  e$_{11}$  } \right]&= \text{k}_{\text{flow}}  ( \text{ [e$_{11}$]$^{\text{up}}$}  - \left[ \text{e$_{11}$} \right] ) \\
 &\hskip2.5ex - \text{k$_{\text{e$_{11}$}}^{\text{on,*}}$}  \left[ \text{e$_{11}$} \right]  \left[ \text{p}_{ \text{11,*} }^{\text{avail}} \right] + \text{k}_{ \text{e$_{11}$} }^{\text{off,*}}  \left[ \text{e$_{11}^{\text{m,*}}$} \right] \\
 &\hskip2.5ex - \text{k}_{ \text{z$_{9}$:e$_{11}$} }^{\text{+}}  \left[ \text{z$_{9}$} \right]  \left[ \text{e$_{11}$} \right] + ( \text{k}_{ \text{z$_{9}$:e$_{11}$} }^{\text{-}} + \text{k}_{ \text{z$_{9}$:e$_{11}$} }^{\text{cat}} )  \left[ \text{z$_{9}$:e$_{11}$} \right] \\
 &\hskip2.5ex - \text{k}_{ \text{z$_{11}$:e$_{11}$} }^{\text{+}}  \left[ \text{z$_{11}$} \right]  \left[ \text{e$_{11}$} \right] + ( \text{k}_{ \text{z$_{11}$:e$_{11}$} }^{\text{-}} + \text{k}_{ \text{z$_{11}$:e$_{11}$} }^{\text{cat}} )  \left[ \text{z$_{11}$:e$_{11}$} \right] \\
 &\hskip2.5ex + \text{k}_{ \text{e$_{11\text{h}}$:e$_{11\text{h}}$} }^{\text{cat}}  \left[ \text{e$_{11\text{h}}$:e$_{11\text{h}}$} \right] \\
 &\hskip2.5ex - \text{k}_{ \text{e$_{11\text{h}}$:e$_{11}$} }^{\text{+}}  \left[ \text{e$_{11\text{h}}$} \right]  \left[ \text{e$_{11}$} \right] + ( \text{k}_{ \text{e$_{11\text{h}}$:e$_{11}$} }^{\text{-}} + 2  \text{k}_{ \text{e$_{11\text{h}}$:e$_{11}$} }^{\text{cat}} )  \left[ \text{e$_{11\text{h}}$:e$_{11}$} \right] \\
 &\hskip2.5ex + \text{k}_{ \text{e$_{11\text{h}}$:e$_{\text{2}}$} }^{\text{cat}}  \left[ \text{e$_{11\text{h}}$:e$_{\text{2}}$} \right] \\
 &\hskip2.5ex - \text{k}_{ \text{e$_{11}$:AT} }  \left[ \text{AT} \right]  \left[ \text{e$_{11}$} \right] \\
 &\hskip2.5ex - \text{k}_{ \text{e$_{11}$:ATH} }  \left[ \text{ATH} \right]  \left[ \text{e$_{11}$} \right]&
 \end{flalign*} 

 \begin{flalign*} 
 \frac{d}{dt}\left[ \text{  z$_{11}^{\text{m}}$  } \right]&= \text{k}_{ \text{z$_{11}$} }^{\text{on}}  \left[ \text{z$_{11}$} \right]  \left[ \text{p}_{ \text{11} }^{\text{avail}} \right] - \text{k}_{ \text{z$_{11}$} }^{\text{off}}  \left[ \text{z$_{11}^{\text{m}}$} \right]\\
 &\hskip2.5ex  - \text{k}_{ \text{z$_{11}^{\text{m}}$:e$_{11\text{h}}^{\text{m}}$} }^{\text{+}}  \left[ \text{z$_{11}^{\text{m}}$} \right]  \left[ \text{e$_{11\text{h}}^{\text{m}}$} \right] + \text{k}_{ \text{z$_{11}^{\text{m}}$:e$_{11\text{h}}^{\text{m}}$} }^{\text{-}}  \left[ \text{z$_{11}^{\text{m}}$:e$_{11\text{h}}^{\text{m}}$} \right] \\
 &\hskip2.5ex  - \text{k}_{ \text{z$_{11}^{\text{m}}$:e$_{11}^{\text{m,*}}$} }^{\text{+}}  \left[ \text{z$_{11}^{\text{m}}$} \right]  \left[ \text{e$_{11}^{\text{m,*}}$} \right] + \text{k}_{ \text{z$_{11}^{\text{m}}$:e$_{11}^{\text{m,*}}$} }^{\text{-}}  \left[ \text{z$_{11}^{\text{m}}$:e$_{11}^{\text{m,*}}$} \right] \\
 &\hskip2.5ex  - \text{k}_{ \text{z$_{11}^{\text{m}}$:e$_2^{\text{m}}$} }^{\text{+}}  \left[ \text{z$_{11}^{\text{m}}$} \right]  \left[ \text{e$_2^{\text{m}}$} \right] + \text{k}_{ \text{z$_{11}^{\text{m}}$:e$_2^{\text{m}}$} }^{\text{-}}  \left[ \text{z$_{11}^{\text{m}}$:e$_2^{\text{m}}$} \right]&
 \end{flalign*} 

 \begin{flalign*} 
 \frac{d}{dt}\left[ \text{  e$_{11}^{\text{m,*}}$  } \right]&= \left[ \text{k$_{\text{e$_{11}$}}^{\text{on,*}}$} \right]  \left[ \text{e$_{11}$} \right]  \left[ \text{p}_{ \text{11,*} }^{\text{avail}} \right] - \text{k}_{ \text{e$_{11}$} }^{\text{off,*}}  \left[ \text{e$_{11}^{\text{m,*}}$} \right]\\
 &\hskip2.5ex - \text{k}_{ \text{z$_9^{\text{m}}$:e$_{11}^{\text{m,*}}$} }^{\text{+}}  \left[ \text{e$_{11}^{\text{m,*}}$} \right]  \left[ \text{z$_9^{\text{m}}$} \right] \\
 &\hskip2.5ex + ( \text{k}_{ \text{z$_9^{\text{m}}$:e$_{11}^{\text{m,*}}$} }^{\text{-}} + \text{k}_{ \text{z$_9^{\text{m}}$:e$_{11}^{\text{m,*}}$} }^{\text{cat}} )  \left[ \text{z$_9^{\text{m}}$:e$_{11}^{\text{m,*}}$} \right] \\
 &\hskip2.5ex - \text{k}_{ \text{z$_{11}^{\text{m}}$:e$_{11}^{\text{m,*}}$} }^{\text{+}}  \left[ \text{z$_{11}^{\text{m}}$} \right]  \left[ \text{e$_{11}^{\text{m,*}}$} \right] \\
 &\hskip2.5ex + ( \text{k}_{ \text{z$_{11}^{\text{m}}$:e$_{11}^{\text{m,*}}$} }^{\text{-}} + \text{k}_{ \text{z$_{11}^{\text{m}}$:e$_{11}^{\text{m,*}}$} }^{\text{cat}} )  \left[ \text{z$_{11}^{\text{m}}$:e$_{11}^{\text{m,*}}$} \right] \\
 &\hskip2.5ex + \text{k}_{ \text{e$_{11\text{h}}^{\text{m,*}}$:e$_{11\text{h}}^{\text{m}}$} }^{\text{cat}}  \left[ \text{e$_{11\text{h}}^{\text{m,*}}$:e$_{11\text{h}}^{\text{m}}$} \right] \\
 &\hskip2.5ex - \text{k}_{ \text{e$_{11\text{h}}^{\text{m,*}}$:e$_{11}^{\text{m,*}}$} }^{\text{+}}  \left[ \text{e$_{11\text{h}}^{\text{m,*}}$} \right]  \left[ \text{e$_{11}^{\text{m,*}}$} \right] \\
 &\hskip2.5ex + ( \text{k}_{ \text{e$_{11\text{h}}^{\text{m,*}}$:e$_{11}^{\text{m,*}}$} }^{\text{-}} + 2  \text{k}_{ \text{e$_{11\text{h}}^{\text{m,*}}$:e$_{11}^{\text{m,*}}$} }^{\text{cat}} )  \left[ \text{e$_{11\text{h}}^{\text{m,*}}$:e$_{11\text{h}}^{\text{m}}$} \right] \\
 &\hskip2.5ex + \text{k}_{ \text{e$_{11\text{h}}^{\text{m,*}}$:e$_2^{\text{m}}$} }^{\text{cat}}  \left[ \text{e$_{11\text{h}}^{\text{m,*}}$:e$_2^{\text{m}}$} \right] \\
 &\hskip2.5ex - \text{k}_{ \text{e$_{11}$:AT} }  \left[ \text{AT} \right]  \left[ \text{e$_{11}^{\text{m,*}}$} \right] \\
 &\hskip2.5ex - \text{k}_{ \text{e$_{11}$:ATH} }  \left[ \text{ATH} \right]  \left[ \text{e$_{11}^{\text{m,*}}$} \right]&
 \end{flalign*} 

 \begin{flalign*} 
 \frac{d}{dt}\left[ \text{  z$_{11}^{\text{m}}$:e$_2^{\text{m}}$  } \right]&= \text{k}_{ \text{z$_{11}^{\text{m}}$:e$_2^{\text{m}}$} }^{\text{+}}  \left[ \text{z$_{11}^{\text{m}}$} \right]  \left[ \text{e$_2^{\text{m}}$} \right] - ( \text{k}_{ \text{z$_{11}^{\text{m}}$:e$_2^{\text{m}}$} }^{\text{-}} + \text{k}_{ \text{z$_{11}^{\text{m}}$:e$_2^{\text{m}}$} }^{\text{cat}} )  \left[ \text{z$_{11}^{\text{m}}$:e$_2^{\text{m}}$} \right]&
 \end{flalign*} 

 \begin{flalign*} 
 \frac{d}{dt}\left[ \text{  z$_9^{\text{m}}$:e$_{11}^{\text{m,*}}$  } \right]&= \text{k}_{ \text{z$_9^{\text{m}}$:e$_{11}^{\text{m,*}}$} }^{\text{+}}  \left[ \text{z$_9^{\text{m}}$} \right]  \left[ \text{e$_{11}^{\text{m,*}}$} \right]\\
 &\hskip2.5ex    - ( \text{k}_{ \text{z$_9^{\text{m}}$:e$_{11}^{\text{m,*}}$} }^{\text{-}} + \text{k}_{ \text{z$_9^{\text{m}}$:e$_{11}^{\text{m,*}}$} }^{\text{cat}} )  \left[ \text{z$_9^{\text{m}}$:e$_{11}^{\text{m,*}}$} \right]&
 \end{flalign*} 

 \begin{flalign*} 
 \frac{d}{dt}\left[ \text{  z$_{11}$:e$_{\text{2}}$  } \right]&= \text{k}_{ \text{\text{flow}}}  (\text{[z$_{11}$:e$_{\text{2}}$]$^{\text{up}}$} - \left[ \text{z$_{11}$:e$_{\text{2}}$} \right] ) + \text{k}_{\text{z$_{11}$:e$_{\text{2}}$}}  \left[ \text{z$_{11}$} \right]  \left[ \text{e$_{\text{2}}$} \right]\\
 &\hskip2.5ex  - ( \text{k}_{ \text{z$_{11}$:e$_{\text{2}}$} }^{\text{-}} + \text{k}_{ \text{z$_{11}$:e$_{\text{2}}$} }^{\text{cat}} )  \left[ \text{z$_{11}$:e$_{\text{2}}$} \right]&
 \end{flalign*} 

 \begin{flalign*} 
 \frac{d}{dt}\left[ \text{  z$_{9}$:e$_{11}$  } \right]&= \text{k}_{ \text{\text{flow}}}  ( \text{[z$_{9}$:e$_{11}$]$^{\text{up}}$} - \left[ \text{z$_{9}$:e$_{11}$} \right] ) + \text{k}_{\text{z$_{9}$:e$_{11}$}} \left[ \text{z$_{9}$} \right]  \left[ \text{e$_{11}$} \right]\\
 &\hskip2.5ex  - ( \text{k}_{ \text{z$_{9}$:e$_{11}$} }^{\text{-}} + \text{k}_{ \text{z$_{9}$:e$_{11}$} }^{\text{cat}} )  \left[ \text{z$_{9}$:e$_{11}$} \right]&
 \end{flalign*} 

 \begin{flalign*} 
 \frac{d}{dt}\left[ \text{  z$_{11}$:e$_{11}$  } \right]&= \text{k}_{ \text{\text{flow}}}  ( \text{[z$_{11}$:e$_{11}$]$^{\text{up}}$} - \left[ \text{z$_{11}$:e$_{11}$} \right] ) + \text{k}_{\text{z$_{11}$:e$_{11}$}}  \left[ \text{z$_{11}$} \right]  \left[ \text{e$_{11}$} \right]\\
 &\hskip2.5ex - ( \text{k}_{ \text{z$_{11}$:e$_{11}$} }^{\text{-}} + \text{k}_{ \text{z$_{11}$:e$_{11}$} }^{\text{cat}} )  \left[ \text{z$_{11}$:e$_{11}$} \right]&
 \end{flalign*} 

 \begin{flalign*} 
 \frac{d}{dt}\left[ \text{  e$_{11\text{h}}$  } \right]&= - \text{ke$_{11\text{h}}$$^{\text{on,*}}$}  \left[ \text{e$_{11\text{h}}$} \right]  \left[ \text{p}_{ \text{11,*} }^{\text{avail}} \right] + \text{k}_{ \text{e$_{11\text{h}}$} }^{\text{off,*}}  \left[ \text{e$_{11\text{h}}^{\text{m,*}}$} \right]\\
 &\hskip2.5ex  - \text{k}_{ \text{e$_{11\text{h}}$} }^{\text{on}}  \left[ \text{e$_{11\text{h}}$} \right]  \left[ \text{p}_{ \text{11} }^{\text{avail}} \right] + \text{k}_{ \text{e$_{11\text{h}}$} }^{\text{off}}  \left[ \text{e$_{11\text{h}}^{\text{m}}$} \right] \\
 &\hskip2.5ex  - \text{k}_{ \text{z$_{9}$:e$_{11\text{h}}$} }^{\text{+}}  \left[ \text{z$_{9}$} \right]  \left[ \text{e$_{11\text{h}}$} \right] + ( \text{k}_{ \text{z$_{9}$:e$_{11\text{h}}$} }^{\text{-}} + \text{k}_{ \text{z$_{9}$:e$_{11\text{h}}$} }^{\text{cat}} )  \left[ \text{z$_{9}$:e$_{11\text{h}}$} \right] \\
 &\hskip2.5ex  - \text{k}_{ \text{z$_{11}$:e$_{11\text{h}}$} }^{\text{+}}  \left[ \text{z$_{11}$} \right]  \left[ \text{e$_{11\text{h}}$} \right] + ( \text{k}_{ \text{z$_{11}$:e$_{11\text{h}}$} }^{\text{-}} + 2  \text{k}_{ \text{z$_{11}$:e$_{11\text{h}}$} }^{\text{cat}} )  \left[ \text{z$_{11}$:e$_{11\text{h}}$} \right] \\
 &\hskip2.5ex  + \text{k}_{ \text{z$_{11}$:e$_{11}$} }^{\text{cat}}  \left[ \text{z$_{11}$:e$_{11}$} \right] + \text{k}_{ \text{z$_{11}$:e$_{\text{2}}$} }^{\text{cat}}  \left[ \text{z$_{11}$:e$_{\text{2}}$} \right] \\
 &\hskip2.5ex  - 2  \text{k}_{ \text{e$_{11\text{h}}$:e$_{11\text{h}}$} }^{\text{+}}  \left[ \text{e$_{11\text{h}}$} \right]  \left[ \text{e$_{11\text{h}}$} \right] \\
 &\hskip2.5ex  + ( 2  \text{k}_{ \text{e$_{11\text{h}}$:e$_{11\text{h}}$} }^{\text{-}} + \text{k}_{ \text{e$_{11\text{h}}$:e$_{11\text{h}}$} }^{\text{cat}} )  \left[ \text{e$_{11\text{h}}$:e$_{11\text{h}}$} \right] \\
 &\hskip2.5ex  - \text{k}_{ \text{e$_{11\text{h}}$:e$_{11}$} }^{\text{+}}  \left[ \text{e$_{11\text{h}}$} \right]  \left[ \text{e$_{11}$} \right] + \text{k}_{ \text{e$_{11\text{h}}$:e$_{11}$} }^{\text{-}}  \left[ \text{e$_{11\text{h}}$:e$_{11}$} \right] \\
 &\hskip2.5ex  - \text{k}_{ \text{e$_{11\text{h}}$:e$_{\text{2}}$} }^{\text{+}}  \left[ \text{e$_{11\text{h}}$} \right]  \left[ \text{e$_{\text{2}}$} \right] + \text{k}_{ \text{e$_{11\text{h}}$:e$_{\text{2}}$} }^{\text{-}}  \left[ \text{e$_{11\text{h}}$:e$_{\text{2}}$} \right] \\
 &\hskip2.5ex  + \text{k}_{\text{flow}}  ( \text{ [e$_{11\text{h}}$]$^{\text{up}}$}  - \left[ \text{e$_{11\text{h}}$} \right] ) \\
 &\hskip2.5ex  - \text{k}_{ \text{e$_{11}$:AT} }  \left[ \text{AT} \right]  \left[ \text{e$_{11\text{h}}$} \right] \\
 &\hskip2.5ex  - \text{k}_{ \text{e$_{11}$:ATH} }  \left[ \text{ATH} \right]  \left[ \text{e$_{11\text{h}}$} \right]&
 \end{flalign*} 

 \begin{flalign*} 
 \frac{d}{dt}\left[ \text{  e$_{11\text{h}}^{\text{m}}$  } \right]&= \text{k}_{ \text{e$_{11\text{h}}$} }^{\text{on}}  \left[ \text{e$_{11\text{h}}$} \right]  \left[ \text{p}_{ \text{11} }^{\text{avail}} \right] - \text{k}_{ \text{e$_{11\text{h}}$} }^{\text{off}}  \left[ \text{e$_{11\text{h}}^{\text{m}}$} \right]\\
 &\hskip2.5ex  - \text{k}_{ \text{z$_9^{\text{m}}$:e$_{11\text{h}}^{\text{m}}$} }^{\text{+}}  \left[ \text{z$_9^{\text{m}}$} \right]  \left[ \text{e$_{11\text{h}}^{\text{m}}$} \right] \\
 &\hskip2.5ex  + ( \text{k}_{ \text{z$_9^{\text{m}}$:e$_{11\text{h}}^{\text{m}}$} }^{\text{-}} + \text{k}_{ \text{z$_9^{\text{m}}$:e$_{11\text{h}}^{\text{m}}$} }^{\text{cat}} )  \left[ \text{z$_9^{\text{m}}$:e$_{11\text{h}}^{\text{m}}$} \right] \\
 &\hskip2.5ex  - \text{k}_{ \text{z$_{11}^{\text{m}}$:e$_{11\text{h}}^{\text{m}}$} }^{\text{+}}  \left[ \text{z$_{11}^{\text{m}}$} \right]  \left[ \text{e$_{11\text{h}}^{\text{m}}$} \right] \\
 &\hskip2.5ex  + ( \text{k}_{ \text{z$_{11}^{\text{m}}$:e$_{11\text{h}}^{\text{m}}$} }^{\text{-}} + 2  \text{k}_{ \text{z$_{11}^{\text{m}}$:e$_{11\text{h}}^{\text{m}}$} }^{\text{cat}} )  \left[ \text{z$_{11}^{\text{m}}$:e$_{11\text{h}}^{\text{m}}$} \right] \\
 &\hskip2.5ex  + \text{k}_{ \text{z$_{11}^{\text{m}}$:e$_{11}^{\text{m,*}}$} }^{\text{cat}}  \left[ \text{z$_{11}^{\text{m}}$:e$_{11}^{\text{m,*}}$} \right] + \text{k}_{ \text{z$_{11}^{\text{m}}$:e$_2^{\text{m}}$} }^{\text{cat}}  \left[ \text{z$_{11}^{\text{m}}$:e$_2^{\text{m}}$} \right] \\
 &\hskip2.5ex  - \text{k}_{ \text{e$_{11\text{h}}^{\text{m,*}}$:e$_{11\text{h}}^{\text{m}}$} }^{\text{+}}  \left[ \text{e$_{11\text{h}}^{\text{m,*}}$} \right]  \left[ \text{e$_{11\text{h}}^{\text{m}}$} \right] \\
 &\hskip2.5ex  + ( \text{k}_{ \text{e$_{11\text{h}}^{\text{m,*}}$:e$_{11\text{h}}^{\text{m}}$} }^{\text{-}} + \text{k}_{ \text{e$_{11\text{h}}^{\text{m,*}}$:e$_{11\text{h}}^{\text{m}}$} }^{\text{cat}} )  \left[ \text{e$_{11\text{h}}^{\text{m,*}}$:e$_{11\text{h}}^{\text{m}}$} \right] \\
 &\hskip2.5ex  - \text{k}_{ \text{e$_{11}$:AT} }  \left[ \text{AT} \right]  \left[ \text{e$_{11\text{h}}^{\text{m}}$} \right] \\
 &\hskip2.5ex  - \text{k}_{ \text{e$_{11}$:ATH} }  \left[ \text{ATH} \right]  \left[ \text{e$_{11\text{h}}^{\text{m}}$} \right]&
 \end{flalign*} 

 \begin{flalign*} 
 \frac{d}{dt}\left[ \text{  e$_{11\text{h}}^{\text{m,*}}$  } \right]&= \text{ke$_{11\text{h}}$$^{\text{on,*}}$}  \left[ \text{e$_{11\text{h}}$} \right]  \left[ \text{p}_{ \text{11,*} }^{\text{avail}} \right] - \text{k}_{ \text{e$_{11\text{h}}$} }^{\text{off,*}}  \left[ \text{e$_{11\text{h}}^{\text{m,*}}$} \right]\\
 &\hskip2.5ex - \text{k}_{ \text{e$_{11\text{h}}^{\text{m,*}}$:e$_{11\text{h}}^{\text{m}}$} }^{\text{+}}  \left[ \text{e$_{11\text{h}}^{\text{m,*}}$} \right]  \left[ \text{e$_{11\text{h}}^{\text{m}}$} \right] + \text{k}_{ \text{e$_{11\text{h}}^{\text{m,*}}$:e$_{11\text{h}}^{\text{m}}$} }^{\text{-}}  \left[ \text{e$_{11\text{h}}^{\text{m,*}}$:e$_{11\text{h}}^{\text{m}}$} \right] \\
 &\hskip2.5ex - \text{k}_{ \text{e$_{11\text{h}}^{\text{m,*}}$:e$_{11}^{\text{m,*}}$} }^{\text{+}}  \left[ \text{e$_{11\text{h}}^{\text{m,*}}$} \right]  \left[ \text{e$_{11}^{\text{m,*}}$} \right] + \text{k}_{ \text{e$_{11\text{h}}^{\text{m,*}}$:e$_{11}^{\text{m,*}}$} }^{\text{-}}  \left[ \text{e$_{11\text{h}}^{\text{m,*}}$:e$_{11}^{\text{m,*}}$} \right] \\
 &\hskip2.5ex - \text{k}_{ \text{e$_{11\text{h}}^{\text{m,*}}$:e$_2^{\text{m}}$} }^{\text{+}}  \left[ \text{e$_{11\text{h}}^{\text{m,*}}$} \right]  \left[ \text{e$_2^{\text{m}}$} \right] + \text{k}_{ \text{e$_{11\text{h}}^{\text{m,*}}$:e$_2^{\text{m}}$} }^{\text{-}}  \left[ \text{e$_{11\text{h}}^{\text{m,*}}$:e$_2^{\text{m}}$} \right]&
 \end{flalign*} 

 \begin{flalign*} 
 \frac{d}{dt}\left[ \text{  z$_{9}$:e$_{11\text{h}}$  } \right]&= \text{k}_{\text{flow}}  ( \text{ [z$_{9}$:e$_{11\text{h}}$]$^{\text{up}}$}  - \left[ \text{z$_{9}$:e$_{11\text{h}}$} \right] ) \\
 &\hskip2.5ex + \text{k}_{ \text{z$_{9}$:e$_{11\text{h}}$} }^{\text{+}}  \left[ \text{z$_{9}$} \right]  \left[ \text{e$_{11\text{h}}$} \right] - ( \text{k}_{ \text{z$_{9}$:e$_{11\text{h}}$} }^{\text{-}} + \text{k}_{ \text{z$_{9}$:e$_{11\text{h}}$} }^{\text{cat}} )  \left[ \text{z$_{9}$:e$_{11\text{h}}$} \right]&
 \end{flalign*} 

 \begin{flalign*} 
 \frac{d}{dt}\left[ \text{  z$_9^{\text{m}}$:e$_{11\text{h}}^{\text{m}}$  } \right]&= \text{k}_{ \text{z$_9^{\text{m}}$:e$_{11\text{h}}^{\text{m}}$} }^{\text{+}}  \left[ \text{z$_9^{\text{m}}$} \right]  \left[ \text{e$_{11\text{h}}^{\text{m}}$} \right]\\
 &\hskip2.5ex - ( \text{k}_{ \text{z$_9^{\text{m}}$:e$_{11\text{h}}^{\text{m}}$} }^{\text{-}} + \text{k}_{ \text{z$_9^{\text{m}}$:e$_{11\text{h}}^{\text{m}}$} }^{\text{cat}} )  \left[ \text{z$_9^{\text{m}}$:e$_{11\text{h}}^{\text{m}}$} \right]&
 \end{flalign*} 

 \begin{flalign*} 
 \frac{d}{dt}\left[ \text{  z$_{11}$:e$_{11\text{h}}$  } \right]&= \text{k}_{\text{flow}}  ( \text{ [z$_{11}$:e$_{11\text{h}}$]$^{\text{up}}$}  - \left[ \text{z$_{11}$:e$_{11\text{h}}$} \right] ) \\
 &\hskip2.5ex + \text{k}_{ \text{z$_{11}$:e$_{11\text{h}}$} }^{\text{+}}  \left[ \text{z$_{11}$} \right]  \left[ \text{e$_{11\text{h}}$} \right] - ( \text{k}_{ \text{z$_{11}$:e$_{11\text{h}}$} }^{\text{-}} + \text{k}_{ \text{z$_{11}$:e$_{11\text{h}}$} }^{\text{cat}} )  \left[ \text{z$_{11}$:e$_{11\text{h}}$} \right]&
 \end{flalign*} 

 \begin{flalign*} 
 \frac{d}{dt}\left[ \text{  e$_{11\text{h}}$:e$_{11\text{h}}$  } \right]&= \text{k}_{\text{flow}}  ( \text{ [e$_{11\text{h}}$:e$_{11\text{h}}$]$^{\text{up}}$}  - \left[ \text{e$_{11\text{h}}$:e$_{11\text{h}}$} \right] ) \\
 &\hskip2.5ex + \text{k}_{ \text{e$_{11\text{h}}$:e$_{11\text{h}}$} }^{\text{+}}  \left[ \text{e$_{11\text{h}}$} \right]  \left[ \text{e$_{11\text{h}}$} \right] - ( \text{k}_{ \text{e$_{11\text{h}}$:e$_{11\text{h}}$} }^{\text{-}} + \text{k}_{ \text{e$_{11\text{h}}$:e$_{11\text{h}}$} }^{\text{cat}} )  \left[ \text{e$_{11\text{h}}$:e$_{11\text{h}}$} \right]&
 \end{flalign*} 

 \begin{flalign*} 
 \frac{d}{dt}\left[ \text{  e$_{11\text{h}}$:e$_{11}$  } \right]&= \text{k}_{\text{flow}}  ( \text{ [e$_{11\text{h}}$:e$_{11}$]$^{\text{up}}$}  - \left[ \text{e$_{11\text{h}}$:e$_{11}$} \right] ) \\
 &\hskip2.5ex  + \text{k}_{ \text{e$_{11\text{h}}$:e$_{11}$} }^{\text{+}}  \left[ \text{e$_{11\text{h}}$} \right]  \left[ \text{e$_{11}$} \right] - ( \text{k}_{ \text{e$_{11\text{h}}$:e$_{11}$} }^{\text{-}} + \text{k}_{ \text{e$_{11\text{h}}$:e$_{11}$} }^{\text{cat}} )  \left[ \text{e$_{11\text{h}}$:e$_{11}$} \right]&
 \end{flalign*} 

 \begin{flalign*} 
 \frac{d}{dt}\left[ \text{  e$_{11\text{h}}$:e$_{\text{2}}$  } \right]&= \text{k}_{\text{flow}}  ( \text{ [e$_{11\text{h}}$:e$_{\text{2}}$]$^{\text{up}}$}  - \left[ \text{e$_{11\text{h}}$:e$_{\text{2}}$} \right] ) \\
 &\hskip2.5ex  + \text{k}_{ \text{e$_{11\text{h}}$:e$_{\text{2}}$} }^{\text{+}}  \left[ \text{e$_{11\text{h}}$} \right]  \left[ \text{e$_{\text{2}}$} \right] - ( \text{k}_{ \text{e$_{11\text{h}}$:e$_{\text{2}}$} }^{\text{-}} + \text{k}_{ \text{e$_{11\text{h}}$:e$_{\text{2}}$} }^{\text{cat}} )  \left[ \text{e$_{11\text{h}}$:e$_{\text{2}}$} \right]&
 \end{flalign*} 

 \begin{flalign*} 
 \frac{d}{dt}\left[ \text{  z$_{11}^{\text{m}}$:e$_{11\text{h}}^{\text{m}}$  } \right]&= \text{k}_{ \text{z$_{11}^{\text{m}}$:e$_{11\text{h}}^{\text{m}}$} }^{\text{+}}  \left[ \text{z$_{11}^{\text{m}}$} \right]  \left[ \text{e$_{11\text{h}}^{\text{m}}$} \right]\\
 &\hskip2.5ex  - ( \text{k}_{ \text{z$_{11}^{\text{m}}$:e$_{11\text{h}}^{\text{m}}$} }^{\text{-}} + \text{k}_{ \text{z$_{11}^{\text{m}}$:e$_{11\text{h}}^{\text{m}}$} }^{\text{cat}} )  \left[ \text{z$_{11}^{\text{m}}$:e$_{11\text{h}}^{\text{m}}$} \right]&
 \end{flalign*} 

 \begin{flalign*} 
 \frac{d}{dt}\left[ \text{  z$_{11}^{\text{m}}$:e$_{11}^{\text{m,*}}$  } \right]&= \text{k}_{ \text{z$_{11}^{\text{m}}$:e$_{11}^{\text{m,*}}$} }^{\text{+}}  \left[ \text{z$_{11}^{\text{m}}$} \right]  \left[ \text{e$_{11}^{\text{m,*}}$} \right]\\
 &\hskip2.5ex   - ( \text{k}_{ \text{z$_{11}^{\text{m}}$:e$_{11}^{\text{m,*}}$} }^{\text{-}} + \text{k}_{ \text{z$_{11}^{\text{m}}$:e$_{11}^{\text{m,*}}$} }^{\text{cat}} )  \left[ \text{z$_{11}^{\text{m}}$:e$_{11}^{\text{m,*}}$} \right]&
 \end{flalign*} 

 \begin{flalign*} 
 \frac{d}{dt}\left[ \text{  e$_{11\text{h}}^{\text{m,*}}$:e$_{11\text{h}}^{\text{m}}$  } \right]&= \text{k}_{ \text{e$_{11\text{h}}^{\text{m,*}}$:e$_{11\text{h}}^{\text{m}}$} }^{\text{+}}  \left[ \text{e$_{11\text{h}}^{\text{m,*}}$} \right]  \left[ \text{e$_{11\text{h}}^{\text{m}}$} \right]\\
 &\hskip2.5ex   - ( \text{k}_{ \text{e$_{11\text{h}}^{\text{m,*}}$:e$_{11\text{h}}^{\text{m}}$} }^{\text{-}} + \text{k}_{ \text{e$_{11\text{h}}^{\text{m,*}}$:e$_{11\text{h}}^{\text{m}}$} }^{\text{cat}} )  \left[ \text{e$_{11\text{h}}^{\text{m,*}}$:e$_{11\text{h}}^{\text{m}}$} \right]&
 \end{flalign*} 

 \begin{flalign*} 
 \frac{d}{dt}\left[ \text{  e$_{11\text{h}}^{\text{m,*}}$:e$_{11}^{\text{m,*}}$  } \right]&= \text{k}_{ \text{e$_{11\text{h}}^{\text{m,*}}$:e$_{11}^{\text{m,*}}$} }^{\text{+}}  \left[ \text{e$_{11\text{h}}^{\text{m,*}}$} \right]  \left[ \text{e$_{11}^{\text{m,*}}$} \right]\\
 &\hskip2.5ex   - ( \text{k}_{ \text{e$_{11\text{h}}^{\text{m,*}}$:e$_{11}^{\text{m,*}}$} }^{\text{-}} + \text{k}_{ \text{e$_{11\text{h}}^{\text{m,*}}$:e$_{11}^{\text{m,*}}$} }^{\text{cat}} )  \left[ \text{e$_{11\text{h}}^{\text{m,*}}$:e$_{11}^{\text{m,*}}$} \right]&
 \end{flalign*} 

 \begin{flalign*} 
 \frac{d}{dt}\left[ \text{  e$_{11\text{h}}^{\text{m,*}}$:e$_2^{\text{m}}$  } \right]&= \text{k}_{ \text{e$_{11\text{h}}^{\text{m,*}}$:e$_2^{\text{m}}$} }^{\text{+}}  \left[ \text{e$_{11\text{h}}^{\text{m,*}}$} \right]  \left[ \text{e$_2^{\text{m}}$} \right]\\
 &\hskip2.5ex   - ( \text{k}_{ \text{e$_{11\text{h}}^{\text{m,*}}$:e$_2^{\text{m}}$} }^{\text{-}} + \text{k}_{ \text{e$_{11\text{h}}^{\text{m,*}}$:e$_2^{\text{m}}$} }^{\text{cat}} )  \left[ \text{e$_{11\text{h}}^{\text{m,*}}$:e$_2^{\text{m}}$} \right]&
 \end{flalign*} 

 \begin{flalign*} 
 \frac{d}{dt}\left[ \text{  e$_{5\text{h}}^{\text{m}}$  } \right]&= \text{k}_{ \text{z$_5^{\text{m}}$:e$_{10}^{\text{m}}$} }^{\text{cat}}  \left[ \text{z$_5^{\text{m}}$:e$_{10}^{\text{m}}$} \right] + \text{k}_{ \text{5} }^{\text{on}}  \left[ \text{e$_{\text{5h}}$} \right]  \left[ \text{p}_{\text{5}}^{\text{avail}} \right] \\
 &\hskip2.5ex  - \text{k}_{ \text{5} }^{\text{off}}  \left[ \text{e$_{5\text{h}}^{\text{m}}$} \right] - \text{k}_{ \text{e$_{5\text{h}}^{\text{m}}$:e$_{10}^{\text{m}}$} }^{\text{+}}  \left[ \text{e$_{10}^{\text{m}}$} \right]  \left[ \text{e$_{5\text{h}}^{\text{m}}$} \right] + \text{k}_{ \text{e$_{5\text{h}}^{\text{m}}$:e$_{10}^{\text{m}}$} }^{\text{-}}  \left[ \text{PROh} \right] \\
 &\hskip2.5ex  - \text{k}_{ \text{e$_{5\text{h}}^{\text{m}}$:e$_2^{\text{m}}$} }^{\text{+}}  \left[ \text{e$_2^{\text{m}}$} \right]  \left[ \text{e$_{5\text{h}}^{\text{m,*}}$} \right] + \text{k}_{ \text{e$_{5\text{h}}^{\text{m}}$:e$_2^{\text{m}}$} }^{\text{-}}  \left[ \text{e$_{5\text{h}}^{\text{m}}$:e$_2^{\text{m}}$} \right] \\
 &\hskip2.5ex  - \text{k}_{ \text{TFPI:e$_{5\text{h}}^{\text{m}}$} }^{\text{+}}  \left[ \text{e$_{5\text{h}}^{\text{m}}$} \right]  \left[ \text{TFPI} \right] + \text{k}_{ \text{TFPI:e$_{5\text{h}}^{\text{m}}$} }^{\text{-}}  \left[ \text{TFPI:e$_{5\text{h}}^{\text{m}}$} \right] \\
 &\hskip2.5ex  - \text{k}_{ \text{e$_{5\text{h}}^{\text{m}}$:APC} }^{\text{+}}  \left[ \text{e$_{5\text{h}}^{\text{m}}$} \right]  \left[ \text{APC} \right] + \text{k}_{ \text{e$_{5\text{h}}^{\text{m}}$:APC} }^{\text{-}}  \left[ \text{APC:e$_{5\text{h}}^{\text{m}}$} \right] \\
 &\hskip2.5ex  - \text{k}_{ \text{(TFPI:e$_{10}$):e$_{5\text{h}}^{\text{m}}$} }^{\text{+}}  \left[ \text{TFPI:e$_{10}^{\text{m}}$} \right]  \left[ \text{e$_{5\text{h}}^{\text{m}}$} \right] + \text{k}_{ \text{(TFPI:e$_{10}$):e$_{5\text{h}}^{\text{m}}$} }^{\text{-}}  \left[ \text{e$_{10}^{\text{m}}$:TFPI:e$_{5\text{h}}^{\text{m}}$} \right] &
 \end{flalign*} 

 \begin{flalign*} 
 \frac{d}{dt}\left[ \text{  e$_{\text{5h}}$  } \right]&= - \text{k}_{ \text{5} }^{\text{on}}  \left[ \text{e$_{\text{5h}}$} \right]  \left[ \text{p}_{ \text{5} }^{\text{avail}}  \right]+ \text{k}_{ \text{5} }^{\text{off}}  \left[ \text{e$_{5\text{h}}^{\text{m}}$} \right]\\
 &\hskip2.5ex  + \text{k}_{\text{flow}}  ( \text{ [e$_{\text{5h}}$]$^{\text{up}}$}  - \left[ \text{e$_{\text{5h}}$} \right] ) \\
 &\hskip2.5ex  + \text{h5}\cdot \text{n5}  \left[ \frac{d}{dt} \left( [\text{PL}^{\text{as}}] + [\text{PL}^{\text{av}}] \right) \right]  \\
 &\hskip2.5ex  - \text{k}_{ \text{e$_{\text{5h}}$:e$_{\text{2}}$} }^{\text{+}}  \left[ \text{e$_{\text{2}}$} \right]  \left[ \text{e$_{\text{5h}}$} \right] + \text{k}_{ \text{e$_{\text{5h}}$:e$_{\text{2}}$} }^{\text{-}}  \left[ \text{e$_{\text{5h}}$:e$_{\text{2}}$} \right] \\
 &\hskip2.5ex  - \text{k}_{ \text{e$_{\text{5h}}$:APC} }^{\text{+}}  \left[ \text{APC} \right]  \left[ \text{e$_{\text{5h}}$} \right] + \text{k}_{ \text{e$_{\text{5h}}$:APC} }^{\text{-}}  \left[ \text{APC:e$_{\text{5h}}$} \right] \\
 &\hskip2.5ex  - \text{k}_{ \text{TFPI:e$_{\text{5h}}$} }^{\text{+}}  \left[ \text{e$_{\text{5h}}$} \right]  \left[ \text{TFPI} \right] + \text{k}_{ \text{TFPI:e$_{\text{5h}}$} }^{\text{-}}  \left[ \text{TFPI:e$_{\text{5h}}$} \right] \\
 &\hskip2.5ex  - \text{k}_{ \text{(TFPI:e$_{10}$):e$_{\text{5h}}$} }^{\text{+}}  \left[ \text{TFPI:e$_{10}$} \right]  \left[ \text{e$_{\text{5h}}$} \right] + \text{k}_{ \text{(TFPI:e$_{10}$):e$_{\text{5h}}$} }^{\text{-}}  \left[ \text{e$_{10}$:TFPI:e$_{\text{5h}}$} \right] \\
 &\hskip2.5ex  - \text{k}_{ \text{(TFPI:e$_{10}^{\text{m}}$):E5h} }^{\text{+}}  \left[ \text{TFPI:e$_{10}^{\text{m}}$} \right]  \left[ \text{e$_{\text{5h}}$} \right] + \text{k}_{ \text{(TFPI:e$_{10}^{\text{m}}$):E5h} }^{\text{-}}  \left[ \text{e$_{10}^{\text{m}}$:TFPI:e$_{\text{5h}}$} \right] &
 \end{flalign*} 

 \begin{flalign*} 
 \frac{d}{dt}\left[ \text{  PROh  } \right]&= + \text{k}_{ \text{e$_{5\text{h}}^{\text{m}}$:e$_{10}^{\text{m}}$} }^{\text{+}}  \left[ \text{e$_{10}^{\text{m}}$} \right]  \left[ \text{e$_{5\text{h}}^{\text{m}}$} \right] - \text{k}_{ \text{e$_{5\text{h}}^{\text{m}}$:e$_{10}^{\text{m}}$} }^{\text{-}}  \left[ \text{PROh} \right]\\
 &\hskip2.5ex  - \text{k}_{ \text{z$_2^{\text{m}}$:PROh} }^{\text{+}}  \left[ \text{PROh} \right]  \left[ \text{z$_2^{\text{m}}$} \right] + \text{k}_{ \text{z$_2^{\text{m}}$:PROh} }^{\text{-}}  \left[ \text{z$_2^{\text{m}}$:PROh} \right] + \text{k}_{ \text{z$_2^{\text{m}}$:PROh} }^{\text{cat}}  \left[ \text{z$_2^{\text{m}}$:PROh} \right] \\
 &\hskip2.5ex  - \text{k}_{ \text{TFPI:PROhv10} }^{\text{+}}  \left[ \text{PROh} \right]  \left[ \text{TFPI} \right] + \text{k}_{ \text{TFPI:PROhv10} }^{\text{-}}  \left[ \text{TFPI:PROhv10} \right] \\
 &\hskip2.5ex  - \text{k}_{ \text{TFPI:PROhv5} }^{\text{+}}  \left[ \text{PROh} \right]  \left[ \text{TFPI} \right] + \text{k}_{ \text{TFPI:PROhv5} }^{\text{-}}  \left[ \text{TFPI:PROhv5} \right] \\
 &\hskip2.5ex  - \text{k}_{ \text{PROh:e$_2^{\text{m}}$} }^{\text{+}}  \left[ \text{PROh} \right]  \left[ \text{e$_2^{\text{m}}$} \right] + \text{k}_{ \text{PROh:e$_2^{\text{m}}$} }^{\text{-}}  \left[ \text{PROh:e$_2^{\text{m}}$} \right] &
 \end{flalign*} 

 \begin{flalign*} 
 \frac{d}{dt}\left[ \text{  z$_2^{\text{m}}$:PROh  } \right]&= + \text{k}_{ \text{z$_2^{\text{m}}$:PROh} }^{\text{+}}  \left[ \text{PROh} \right]  \left[ \text{z$_2^{\text{m}}$} \right] - \text{k}_{ \text{z$_2^{\text{m}}$:PROh} }^{\text{-}}  \left[ \text{z$_2^{\text{m}}$:PROh} \right]\\
 &\hskip2.5ex  - \text{k}_{ \text{z$_2^{\text{m}}$:PROh} }^{\text{cat}}  \left[ \text{z$_2^{\text{m}}$:PROh} \right]&
 \end{flalign*} 

 \begin{flalign*} 
 \frac{d}{dt}\left[ \text{  e$_{5\text{h}}^{\text{m}}$:e$_2^{\text{m}}$  } \right]&= + \text{k}_{ \text{e$_{5\text{h}}^{\text{m}}$:e$_2^{\text{m}}$} }^{\text{+}}  \left[ \text{e$_2^{\text{m}}$} \right]  \left[ \text{e$_{5\text{h}}^{\text{m,*}}$} \right] - \text{k}_{ \text{e$_{5\text{h}}^{\text{m}}$:e$_2^{\text{m}}$} }^{\text{-}}  \left[ \text{e$_{5\text{h}}^{\text{m}}$:e$_2^{\text{m}}$} \right]\\
 &\hskip2.5ex  - \text{k}_{ \text{e$_{5\text{h}}^{\text{m}}$:e$_2^{\text{m}}$} }^{\text{cat}}  \left[ \text{e$_{5\text{h}}^{\text{m}}$:e$_2^{\text{m}}$} \right]&
 \end{flalign*} 

 \begin{flalign*} 
 \frac{d}{dt}\left[ \text{  e$_{\text{5h}}$:e$_{\text{2}}$  } \right]&= + \text{k}_{ \text{e$_{\text{5h}}$:e$_{\text{2}}$} }^{\text{+}}  \left[ \text{e$_{\text{2}}$} \right]  \left[ \text{e$_{\text{5h}}$} \right] - \text{k}_{ \text{e$_{\text{5h}}$:e$_{\text{2}}$} }^{\text{-}}  \left[ \text{e$_{\text{5h}}$:e$_{\text{2}}$} \right]\\
 &\hskip2.5ex  - \text{k}_{ \text{e$_{\text{5h}}$:e$_{\text{2}}$} }^{\text{cat}}  \left[ \text{e$_{\text{5h}}$:e$_{\text{2}}$} \right] \\
 &\hskip2.5ex  + \text{k}_{\text{flow}}  ( \text{ [e$_{\text{5h}}$:e$_{\text{2}}$]$^{\text{up}}$}  - \left[ \text{e$_{\text{5h}}$:e$_{\text{2}}$} \right] )  &
 \end{flalign*} 

 \begin{flalign*} 
 \frac{d}{dt}\left[ \text{  TFPI:e$_{5\text{h}}^{\text{m}}$  } \right]&= + \text{k}_{ \text{TFPI:e$_{5\text{h}}^{\text{m}}$} }^{\text{+}}  \left[ \text{e$_{5\text{h}}^{\text{m}}$} \right]  \left[ \text{TFPI} \right] - \text{k}_{ \text{TFPI:e$_{5\text{h}}^{\text{m}}$} }^{\text{-}}  \left[ \text{TFPI:e$_{5\text{h}}^{\text{m}}$} \right]\\
 &\hskip2.5ex  - \text{k}_{ \text{TFPI:e$_{5\text{h}}^{\text{m}}$:e$_{10}^{\text{m}}$} }^{\text{+}}  \left[ \text{TFPI:e$_{5\text{h}}^{\text{m}}$} \right]  \left[ \text{e$_{10}^{\text{m}}$} \right] \\
 &\hskip2.5ex  + \text{k}_{ \text{TFPI:e$_{5\text{h}}^{\text{m}}$:e$_{10}^{\text{m}}$} }^{\text{-}}  \left[ \text{e$_{10}^{\text{m}}$:TFPI:e$_{5\text{h}}^{\text{m}}$} \right] \\
 &\hskip2.5ex  + \text{k}_{ \text{5} }^{\text{on}}  \left[ \text{TFPI:e$_{\text{5h}}$} \right]  \left[ \text{p}_{ \text{5} }^{\text{avail}}  \right]- \text{k}_{ \text{5} }^{\text{off}}  \left[ \text{TFPI:e$_{5\text{h}}^{\text{m}}$} \right] \\
 &\hskip2.5ex  - \text{k}_{ \text{TFPI:e$_{5\text{h}}^{\text{m}}$:e$_{10}$} }^{\text{+}}  \left[ \text{TFPI:e$_{5\text{h}}^{\text{m}}$} \right]  \left[ \text{e$_{10}$} \right] \\
 &\hskip2.5ex  + \text{k}_{ \text{TFPI:e$_{5\text{h}}^{\text{m}}$:e$_{10}$} }^{\text{-}}  \left[ \text{e$_{10}$:TFPI:e$_{5\text{h}}^{\text{m}}$} \right]&
 \end{flalign*} 

 \begin{flalign*} 
 \frac{d}{dt}\left[ \text{  APC:e$_{5\text{h}}^{\text{m}}$  } \right]&= + \text{k}_{ \text{e$_{5\text{h}}^{\text{m}}$:APC} }^{\text{+}}  \left[ \text{e$_{5\text{h}}^{\text{m}}$} \right]  \left[ \text{APC} \right] - \text{k}_{ \text{e$_{5\text{h}}^{\text{m}}$:APC} }^{\text{-}}  \left[ \text{APC:e$_{5\text{h}}^{\text{m}}$} \right]\\
 &\hskip2.5ex  - \text{k}_{ \text{e$_{5\text{h}}^{\text{m}}$:APC} }^{\text{cat}}  \left[ \text{APC:e$_{5\text{h}}^{\text{m}}$} \right]&
 \end{flalign*} 

 \begin{flalign*} 
 \frac{d}{dt}\left[ \text{  APC:e$_{\text{5h}}$  } \right]&= + \text{k}_{ \text{e$_{\text{5h}}$:APC} }^{\text{+}}  \left[ \text{APC} \right]  \left[ \text{e$_{\text{5h}}$} \right] - \text{k}_{ \text{e$_{\text{5h}}$:APC} }^{\text{-}}  \left[ \text{APC:e$_{\text{5h}}$} \right]\\
 &\hskip2.5ex  - \text{k}_{ \text{e$_{\text{5h}}$:APC} }^{\text{cat}}  \left[ \text{APC:e$_{\text{5h}}$} \right] \\
 &\hskip2.5ex  + \text{k}_{\text{flow}}  ( \text{ [APC:e$_{\text{5h}}$]$^{\text{up}}$}  - \left[ \text{APC:e$_{\text{5h}}$} \right] )  &
 \end{flalign*} 

 \begin{flalign*} 
 \frac{d}{dt}\left[ \text{  TFPI:e$_{\text{5h}}$  } \right]&= + \text{k}_{ \text{TFPI:e$_{\text{5h}}$} }^{\text{+}}  \left[ \text{e$_{\text{5h}}$} \right]  \left[ \text{TFPI} \right] - \text{k}_{ \text{TFPI:e$_{\text{5h}}$} }^{\text{-}}  \left[ \text{TFPI:e$_{\text{5h}}$} \right]\\
 &\hskip2.5ex  + \text{k}_{\text{flow}}  ( \text{ [TFPI:e$_{\text{5h}}$]$^{\text{up}}$}  - \left[ \text{TFPI:e$_{\text{5h}}$} \right] ) \\
 &\hskip2.5ex  - \text{k}_{ \text{(TFPI:e$_{\text{5h}}$):e$_{10}$} }^{\text{+}}  \left[ \text{TFPI:e$_{\text{5h}}$} \right]  \left[ \text{e$_{10}$} \right] + \text{k}_{ \text{(TFPI:e$_{\text{5h}}$):e$_{10}$} }^{\text{-}}  \left[ \text{e$_{10}$:TFPI:e$_{\text{5h}}$} \right] \\
 &\hskip2.5ex  - \text{k}_{ \text{5} }^{\text{on}}  \left[ \text{TFPI:e$_{\text{5h}}$} \right]  \left[ \text{p}_{ \text{5} }^{\text{avail}}  \right]+ \text{k}_{ \text{5} }^{\text{off}}  \left[ \text{TFPI:e$_{5\text{h}}^{\text{m}}$} \right]&
 \end{flalign*} 

 \begin{flalign*} 
 \frac{d}{dt}\left[ \text{  \text{TFPI:e$_{10}^{\text{m}}$}  } \right]&= + \text{k}_{ \text{TFPI:e$_{10}^{\text{m}}$} }^{\text{+}}  \left[ \text{TFPI} \right]  \left[ \text{e$_{10}^{\text{m}}$} \right] - \text{k}_{ \text{TFPI:e$_{10}^{\text{m}}$} }^{\text{-}}  \left[ \text{TFPI:e$_{10}^{\text{m}}$} \right]\\
 &\hskip2.5ex  - \text{k}_{ \text{(TFPI:e$_{10}$):e$_{5\text{h}}^{\text{m}}$} }^{\text{+}}  \left[ \text{TFPI:e$_{10}^{\text{m}}$} \right]  \left[ \text{e$_{5\text{h}}^{\text{m}}$} \right] + \text{k}_{ \text{(TFPI:e$_{10}$):e$_{5\text{h}}^{\text{m}}$} }^{\text{-}}  \left[ \text{e$_{10}^{\text{m}}$:TFPI:e$_{5\text{h}}^{\text{m}}$} \right] \\
 &\hskip2.5ex  + \text{k}_{ \text{10} }^{\text{on}}  \left[ \text{TFPI:e$_{10}$} \right]  \left[ \text{p}_{ \text{10} }^{\text{avail}} \right] - \text{k}_{ \text{10} }^{\text{off}}  \left[ \text{TFPI:e$_{10}^{\text{m}}$} \right] \\
 &\hskip2.5ex  - \text{k}_{ \text{(TFPI:e$_{10}^{\text{m}}$):E5h} }^{\text{+}}  \left[ \text{TFPI:e$_{10}^{\text{m}}$} \right]  \left[ \text{e$_{\text{5h}}$} \right] + \text{k}_{ \text{(TFPI:e$_{10}^{\text{m}}$):E5h} }^{\text{-}}  \left[ \text{e$_{10}^{\text{m}}$:TFPI:e$_{\text{5h}}$} \right]&
 \end{flalign*} 

 \begin{flalign*} 
 \frac{d}{dt}\left[ \text{  TFPI:PROhv10  } \right]&= + \text{k}_{ \text{TFPI:PROhv10} }^{\text{+}}  \left[ \text{PROh} \right]  \left[ \text{TFPI} \right] - \text{k}_{ \text{TFPI:PROhv10} }^{\text{-}}  \left[ \text{TFPI:PROhv10} \right]&
 \end{flalign*} 

 \begin{flalign*} 
 \frac{d}{dt}\left[ \text{  TFPI:PROhv5  } \right]&= + \text{k}_{ \text{TFPI:PROhv5} }^{\text{+}}  \left[ \text{PROh} \right]  \left[ \text{TFPI} \right] - \text{k}_{ \text{TFPI:PROhv5} }^{\text{-}}  \left[ \text{TFPI:PROhv5} \right]&
 \end{flalign*} 

 \begin{flalign*} 
 \frac{d}{dt}\left[ \text{  e$_{10}^{\text{m}}$:TFPI:e$_{5\text{h}}^{\text{m}}$  } \right]&= + \text{k}_{ \text{(TFPI:e$_{10}$):e$_{5\text{h}}^{\text{m}}$} }^{\text{+}}  \left[ \text{TFPI:e$_{10}^{\text{m}}$} \right]  \left[ \text{e$_{5\text{h}}^{\text{m}}$} \right] - \text{k}_{ \text{(TFPI:e$_{10}$):e$_{5\text{h}}^{\text{m}}$} }^{\text{-}}  \left[ \text{e$_{10}^{\text{m}}$:TFPI:e$_{5\text{h}}^{\text{m}}$} \right]\\
 &\hskip2.5ex   + \text{k}_{ \text{TFPI:e$_{5\text{h}}^{\text{m}}$:e$_{10}^{\text{m}}$} }^{\text{+}}  \left[ \text{TFPI:e$_{5\text{h}}^{\text{m}}$} \right]  \left[ \text{e$_{10}^{\text{m}}$} \right] \\
 &\hskip2.5ex   - \text{k}_{ \text{TFPI:e$_{5\text{h}}^{\text{m}}$:e$_{10}^{\text{m}}$} }^{\text{-}}  \left[ \text{e$_{10}^{\text{m}}$:TFPI:e$_{5\text{h}}^{\text{m}}$} \right] \\
 &\hskip2.5ex   + \text{k$_{10}^{\text{on,TFPI}}$}  \left[ \text{e$_{10}$:TFPI:e$_{5\text{h}}^{\text{m}}$} \right]  \left[ \text{p}_{ \text{10} }^{\text{avail}} \right] - \text{k$_{10}^{\text{off,TFPI}}$}  \left[ \text{e$_{10}^{\text{m}}$:TFPI:e$_{5\text{h}}^{\text{m}}$} \right] \\
 &\hskip2.5ex   + \text{k$_{5}^{\text{on,TFPI}}$}  \left[ \text{e$_{10}^{\text{m}}$:TFPI:e$_{\text{5h}}$} \right]  \left[ \text{p}_{ \text{5} }^{\text{avail}}  \right]- \text{k$_{5}^{\text{off,TFPI}}$}  \left[ \text{e$_{10}^{\text{m}}$:TFPI:e$_{5\text{h}}^{\text{m}}$} \right]&
 \end{flalign*} 

 \begin{flalign*} 
 \frac{d}{dt}\left[ \text{  e$_{10}$:TFPI:e$_{\text{5h}}$  } \right]&= + \text{k}_{\text{flow}}  ( \text{ [e$_{10}$:TFPI:e$_{\text{5h}}$]$^{\text{up}}$}  - \left[ \text{e$_{10}$:TFPI:e$_{\text{5h}}$} \right] ) \\
 &\hskip2.5ex   + \text{k}_{ \text{(TFPI:e$_{10}$):e$_{\text{5h}}$} }^{\text{+}}  \left[ \text{TFPI:e$_{10}$} \right]  \left[ \text{e$_{\text{5h}}$} \right] - \text{k}_{ \text{(TFPI:e$_{10}$):e$_{\text{5h}}$} }^{\text{-}}  \left[ \text{e$_{10}$:TFPI:e$_{\text{5h}}$} \right] \\
 &\hskip2.5ex   + \text{k}_{ \text{(TFPI:e$_{\text{5h}}$):e$_{10}$} }^{\text{+}}  \left[ \text{TFPI:e$_{\text{5h}}$} \right]  \left[ \text{e$_{10}$} \right] \\
 &\hskip2.5ex   - \text{k}_{ \text{(TFPI:e$_{\text{5h}}$):e$_{10}$} }^{\text{-}}  \left[ \text{e$_{10}$:TFPI:e$_{\text{5h}}$} \right] \\
 &\hskip2.5ex   - \text{k$_{5}^{\text{on,TFPI}}$}  \left[ \text{e$_{10}$:TFPI:e$_{\text{5h}}$} \right]  \left[ \text{p}_{ \text{5} }^{\text{avail}}  \right]+ \text{k$_{5}^{\text{off,TFPI}}$}  \left[ \text{e$_{10}$:TFPI:e$_{5\text{h}}^{\text{m}}$} \right] \\
 &\hskip2.5ex   - \text{k$_{10}^{\text{on,TFPI}}$}  \left[ \text{e$_{10}$:TFPI:e$_{\text{5h}}$} \right]  \left[ \text{p}_{ \text{10} }^{\text{avail}} \right] + \text{k$_{10}^{\text{off,TFPI}}$}  \left[ \text{e$_{10}^{\text{m}}$:TFPI:e$_{\text{5h}}$} \right]&
 \end{flalign*} 

 \begin{flalign*} 
 \frac{d}{dt}\left[ \text{  e$_{10}$:TFPI:e$_{5\text{h}}^{\text{m}}$  } \right]&= + \text{k$_{5}^{\text{on,TFPI}}$}  \left[ \text{e$_{10}$:TFPI:e$_{\text{5h}}$} \right]  \left[ \text{p}_{ \text{5} }^{\text{avail}}  \right]- \text{k$_{5}^{\text{off,TFPI}}$}  \left[ \text{e$_{10}$:TFPI:e$_{5\text{h}}^{\text{m}}$} \right]\\
 &\hskip2.5ex   - \text{k$_{10}^{\text{on,TFPI}}$}  \left[ \text{e$_{10}$:TFPI:e$_{5\text{h}}^{\text{m}}$} \right]  \left[ \text{p}_{ \text{10} }^{\text{avail}} \right] + \text{k$_{10}^{\text{off,TFPI}}$}  \left[ \text{e$_{10}^{\text{m}}$:TFPI:e$_{5\text{h}}^{\text{m}}$} \right] \\
 &\hskip2.5ex   + \text{k}_{ \text{TFPI:e$_{5\text{h}}^{\text{m}}$:e$_{10}$} }^{\text{+}}  \left[ \text{TFPI:e$_{5\text{h}}^{\text{m}}$} \right]  \left[ \text{e$_{10}$} \right] \\
 &\hskip2.5ex   - \text{k}_{ \text{TFPI:e$_{5\text{h}}^{\text{m}}$:e$_{10}$} }^{\text{-}}  \left[ \text{e$_{10}$:TFPI:e$_{5\text{h}}^{\text{m}}$} \right]&
 \end{flalign*} 

 \begin{flalign*} 
 \frac{d}{dt}\left[ \text{  e$_{10}^{\text{m}}$:TFPI:e$_{\text{5h}}$  } \right]&= + \text{k$_{10}^{\text{on,TFPI}}$}  \left[ \text{e$_{10}$:TFPI:e$_{\text{5h}}$} \right]  \left[ \text{p}_{ \text{10} }^{\text{avail}} \right] - \text{k$_{10}^{\text{off,TFPI}}$}  \left[ \text{e$_{10}^{\text{m}}$:TFPI:e$_{\text{5h}}$} \right]\\
 &\hskip2.5ex   - \text{k$_{5}^{\text{on,TFPI}}$}  \left[ \text{e$_{10}^{\text{m}}$:TFPI:e$_{\text{5h}}$} \right]  \left[ \text{p}_{ \text{5} }^{\text{avail}}  \right]+ \text{k$_{5}^{\text{off,TFPI}}$}  \left[ \text{e$_{10}^{\text{m}}$:TFPI:e$_{5\text{h}}^{\text{m}}$} \right] \\
 &\hskip2.5ex   + \text{k}_{ \text{(TFPI:e$_{10}^{\text{m}}$):E5h} }^{\text{+}}  \left[ \text{TFPI:e$_{10}^{\text{m}}$} \right]  \left[ \text{e$_{\text{5h}}$} \right] - \text{k}_{ \text{(TFPI:e$_{10}^{\text{m}}$):E5h} }^{\text{-}}  \left[ \text{e$_{10}^{\text{m}}$:TFPI:e$_{\text{5h}}$} \right]&
 \end{flalign*} 

 \begin{flalign*} 
 \frac{d}{dt}\left[ \text{  TFPIb  } \right]&= - \text{k}_{ \text{TFPIb$^{\text{m}}$} }^{\text{+}}  \left[ \text{e$_{10}$$^{\text{ec}}$} \right]  \left[ \text{TFPIb} \right] + \text{k}_{ \text{TFPIb$^{\text{m}}$} }^{\text{-}}  \left[ \text{TFPIb$^{\text{m}}$} \right]&
 \end{flalign*} 

 \begin{flalign*} 
 \frac{d}{dt}\left[ \text{  TFPIb$^{\text{m}}$  } \right]&= + \text{k}_{ \text{TFPIb$^{\text{m}}$} }^{\text{+}}  \left[ \text{e$_{10}$$^{\text{ec}}$} \right]  \left[ \text{TFPIb} \right] - \text{k}_{ \text{TFPIb$^{\text{m}}$} }^{\text{-}}  \left[ \text{TFPIb$^{\text{m}}$} \right]&
 \end{flalign*} 

 \begin{flalign*} 
 \frac{d}{dt}\left[ \text{  PROh:e$_2^{\text{m}}$  } \right]&= + \text{k}_{ \text{PROh:e$_2^{\text{m}}$} }^{\text{+}}  \left[ \text{PROh} \right]  \left[ \text{e$_2^{\text{m}}$} \right]\\
 &\hskip2.5ex   - \text{k}_{ \text{PROh:e$_2^{\text{m}}$} }^{\text{-}}  \left[ \text{PROh:e$_2^{\text{m}}$} \right] - \text{k}_{ \text{PROh:e$_2^{\text{m}}$} }^{\text{cat}}  \left[ \text{PROh:e$_2^{\text{m}}$} \right]&
 \end{flalign*} 

 \begin{flalign*} 
 \frac{d}{dt}\left[ \text{  e$_9^{\text{m}}$:AT  } \right]&= + \text{k}_{ \text{e$_9^{\text{m}}$:AT} }  \left[ \text{AT} \right]  \left[ \text{e$_9^{\text{m}}$} \right]\\
 &\hskip2.5ex   - \text{k}_{ \text{9:AT} }^{\text{off}}  \left[ \text{e$_9^{\text{m}}$:AT} \right] + \text{k}_{ \text{9:AT} }^{\text{on}}  \left[ \text{p}_{ \text{9} }^{\text{avail}} \right]  \left[ \text{e$_{9}$:AT} \right]&
 \end{flalign*} 

 \begin{flalign*} 
 \frac{d}{dt}\left[ \text{  e$_{10}^{\text{m}}$:AT  } \right]&= + \text{k}_{ \text{ \text{e$_{10}^{\text{m}}$:AT} }}^{\text{off}}  \left[\text{AT}\right]  \left[\text{e$_{10}^{\text{m}}$}\right] - \text{k}_{\text{10:AT}}   \left[ \text{e$_{10}^{\text{m}}$:AT} \right]\\
 &\hskip2.5ex   + \text{k}_{ \text{10:AT} }^{\text{on}}  \left[ \text{p}_{ \text{10} }^{\text{avail}} \right]  \left[ \text{e$_{10}$:AT} \right]&
 \end{flalign*} 

 \begin{flalign*} 
 \frac{d}{dt}\left[ \text{  e$_2^{\text{m}}$:AT  } \right]&= + \text{k}_{ \text{ \text{e$_{\text{2}}$:AT} }}^{\text{on}}  \left[\text{AT}\right]  \left[\text{e$_2^{\text{m}}$}\right] - \text{k}_{\text{e$_{\text{2}}$:AT}}^{\text{off}}   \left[ \text{e$_2^{\text{m}}$:AT} \right]\\
 &\hskip2.5ex   + \text{k}_{ \text{2:AT} }^{\text{on}}  \left[ \text{p}_{ \text{2,*} }^{\text{avail}} \right]  \left[ \text{e$_{\text{2}}$:AT} \right]&
 \end{flalign*} 

 \begin{flalign*} 
 \frac{d}{dt}\left[ \text{  e$_{11}$:AT  } \right]&= + \text{k}_{ \text{e$_{11}$:AT} }  \left[ \text{AT} \right]  \left[ \text{e$_{11}$} \right] - \text{k}_{ \text{e$_{11}$:AT} }  \left[ \text{AT} \right]  \left[ \text{e$_{11}$:AT} \right]\\
 &\hskip2.5ex   + \text{k$_{\text{11:AT}}^{\text{off,*}}$}  \left[ \text{e$_{11}^{\text{m,*}}$:AT} \right] - \left[ \text{k$_{\text{11:AT}}^{\text{on,*}}$} \right]  \left[ \text{p}_{ \text{11,*} }^{\text{avail}} \right]  \left[ \text{e$_{11}$:AT} \right]&
 \end{flalign*} 

 \begin{flalign*} 
 \frac{d}{dt}\left[ \text{  AT:e$_{11}$:AT  } \right]&= + \text{k}_{ \text{e$_{11}$:AT} }  \left[ \text{AT} \right]  \left[ \text{e$_{11}$:AT} \right]&
 \end{flalign*} 

 \begin{flalign*} 
 \frac{d}{dt}\left[ \text{  e$_{11}^{\text{m,*}}$:AT  } \right]&= + \text{k}_{ \text{e$_{11}$:AT} }  \left[ \text{AT} \right]  \left[ \text{e$_{11}^{\text{m,*}}$} \right]\\
 &\hskip2.5ex   - \text{k$_{\text{11:AT}}^{\text{off,*}}$}  \left[ \text{e$_{11}^{\text{m,*}}$:AT} \right] + \left[ \text{k$_{\text{11:AT}}^{\text{on,*}}$} \right]  \left[ \text{p}_{ \text{11,*} }^{\text{avail}} \right]  \left[ \text{e$_{11}$:AT} \right]&
 \end{flalign*} 

 \begin{flalign*} 
 \frac{d}{dt}\left[ \text{  e$_{11\text{h}}$:AT  } \right]&= + \text{k}_{ \text{e$_{11}$:AT} }  \left[ \text{AT} \right]  \left[ \text{e$_{11\text{h}}$} \right]\\
 &\hskip2.5ex   + \text{k$_{\text{11:AT}}^{\text{off}}$}  \left[ \text{e$_{11\text{h}}^{\text{m}}$:AT} \right] - \text{k$_{\text{11:AT}}^{\text{on}}$}  \left[ \text{p}_{ \text{11} }^{\text{avail}} \right]  \left[ \text{e$_{11\text{h}}$:AT} \right]&
 \end{flalign*} 

 \begin{flalign*} 
 \frac{d}{dt}\left[ \text{  e$_{11\text{h}}^{\text{m}}$:AT  } \right]&= + \text{k}_{ \text{e$_{11}$:AT} }  \left[ \text{AT} \right]  \left[ \text{e$_{11\text{h}}^{\text{m}}$} \right]\\
 &\hskip2.5ex   - \text{k$_{\text{11:AT}}^{\text{off}}$}  \left[ \text{e$_{11\text{h}}^{\text{m}}$:AT} \right] + \text{k$_{\text{11:AT}}^{\text{on}}$}  \left[ \text{p}_{ \text{11} }^{\text{avail}} \right]  \left[ \text{e$_{11\text{h}}$:AT} \right]&
 \end{flalign*} 

 \begin{flalign*} 
 \frac{d}{dt}\left[ \text{  e$_{9}$:AT  } \right]&= + \text{k}_{\text{flow}}  ( \text{ [e$_{9}$:AT]$^{\text{up}}$}  - \left[ \text{e$_{9}$:AT} \right] ) \\
 &\hskip2.5ex   - \text{k}_{ \text{9:AT} }^{\text{on}}  \left[ \text{p}_{ \text{9} }^{\text{avail}} \right]  \left[ \text{e$_{9}$:AT} \right] + \text{k}_{ \text{9:AT} }^{\text{off}}  \left[ \text{e$_9^{\text{m}}$:AT} \right] \\
 &\hskip2.5ex   + \text{k}_{ \text{ \text{e$_{9}$:AT} }  \left[ \text{AT} \right]  \left[ \text{e$_{9}$} \right] - \text{k}_{9:AT} }^{\text{on}}  \left[ \text{p}_{ \text{9,*} }^{\text{avail}} \right]  \left[ \text{e$_{9}$:AT} \right] \\
 &\hskip2.5ex   + \text{k}_{ \text{9:AT} }^{\text{off}}  \left[ \text{e$_{9}^{\text{m,*}}$:AT} \right]&
 \end{flalign*} 

 \begin{flalign*} 
 \frac{d}{dt}\left[ \text{  e$_{10}$:AT  } \right]&= + \text{k}_{ \text{10:AT} }^{\text{off}}  \left[ \text{e$_{10}^{\text{m}}$:AT} \right] - \text{k}_{ \text{10:AT} }^{\text{on}}  \left[ \text{p}_{ \text{10} }^{\text{avail}} \right]  \left[ \text{e$_{10}$:AT} \right]\\
 &\hskip2.5ex   + \text{k}_{ \text{e$_{10}$:AT} }  \left[ \text{AT} \right]  \left[ \text{e$_{10}$} \right] + \text{k}_{\text{flow}}  ( \text{ [e$_{10}$:AT]$^{\text{up}}$}  - \left[ \text{e$_{10}$:AT} \right] )  &
 \end{flalign*} 

 \begin{flalign*} 
 \frac{d}{dt}\left[ \text{  e$_{\text{2}}$:AT  } \right]&= + \text{k}_{ \text{2:AT} }^{\text{off}}  \left[ \text{e$_2^{\text{m}}$:AT$^{\text{*}}$} \right] - \text{k}_{ \text{2:AT} }^{\text{on}}  \left[ \text{p}_{ \text{2,*} }^{\text{avail}} \right]  \left[ \text{e$_{\text{2}}$:AT$^{\text{*}}$} \right]\\
 &\hskip2.5ex   + \text{k}_{ \text{e$_{\text{2}}$:AT} }  \left[ \text{AT} \right]  \left[ \text{e$_{\text{2}}$} \right] + \text{k}_{\text{flow}}  ( \text{ [e$_{\text{2}}$:AT]$^{\text{up}}$}  - \left[ \text{e$_{\text{2}}$:AT} \right] )  &
 \end{flalign*} 

 \begin{flalign*} 
 \frac{d}{dt}\left[ \text{  e$_{9}^{\text{m,*}}$:AT  } \right]&= + \text{k}_{\text{\text{e$_9^{\text{m}}$:AT} }}  \left[ \text{AT} \right]  \left[\text{e$_{9}^{\text{m,*}}$}\right] - \text{k}_{9:AT}^{\text{off}}  \left[ \text{e$_{9}^{\text{m,*}}$:AT} \right]\\
 &\hskip2.5ex   + \text{k}_{ \text{9:AT} }^{\text{on}}  \left[ \text{p}_{ \text{9,*} }^{\text{avail}} \right]  \left[ \text{e$_{9}$:AT} \right]&
 \end{flalign*} 

 \begin{flalign*} 
 \frac{d}{dt}\left[ \text{  AT  } \right]&= - \text{k}_{ \text{e$_{9}$:AT} }  \left[ \text{AT} \right]  \left[ \text{e$_{9}$} \right] - \text{k}_{ \text{e$_9^{\text{m}}$:AT} }  \left[ \text{AT} \right]  \left[ \text{e$_9^{\text{m}}$} \right]\\
 &\hskip2.5ex   - \text{k}_{ \text{e$_9^{\text{m}}$:AT} }  \left[ \text{AT} \right]  \left[ \text{e$_{9}^{\text{m,*}}$} \right] - \text{k}_{ \text{e$_{10}$:AT} }  \left[ \text{AT} \right]  \left[ \text{e$_{10}$} \right] \\
 &\hskip2.5ex   - \text{k}_{ \text{e$_{10}^{\text{m}}$:AT} }  \left[ \text{AT} \right]  \left[ \text{e$_{10}^{\text{m}}$} \right] - \text{k}_{ \text{e$_{\text{2}}$:AT} }  \left[ \text{AT} \right]  \left[ \text{e$_{\text{2}}$} \right] \\
 &\hskip2.5ex   - \text{k}_{ \text{e$_2^{\text{m}}$:AT} }  \left[ \text{AT} \right]  \left[ \text{e$_2^{\text{m}}$} \right] \\
 &\hskip2.5ex   - \text{k}_{ \text{e$_{11}$:AT} }  \left[ \text{AT} \right]  \left[ \text{e$_{11}$} \right] - \text{k}_{ \text{e$_{11}$:AT} }  \left[ \text{AT} \right]  \left[ \text{e$_{11}$:AT} \right] \\
 &\hskip2.5ex   - \text{k}_{ \text{e$_{11}$:AT} }  \left[ \text{AT} \right]  \left[ \text{e$_{11}^{\text{m,*}}$} \right] - \text{k}_{ \text{e$_{11}$:AT} }  \left[ \text{AT} \right]  \left[ \text{e$_{11\text{h}}$} \right] \\
 &\hskip2.5ex   - \text{k}_{ \text{e$_{11}$:AT} }  \left[ \text{AT} \right]  \left[ \text{e$_{11\text{h}}^{\text{m}}$} \right] \\
 &\hskip2.5ex   + \text{k}_{\text{flow}}  ( \text{ [AT]$^{\text{up}}$}  - \left[ \text{AT} \right] ) \\
 &\hskip2.5ex   - \text{k}_{ \text{AT:H} }^{\text{+}}  \left[ \text{AT} \right]  \left[ \text{H} \right]  + \text{k}_{ \text{AT:H} }^{\text{-}}  \left[ \text{ATH} \right]&
 \end{flalign*} 

 \begin{flalign*} 
 \frac{d}{dt}\left[ \text{  H  } \right]&= - \text{k}_{ \text{AT:H} }^{\text{+}}  \left[ \text{AT} \right]  \left[ \text{H} \right] + \text{k}_{ \text{AT:H} }^{\text{-}}  \left[ \text{ATH} \right]\\
 &\hskip2.5ex   + \text{k}_{\text{flow}}  ( \text{ [H]$^{\text{up}}$}  - \left[ \text{H} \right] )  &
 \end{flalign*} 

 \begin{flalign*} 
 \frac{d}{dt}\left[ \text{  AT:H  } \right]&= + \text{k}_{ \text{AT:H} }^{\text{+}}  \left[ \text{AT} \right]  \left[ \text{H} \right] - \text{k}_{ \text{AT:H} }^{\text{-}}  \left[ \text{ATH} \right]\\
 &\hskip2.5ex   + \text{k}_{\text{flow}}  ( \text{ [AT:H]$^{\text{up}}$}  - \left[ \text{ATH} \right] ) \\
 &\hskip2.5ex   - \text{k}_{ \text{e$_{10}$:ATH} }  \left[ \text{ATH} \right]  \left[ \text{e$_{10}$} \right] - \text{k}_{ \text{e$_{10}^{\text{m}}$:ATH} }  \left[ \text{ATH} \right]  \left[ \text{e$_{10}^{\text{m}}$} \right] \\
 &\hskip2.5ex   - \text{k}_{ \text{e$_{\text{2}}$:ATH} }  \left[ \text{ATH} \right]  \left[ \text{e$_{\text{2}}$} \right] - \text{k}_{ \text{e$_2^{\text{m}}$:ATH} }  \left[ \text{ATH} \right]  \left[ \text{e$_2^{\text{m}}$} \right] \\
 &\hskip2.5ex   - \text{k}_{ \text{e$_{9}$:ATH} }  \left[ \text{ATH} \right]  \left[ \text{e$_{9}$} \right] - \text{k}_{ \text{e$_9^{\text{m}}$:ATH} }  \left[ \text{ATH} \right]  \left[ \text{e$_9^{\text{m}}$} \right] \\
 &\hskip2.5ex   - \text{k}_{ \text{e$_9^{\text{m}}$:ATH} }  \left[ \text{ATH} \right]  \left[ \text{e$_{9}^{\text{m,*}}$} \right] \\
 &\hskip2.5ex   - \text{k}_{ \text{e$_{11}$:ATH} }  \left[ \text{ATH} \right]  \left[ \text{e$_{11}$} \right] - \text{k}_{ \text{e$_{11}$:ATH} }  \left[ \text{ATH} \right]  \left[ \text{e$_{11}$:ATH} \right] \\
 &\hskip2.5ex   - \text{k}_{ \text{e$_{11}$:ATH} }  \left[ \text{ATH} \right]  \left[ \text{e$_{11}^{\text{m,*}}$} \right] - \text{k}_{ \text{e$_{11}$:ATH} }  \left[ \text{ATH} \right]  \left[ \text{e$_{11\text{h}}$} \right] \\
 &\hskip2.5ex   - \text{k}_{ \text{e$_{11}$:ATH} }  \left[ \text{ATH} \right]  \left[ \text{e$_{11\text{h}}^{\text{m}}$} \right]&
 \end{flalign*} 

 \begin{flalign*} 
 \frac{d}{dt}\left[ \text{  e$_{10}$:ATH  } \right]&= + \text{k}_{ \text{e$_{10}$:ATH} }  \left[ \text{ATH} \right]  \left[ \text{e$_{10}$} \right] + \text{k}_{\text{flow}}  ( \text{ [e$_{10}$:ATH]$^{\text{up}}$}  - \left[ \text{e$_{10}$:ATH} \right] ) \\
 &\hskip2.5ex   - \text{k}_{ \text{10:AT} }^{\text{on}}  \left[ \text{p}_{ \text{10} }^{\text{avail}} \right]  \left[ \text{e$_{10}$:ATH} \right] + \text{k}_{ \text{10:AT} }^{\text{off}}  \left[ \text{e$_{10}^{\text{m}}$:ATH} \right]&
 \end{flalign*} 

 \begin{flalign*} 
 \frac{d}{dt}\left[ \text{  e$_{10}^{\text{m}}$:ATH  } \right]&= + \text{k}_{ \text{e$_{10}^{\text{m}}$:ATH} }  \left[ \text{ATH} \right]  \left[ \text{e$_{10}^{\text{m}}$} \right]\\
 &\hskip2.5ex   + \text{k}_{ \text{10:AT} }^{\text{on}}  \left[ \text{p}_{ \text{10} }^{\text{avail}} \right]  \left[ \text{e$_{10}$:ATH} \right] - \text{k}_{ \text{10:AT} }^{\text{off}}  \left[ \text{e$_{10}^{\text{m}}$:ATH} \right]&
 \end{flalign*} 

 \begin{flalign*} 
 \frac{d}{dt}\left[ \text{  e$_{\text{2}}$:ATH  } \right]&= + \text{k}_{ \text{e$_{\text{2}}$:ATH} }  \left[ \text{ATH} \right]  \left[ \text{e$_{\text{2}}$} \right] + \text{k}_{\text{flow}}  ( \text{ [e$_{\text{2}}$:ATH]$^{\text{up}}$}  - \left[ \text{e$_{\text{2}}$:ATH} \right] ) \\
 &\hskip2.5ex   - \text{k}_{ \text{2:AT} }^{\text{on}}  \left[ \text{p}_{ \text{2,*} }^{\text{avail}} \right]  \left[ \text{e$_{\text{2}}$:ATH} \right] + \text{k}_{ \text{2:AT} }^{\text{off}}  \left[ \text{e$_2^{\text{m}}$:ATH} \right]&
 \end{flalign*} 

 \begin{flalign*} 
 \frac{d}{dt}\left[ \text{  e$_2^{\text{m}}$:ATH  } \right]&= + \text{k}_{ \text{e$_2^{\text{m}}$:ATH} }  \left[ \text{ATH} \right]  \left[ \text{e$_2^{\text{m}}$} \right]\\
 &\hskip2.5ex   + \text{k}_{ \text{2:AT} }^{\text{on}}  \left[ \text{p}_{ \text{2,*} }^{\text{avail}} \right]  \left[ \text{e$_{\text{2}}$:ATH} \right] - \text{k}_{ \text{2:AT} }^{\text{off}}  \left[ \text{e$_2^{\text{m}}$:ATH} \right]&
 \end{flalign*} 

 \begin{flalign*} 
 \frac{d}{dt}\left[ \text{  e$_{9}$:ATH  } \right]&= + \text{k}_{ \text{e$_{9}$:ATH} }  \left[ \text{ATH} \right]  \left[ \text{e$_{9}$} \right] + \text{k}_{\text{flow}}  ( \text{ [e$_{9}$:ATH]$^{\text{up}}$}  - \left[ \text{e$_{9}$:ATH} \right] ) \\
 &\hskip2.5ex   - \text{k}_{ \text{9:AT} }^{\text{on}}  \left[ \text{p}_{ \text{9} }^{\text{avail}} \right]  \left[ \text{e$_{9}$:ATH} \right] + \text{k}_{ \text{9:AT} }^{\text{off}}  \left[ \text{e$_9^{\text{m}}$:ATH} \right] \\
 &\hskip2.5ex   - \text{k}_{ \text{9:AT} }^{\text{on}}  \left[ \text{p}_{ \text{9,*} }^{\text{avail}} \right]  \left[ \text{e$_{9}$:ATH} \right] + \text{k}_{ \text{9:AT} }^{\text{off}}  \left[ \text{e$_{9}^{\text{m,*}}$:ATH} \right]&
 \end{flalign*} 

 \begin{flalign*} 
 \frac{d}{dt}\left[ \text{  e$_9^{\text{m}}$:ATH  } \right]&= + \text{k}_{ \text{e$_9^{\text{m}}$:ATH} }  \left[ \text{ATH} \right]  \left[ \text{e$_9^{\text{m}}$} \right]\\
 &\hskip2.5ex   + \text{k}_{ \text{9:AT} }^{\text{on}}  \left[ \text{p}_{ \text{9} }^{\text{avail}} \right]  \left[ \text{e$_{9}$:ATH} \right] - \text{k}_{ \text{9:AT} }^{\text{off}}  \left[ \text{e$_9^{\text{m}}$:ATH} \right]&
 \end{flalign*} 

 \begin{flalign*} 
 \frac{d}{dt}\left[ \text{  e$_{9}^{\text{m,*}}$:ATH  } \right]&= + \text{k}_{ \text{e$_9^{\text{m}}$:ATH} }  \left[ \text{ATH} \right]  \left[ \text{e$_{9}^{\text{m,*}}$} \right]\\
 &\hskip2.5ex   + \text{k}_{ \text{9:AT} }^{\text{on}}  \left[ \text{p}_{ \text{9,*} }^{\text{avail}} \right]  \left[ \text{e$_{9}$:ATH} \right] - \text{k}_{ \text{9:AT} }^{\text{off}}  \left[ \text{e$_{9}^{\text{m,*}}$:ATH} \right]&
 \end{flalign*} 

 \begin{flalign*} 
 \frac{d}{dt}\left[ \text{  e$_{11}$:ATH  } \right]&= + \text{k}_{ \text{e$_{11}$:ATH} }  \left[ \text{ATH} \right]  \left[ \text{e$_{11}$} \right] - \text{k}_{ \text{e$_{11}$:ATH} }  \left[ \text{ATH} \right]  \left[ \text{e$_{11}$:ATH} \right]\\
 &\hskip2.5ex   + \text{k}_{\text{flow}}  ( \text{ [e$_{11}$:ATH]$^{\text{up}}$}  - \left[ \text{e$_{11}$:ATH} \right] ) \\
 &\hskip2.5ex   - \text{k$_{\text{11:AT}}^{\text{on}}$}  \left[ \text{p}_{ \text{11,*} }^{\text{avail}} \right]  \left[ \text{e$_{11}$:ATH} \right] + \text{k$_{\text{11:AT}}^{\text{off}}$}  \left[ \text{e$_{11}^{\text{m,*}}$:ATH} \right]&
 \end{flalign*} 

 \begin{flalign*} 
 \frac{d}{dt}\left[ \text{  ATH:e$_{11}$:ATH  } \right]&= + \text{k}_{ \text{e$_{11}$:ATH} }  \left[ \text{ATH} \right]  \left[ \text{e$_{11}$:ATH} \right]\\
 &\hskip2.5ex   + \text{k}_{\text{flow}}  ( \text{ [ATH:e$_{11}$:ATH]$^{\text{up}}$}  - \left[ \text{ATH:e$_{11}$:ATH} \right] )  &
 \end{flalign*} 

 \begin{flalign*} 
 \frac{d}{dt}\left[ \text{  e$_{11}^{\text{m,*}}$:ATH  } \right]&= + \text{k}_{ \text{e$_{11}$:ATH} }  \left[ \text{ATH} \right]  \left[ \text{e$_{11}^{\text{m,*}}$} \right]\\
 &\hskip2.5ex   + \text{k$_{\text{11:AT}}^{\text{on}}$}  \left[ \text{p}_{ \text{11,*} }^{\text{avail}} \right]  \left[ \text{e$_{11}$:ATH} \right] - \text{k$_{\text{11:AT}}^{\text{off}}$}  \left[ \text{e$_{11}^{\text{m,*}}$:ATH} \right]&
 \end{flalign*} 

 \begin{flalign*} 
 \frac{d}{dt}\left[ \text{  e$_{11\text{h}}$:ATH  } \right]&= + \text{k}_{ \text{e$_{11}$:ATH} }  \left[ \text{ATH} \right]  \left[ \text{e$_{11\text{h}}$} \right]\\
 &\hskip2.5ex   + \text{k}_{\text{flow}}  ( \text{ [e$_{11\text{h}}$:ATH]$^{\text{up}}$}  - \left[ \text{e$_{11\text{h}}$:ATH} \right] ) \\
 &\hskip2.5ex   + \text{k$_{\text{11:AT}}^{\text{off}}$}  \left[ \text{e$_{11\text{h}}^{\text{m}}$:ATH} \right] \\
 &\hskip2.5ex   - \text{k$_{\text{11:AT}}^{\text{on}}$}  \left[ \text{p}_{ \text{11} }^{\text{avail}} \right]  \left[ \text{e$_{11\text{h}}$:ATH} \right]&
 \end{flalign*} 

 \begin{flalign*} 
 \frac{d}{dt}\left[ \text{  e$_{11\text{h}}^{\text{m}}$:ATH  } \right]&= + \text{k}_{ \text{ \text{e$_{11}$:ATH} }}^{\text{off}}  \left[\text{ATH}\right]  \left[\text{e$_{11\text{h}}$}^{\text{m}}\right] - \text{k}_{11at}   \left[ \text{e$_{11\text{h}}^{\text{m}}$:ATH} \right]\\
 &\hskip2.5ex   + \text{k$_{\text{11:AT}}^{\text{on}}$}  \left[ \text{p}_{ \text{11} }^{\text{avail}} \right]  \left[ \text{e$_{11\text{h}}$:ATH} \right]&
 \end{flalign*} 

 \begin{flalign*} 
 \frac{d}{dt}\left[ \text{  AT$^{\text{ec}}$  } \right]&= + \text{k}_{\text{flow}}  ( \text{ [AT$^{\text{ec}}$]$^{\text{up}}$}  - \left[ \text{AT$^{\text{ec}}$} \right] ) + \text{k}_\text{diff}  ( \left[ \text{AT} \right] - \left[ \text{AT$^{\text{ec}}$} \right] ) \\
 &\hskip2.5ex   - \text{k}_{ \text{e$_{\text{2}}$:AT} }  \left[ \text{AT$^{\text{ec}}$} \right]  \left[ \text{e$_{\text{2}}$$^{\text{ec}}$} \right] - \text{k}_{ \text{e$_{10}$:AT} }  \left[ \text{AT$^{\text{ec}}$} \right]  \left[ \text{e$_{10}$$^{\text{ec}}$} \right] \\
 &\hskip2.5ex   - \text{k}_{ \text{e$_{9}$:AT} }  \left[ \text{AT$^{\text{ec}}$} \right]  \left[ \text{e$_{9}$$^{\text{ec}}$} \right]&
 \end{flalign*} 

 \begin{flalign*} 
 \frac{d}{dt}\left[ \text{  e$_{\text{2}}$out  } \right]&= + \text{k}_{\text{flow}}  \left[ \text{e$_{\text{2}}$} \right]&
 \end{flalign*} 

 \begin{flalign*} 
 \frac{d}{dt}\left[ \text{  TATout  } \right]&= + \text{k}_{\text{flow}}  ( \left[ \text{e$_{\text{2}}$:AT} \right] + \left[ \text{e$_{\text{2}}$:ATH} \right] )  &
 \end{flalign*} 

 \begin{flalign*} 
 \frac{d}{dt}\left[ \text{  e$_{\text{2}}$$^{\text{ec}}$out  } \right]&= + \text{k}_{\text{flow}}  \left[ \text{e$_{\text{2}}$$^{\text{ec}}$} \right]&
 \end{flalign*} 

 \begin{flalign*} 
 \frac{d}{dt}\left[ \text{  ps:TFPI:e$_{\text{5h}}$  } \right]&= \text{k}_{\text{flow}}  ( \text{ [ps:TFPI:e$_{\text{5h}}$]$^{\text{up}}$}  - \left[ \text{ps:TFPI:e$_{\text{5h}}$} \right] ) \\
 &\hskip2.5ex - \text{k}_{ \text{ps} }^{\text{on}}  \left[ \text{ps:TFPI:e$_{\text{5h}}$} \right]  \left[ \text{p}_{ \text{ps} }^{\text{avail}} \right] + \text{k}_{ \text{ps} }^{\text{off}}  \left[ \text{ps$^{\text{m}}$:TFPI:e$_{\text{5h}}$} \right] \\
 &\hskip2.5ex - \text{k}_{\text{5}}^{\text{on,psc}}  \left[ \text{ps:TFPI:e$_{\text{5h}}$} \right]  \left[ \text{p}_{ \text{5} }^{\text{avail}}  \right]+ \text{k}_{\text{5}}^{\text{off,psc}}  \left[ \text{ps:TFPI:e$_{5\text{h}}^{\text{m}}$} \right] \\
 &\hskip2.5ex - \text{k}_{ \text{(ps:TFPI:e$_{\text{5h}}$):e$_{10}$} }^{+}  \left[ \text{ps:TFPI:e$_{\text{5h}}$} \right]  \left[ \text{e$_{10}$} \right] + \text{k}_{ \text{(ps:TFPI:e$_{\text{5h}}$):e$_{10}$} }^{-}  \left[ \text{(ps:TFPI:e$_{\text{5h}}$):e$_{10}$} \right] \\
 &\hskip2.5ex - \text{k}_{ \text{(ps:TFPI:e$_{\text{5h}}$):e$_{10}^{\text{m}}$} }^{+}  \left[ \text{ps:TFPI:e$_{\text{5h}}$} \right]  \left[ \text{e$_{10}^{\text{m}}$} \right] + \text{k}_{ \text{(ps:TFPI:e$_{\text{5h}}$):e$_{10}^{\text{m}}$} }^{-}  \left[ \text{(ps:TFPI:e$_{\text{5h}}$):e$_{10}^{\text{m}}$} \right] &
 \end{flalign*} 

 \begin{flalign*} 
 \frac{d}{dt}\left[ \text{  ps$^{\text{m}}$:TFPI:e$_{\text{5h}}$  } \right]&= \text{k}_{ \text{ps} }^{\text{on}}  \left[ \text{ps:TFPI:e$_{\text{5h}}$} \right]  \left[ \text{p}_{ \text{ps} }^{\text{avail}} \right] - \text{k}_{ \text{ps} }^{\text{off}}  \left[ \text{ps$^{\text{m}}$:TFPI:e$_{\text{5h}}$} \right]\\
 &\hskip2.5ex - \text{k}_{\text{5}}^{\text{on,psc}}  \left[ \text{ps$^{\text{m}}$:TFPI:e$_{\text{5h}}$} \right]  \left[ \text{p}_{ \text{5} }^{\text{avail}}  \right]+ \text{k}_{\text{5}}^{\text{off,psc}}  \left[ \text{ps$^{\text{m}}$:TFPI:e$_{5\text{h}}^{\text{m}}$} \right] \\
 &\hskip2.5ex - \text{k}_{ \text{(ps$^{\text{m}}$:TFPI:e$_{\text{5h}}$):e$_{10}$} }^{+}  \left[ \text{ps$^{\text{m}}$:TFPI:e$_{\text{5h}}$} \right]  \left[ \text{e$_{10}$} \right] + \text{k}_{ \text{(ps$^{\text{m}}$:TFPI:e$_{\text{5h}}$):e$_{10}$} }^{-}  \left[ \text{(ps$^{\text{m}}$:TFPI:e$_{\text{5h}}$):e$_{10}$} \right] \\
 &\hskip2.5ex - \text{k}_{ \text{ps$^{\text{m}}$:TFPI:e$_{\text{5h}}$:e$_{10}^{\text{m}}$} }^{+}  \left[ \text{ps$^{\text{m}}$:TFPI:e$_{\text{5h}}$} \right]  \left[ \text{e$_{10}^{\text{m}}$} \right] + \text{k}_{ \text{ps$^{\text{m}}$:TFPI:e$_{\text{5h}}$:e$_{10}^{\text{m}}$} }^{-}  \left[ \text{(ps$^{\text{m}}$:TFPI:e$_{\text{5h}}$):e$_{10}^{\text{m}}$} \right] &
 \end{flalign*} 

 \begin{flalign*} 
 \frac{d}{dt}\left[ \text{  ps:TFPI:e$_{5\text{h}}^{\text{m}}$  } \right]&= - \text{k}_{ \text{ps} }^{\text{on}}  \left[ \text{ps:TFPI:e$_{5\text{h}}^{\text{m}}$} \right]  \left[ \text{p}_{ \text{ps} }^{\text{avail}} \right] + \text{k}_{ \text{ps} }^{\text{off}}  \left[ \text{ps$^{\text{m}}$:TFPI:e$_{5\text{h}}^{\text{m}}$} \right]\\
 &\hskip2.5ex + \text{k}_{\text{5}}^{\text{on,psc}}  \left[ \text{ps:TFPI:e$_{\text{5h}}$} \right]  \left[ \text{p}_{ \text{5} }^{\text{avail}}  \right]- \text{k}_{\text{5}}^{\text{off,psc}}  \left[ \text{ps:TFPI:e$_{5\text{h}}^{\text{m}}$} \right] \\
 &\hskip2.5ex - \text{k}_{ \text{(ps:TFPI:e$_{5\text{h}}^{\text{m}}$):e$_{10}$} }^{+}  \left[ \text{ps:TFPI:e$_{5\text{h}}^{\text{m}}$} \right]  \left[ \text{e$_{10}$} \right] + \text{k}_{ \text{(ps:TFPI:e$_{5\text{h}}^{\text{m}}$):e$_{10}$} }^{-}  \left[ \text{(ps:TFPI:e$_{5\text{h}}^{\text{m}}$):e$_{10}$} \right] \\
 &\hskip2.5ex - \text{k}_{ \text{(ps:TFPI:e$_{5\text{h}}^{\text{m}}$):e$_{10}^{\text{m}}$} }^{+}  \left[ \text{ps:TFPI:e$_{5\text{h}}^{\text{m}}$} \right]  \left[ \text{e$_{10}^{\text{m}}$} \right] + \text{k}_{ \text{(ps:TFPI:e$_{5\text{h}}^{\text{m}}$):e$_{10}^{\text{m}}$} }^{-}  \left[ \text{(ps:TFPI:e$_{5\text{h}}^{\text{m}}$):e$_{10}^{\text{m}}$} \right] &
 \end{flalign*} 

 \begin{flalign*} 
 \frac{d}{dt}\left[ \text{  ps$^{\text{m}}$:TFPI:e$_{5\text{h}}^{\text{m}}$  } \right]&= \text{k}_{ \text{ps} }^{\text{on}}  \left[ \text{ps:TFPI:e$_{5\text{h}}^{\text{m}}$} \right]  \left[ \text{p}_{ \text{ps} }^{\text{avail}} \right] - \text{k}_{ \text{ps} }^{\text{off}}  \left[ \text{ps$^{\text{m}}$:TFPI:e$_{5\text{h}}^{\text{m}}$} \right]\\
 &\hskip2.5ex + \text{k}_{\text{5}}^{\text{on,psc}}  \left[ \text{ps$^{\text{m}}$:TFPI:e$_{\text{5h}}$} \right]  \left[ \text{p}_{ \text{5} }^{\text{avail}}  \right]- \text{k}_{\text{5}}^{\text{off,psc}}  \left[ \text{ps$^{\text{m}}$:TFPI:e$_{5\text{h}}^{\text{m}}$} \right] \\
 &\hskip2.5ex - \text{k}_{ \text{(ps$^{\text{m}}$:TFPI:e$_{5\text{h}}^{\text{m}}$):e$_{10}$} }^{+}  \left[ \text{ps$^{\text{m}}$:TFPI:e$_{5\text{h}}^{\text{m}}$} \right]  \left[ \text{e$_{10}$} \right] + \text{k}_{ \text{(ps$^{\text{m}}$:TFPI:e$_{5\text{h}}^{\text{m}}$):e$_{10}$} }^{-}  \left[ \text{(ps$^{\text{m}}$:TFPI:e$_{5\text{h}}^{\text{m}}$):e$_{10}$} \right] \\
 &\hskip2.5ex - \text{k}_{ \text{(ps$^{\text{m}}$:TFPI:e$_{5\text{h}}^{\text{m}}$):e$_{10}^{\text{m}}$} }^{+}  \left[ \text{ps$^{\text{m}}$:TFPI:e$_{5\text{h}}^{\text{m}}$} \right]  \left[ \text{e$_{10}^{\text{m}}$} \right] + \text{k}_{ \text{(ps$^{\text{m}}$:TFPI:e$_{5\text{h}}^{\text{m}}$):e$_{10}^{\text{m}}$} }^{-}  \left[ \text{(ps$^{\text{m}}$:TFPI:e$_{5\text{h}}^{\text{m}}$):e$_{10}^{\text{m}}$} \right] &
 \end{flalign*} 

 \begin{flalign*} 
 \frac{d}{dt}\left[ \text{  (ps:TFPI:e$_{\text{5h}}$):e$_{10}$  } \right]&= \text{k}_{\text{flow}}  ( \text{ [(ps:TFPI:e$_{\text{5h}}$):e$_{10}$]$^{\text{up}}$}  - \left[ \text{(ps:TFPI:e$_{\text{5h}}$):e$_{10}$} \right] ) \\
 &\hskip2.5ex - \text{k}_{ \text{ps} }^{\text{on}}  \left[ \text{(ps:TFPI:e$_{\text{5h}}$):e$_{10}$} \right]  \left[ \text{p}_{ \text{ps} }^{\text{avail}} \right] + \text{k}_{ \text{ps} }^{\text{off}}  \left[ \text{(ps$^{\text{m}}$:TFPI:e$_{\text{5h}}$):e$_{10}$} \right] \\
 &\hskip2.5ex - \text{k}_{\text{5}}^{\text{on,psc}}  \left[ \text{(ps:TFPI:e$_{\text{5h}}$):e$_{10}$} \right]  \left[ \text{p}_{ \text{5} }^{\text{avail}}  \right]+ \text{k}_{\text{5}}^{\text{off,psc}}  \left[ \text{(ps:TFPI:e$_{5\text{h}}^{\text{m}}$):e$_{10}$} \right] \\
 &\hskip2.5ex - \text{k}_{ \text{10} }^{\text{on}}  \left[ \text{(ps:TFPI:e$_{\text{5h}}$):e$_{10}$} \right]  \left[ \text{p}_{ \text{10} }^{\text{avail}} \right] + \text{k}_{ \text{10} }^{\text{off}}  \left[ \text{(ps:TFPI:e$_{\text{5h}}$):e$_{10}^{\text{m}}$} \right] \\
 &\hskip2.5ex + \text{k}_{ \text{(ps:TFPI:e$_{\text{5h}}$):e$_{10}$} }^{+}  \left[ \text{ps:TFPI:e$_{\text{5h}}$} \right]  \left[ \text{e$_{10}$} \right] - \text{k}_{ \text{(ps:TFPI:e$_{\text{5h}}$):e$_{10}$} }^{-}  \left[ \text{(ps:TFPI:e$_{\text{5h}}$):e$_{10}$} \right] \\
 &\hskip2.5ex - \text{k}_{ \text{((ps:TFPI:e$_{\text{5h}}$):e$_{10}$):e$_7^{\text{m}}$} }^{+}  \left[ \text{(ps:TFPI:e$_{\text{5h}}$):e$_{10}$} \right]  \left[ \text{e$_7^{\text{m}}$} \right] + \text{k}_{ \text{((ps:TFPI:e$_{\text{5h}}$):e$_{10}$):e$_7^{\text{m}}$} }^{-}  \left[ \text{((ps:TFPI:e$_{\text{5h}}$):e$_{10}$):e$_7^{\text{m}}$} \right]&
 \end{flalign*} 

 \begin{flalign*} 
 \frac{d}{dt}\left[ \text{  (ps$^{\text{m}}$:TFPI:e$_{\text{5h}}$):e$_{10}$  } \right]&=  \text{k}_{ \text{ps} }^{\text{on}}  \left[ \text{(ps:TFPI:e$_{\text{5h}}$):e$_{10}$} \right]  \left[ \text{p}_{ \text{ps} }^{\text{avail}} \right] - \text{k}_{ \text{ps} }^{\text{off}}  \left[ \text{(ps$^{\text{m}}$:TFPI:e$_{\text{5h}}$):e$_{10}$} \right]\\
 &\hskip2.5ex - \text{k}_{\text{5}}^{\text{on,psc}}  \left[ \text{(ps$^{\text{m}}$:TFPI:e$_{\text{5h}}$):e$_{10}$} \right]  \left[ \text{p}_{ \text{5} }^{\text{avail}}  \right]+ \text{k}_{\text{5}}^{\text{off,psc}}  \left[ \text{(ps$^{\text{m}}$:TFPI:e$_{5\text{h}}^{\text{m}}$):e$_{10}$} \right] \\
 &\hskip2.5ex - \text{k}_{ \text{10} }^{\text{on}}  \left[ \text{(ps$^{\text{m}}$:TFPI:e$_{\text{5h}}$):e$_{10}$} \right]  \left[ \text{p}_{ \text{10} }^{\text{avail}} \right] + \text{k}_{ \text{10} }^{\text{off}}  \left[ \text{(ps$^{\text{m}}$:TFPI:e$_{\text{5h}}$):e$_{10}^{\text{m}}$} \right] \\
 &\hskip2.5ex + \text{k}_{ \text{(ps$^{\text{m}}$:TFPI:e$_{\text{5h}}$):e$_{10}$} }^{+}  \left[ \text{ps$^{\text{m}}$:TFPI:e$_{\text{5h}}$} \right]  \left[ \text{e$_{10}$} \right] - \text{k}_{ \text{(ps$^{\text{m}}$:TFPI:e$_{\text{5h}}$):e$_{10}$} }^{-}  \left[ \text{(ps$^{\text{m}}$:TFPI:e$_{\text{5h}}$):e$_{10}$} \right]&
 \end{flalign*} 

 \begin{flalign*} 
 \frac{d}{dt}\left[ \text{  (ps:TFPI:e$_{5\text{h}}^{\text{m}}$):e$_{10}$  } \right]&= - \text{k}_{ \text{ps} }^{\text{on}}  \left[ \text{(ps:TFPI:e$_{5\text{h}}^{\text{m}}$):e$_{10}$} \right]  \left[ \text{p}_{ \text{ps} }^{\text{avail}} \right] + \text{k}_{ \text{ps} }^{\text{off}}  \left[ \text{(ps$^{\text{m}}$:TFPI:e$_{5\text{h}}^{\text{m}}$):e$_{10}$} \right]\\
 &\hskip2.5ex + \text{k}_{\text{5}}^{\text{on,psc}}  \left[ \text{(ps:TFPI:e$_{\text{5h}}$):e$_{10}$} \right]  \left[ \text{p}_{ \text{5} }^{\text{avail}}  \right]- \text{k}_{\text{5}}^{\text{off,psc}}  \left[ \text{(ps:TFPI:e$_{5\text{h}}^{\text{m}}$):e$_{10}$} \right] \\
 &\hskip2.5ex - \text{k}_{ \text{10} }^{\text{on}}  \left[ \text{(ps:TFPI:e$_{5\text{h}}^{\text{m}}$):e$_{10}$} \right]  \left[ \text{p}_{ \text{10} }^{\text{avail}} \right] + \text{k}_{ \text{10} }^{\text{off}}  \left[ \text{(ps:TFPI:e$_{5\text{h}}^{\text{m}}$):e$_{10}^{\text{m}}$} \right] \\
 &\hskip2.5ex + \text{k}_{ \text{(ps:TFPI:e$_{5\text{h}}^{\text{m}}$):e$_{10}$} }^{+}  \left[ \text{ps:TFPI:e$_{5\text{h}}^{\text{m}}$} \right]  \left[ \text{e$_{10}$} \right] - \text{k}_{ \text{(ps:TFPI:e$_{5\text{h}}^{\text{m}}$):e$_{10}$} }^{-}  \left[ \text{(ps:TFPI:e$_{5\text{h}}^{\text{m}}$):e$_{10}$} \right]&
 \end{flalign*} 

 \begin{flalign*} 
 \frac{d}{dt}\left[ \text{  (ps:TFPI:e$_{\text{5h}}$):e$_{10}^{\text{m}}$  } \right]&=  - \text{k}_{ \text{ps} }^{\text{on}}  \left[ \text{(ps:TFPI:e$_{\text{5h}}$):e$_{10}^{\text{m}}$} \right]  \left[ \text{p}_{ \text{ps} }^{\text{avail}} \right] + \text{k}_{ \text{ps} }^{\text{off}}  \left[ \text{(ps$^{\text{m}}$:TFPI:e$_{\text{5h}}$):e$_{10}^{\text{m}}$} \right]\\
 &\hskip2.5ex - \text{k}_{\text{5}}^{\text{on,psc}}  \left[ \text{(ps:TFPI:e$_{\text{5h}}$):e$_{10}^{\text{m}}$} \right]  \left[ \text{p}_{ \text{5} }^{\text{avail}}  \right]+ \text{k}_{\text{5}}^{\text{off,psc}}  \left[ \text{(ps:TFPI:e$_{5\text{h}}^{\text{m}}$):e$_{10}^{\text{m}}$} \right] \\
 &\hskip2.5ex + \text{k}_{ \text{10} }^{\text{on}}  \left[ \text{(ps:TFPI:e$_{\text{5h}}$):e$_{10}$} \right]  \left[ \text{p}_{ \text{10} }^{\text{avail}} \right] - \text{k}_{ \text{10} }^{\text{off}}  \left[ \text{(ps:TFPI:e$_{\text{5h}}$):e$_{10}^{\text{m}}$} \right] \\
 &\hskip2.5ex + \text{k}_{ \text{(ps:TFPI:e$_{\text{5h}}$):e$_{10}^{\text{m}}$} }^{+}  \left[ \text{ps:TFPI:e$_{\text{5h}}$} \right]  \left[ \text{e$_{10}^{\text{m}}$} \right] - \text{k}_{ \text{(ps:TFPI:e$_{\text{5h}}$):e$_{10}^{\text{m}}$} }^{-}  \left[ \text{(ps:TFPI:e$_{\text{5h}}$):e$_{10}^{\text{m}}$} \right] &
 \end{flalign*} 

 \begin{flalign*} 
 \frac{d}{dt}\left[ \text{  ps$^{\text{m}}$:TFPI:e$_{5\text{h}}^{\text{m}}$:e$_{10}$  } \right]&=  \text{k}_{ \text{ps} }^{\text{on}}  \left[ \text{(ps:TFPI:e$_{5\text{h}}^{\text{m}}$):e$_{10}$} \right]  \left[ \text{p}_{ \text{ps} }^{\text{avail}} \right] - \text{k}_{ \text{ps} }^{\text{off}}  \left[ \text{(ps$^{\text{m}}$:TFPI:e$_{5\text{h}}^{\text{m}}$):e$_{10}$} \right]\\
 &\hskip2.5ex + \text{k}_{\text{5}}^{\text{on,psc}}  \left[ \text{(ps$^{\text{m}}$:TFPI:e$_{\text{5h}}$):e$_{10}$} \right]  \left[ \text{p}_{ \text{5} }^{\text{avail}}  \right]- \text{k}_{\text{5}}^{\text{off,psc}}  \left[ \text{(ps$^{\text{m}}$:TFPI:e$_{5\text{h}}^{\text{m}}$):e$_{10}$} \right] \\
 &\hskip2.5ex - \text{k}_{ \text{10} }^{\text{on}}  \left[ \text{(ps$^{\text{m}}$:TFPI:e$_{5\text{h}}^{\text{m}}$):e$_{10}$} \right]  \left[ \text{p}_{ \text{10} }^{\text{avail}} \right] + \text{k}_{ \text{10} }^{\text{off}}  \left[ \text{(ps$^{\text{m}}$:TFPI:e$_{5\text{h}}^{\text{m}}$):e$_{10}^{\text{m}}$} \right] \\
 &\hskip2.5ex + \text{k}_{ \text{(ps$^{\text{m}}$:TFPI:e$_{5\text{h}}^{\text{m}}$):e$_{10}$} }^{+}  \left[ \text{ps$^{\text{m}}$:TFPI:e$_{5\text{h}}^{\text{m}}$} \right]  \left[ \text{e$_{10}$} \right] - \text{k}_{ \text{(ps$^{\text{m}}$:TFPI:e$_{5\text{h}}^{\text{m}}$):e$_{10}$} }^{-}  \left[ \text{(ps$^{\text{m}}$:TFPI:e$_{5\text{h}}^{\text{m}}$):e$_{10}$} \right]&
 \end{flalign*} 

 \begin{flalign*} 
 \frac{d}{dt}\left[ \text{  ps$^{\text{m}}$:TFPI:e$_{\text{5h}}$:e$_{10}^{\text{m}}$  } \right]&=  + \text{k}_{ \text{ps} }^{\text{on}}  \left[ \text{(ps:TFPI:e$_{\text{5h}}$):e$_{10}^{\text{m}}$} \right]  \left[ \text{p}_{ \text{ps} }^{\text{avail}} \right] - \text{k}_{ \text{ps} }^{\text{off}}  \left[ \text{(ps$^{\text{m}}$:TFPI:e$_{\text{5h}}$):e$_{10}^{\text{m}}$} \right]\\
 &\hskip2.5ex - \text{k}_{\text{5}}^{\text{on,psc}}  \left[ \text{(ps$^{\text{m}}$:TFPI:e$_{\text{5h}}$):e$_{10}^{\text{m}}$} \right]  \left[ \text{p}_{ \text{5} }^{\text{avail}}  \right]+ \text{k}_{\text{5}}^{\text{off,psc}}  \left[ \text{(ps$^{\text{m}}$:TFPI:e$_{5\text{h}}^{\text{m}}$):e$_{10}^{\text{m}}$} \right] \\
 &\hskip2.5ex + \text{k}_{ \text{10} }^{\text{on}}  \left[ \text{(ps$^{\text{m}}$:TFPI:e$_{\text{5h}}$):e$_{10}$} \right]  \left[ \text{p}_{ \text{10} }^{\text{avail}} \right] - \text{k}_{ \text{10} }^{\text{off}}  \left[ \text{(ps$^{\text{m}}$:TFPI:e$_{\text{5h}}$):e$_{10}^{\text{m}}$} \right] \\
 &\hskip2.5ex + \text{k}_{ \text{ps$^{\text{m}}$:TFPI:e$_{\text{5h}}$:e$_{10}^{\text{m}}$} }^{+}  \left[ \text{ps$^{\text{m}}$:TFPI:e$_{\text{5h}}$} \right]  \left[ \text{e$_{10}^{\text{m}}$} \right] - \text{k}_{ \text{ps$^{\text{m}}$:TFPI:e$_{\text{5h}}$:e$_{10}^{\text{m}}$} }^{-}  \left[ \text{(ps$^{\text{m}}$:TFPI:e$_{\text{5h}}$):e$_{10}^{\text{m}}$} \right] &
 \end{flalign*} 

 \begin{flalign*} 
 \frac{d}{dt}\left[ \text{  (ps:TFPI:e$_{5\text{h}}^{\text{m}}$):e$_{10}^{\text{m}}$  } \right]&=  - \text{k}_{ \text{ps} }^{\text{on}}  \left[ \text{(ps:TFPI:e$_{5\text{h}}^{\text{m}}$):e$_{10}^{\text{m}}$} \right]  \left[ \text{p}_{ \text{ps} }^{\text{avail}} \right] + \text{k}_{ \text{ps} }^{\text{off}}  \left[ \text{(ps$^{\text{m}}$:TFPI:e$_{5\text{h}}^{\text{m}}$):e$_{10}^{\text{m}}$} \right]\\
 &\hskip2.5ex + \text{k}_{\text{5}}^{\text{on,psc}}  \left[ \text{(ps:TFPI:e$_{\text{5h}}$):e$_{10}^{\text{m}}$} \right]  \left[ \text{p}_{ \text{5} }^{\text{avail}}  \right]- \text{k}_{\text{5}}^{\text{off,psc}}  \left[ \text{(ps:TFPI:e$_{5\text{h}}^{\text{m}}$):e$_{10}^{\text{m}}$} \right] \\
 &\hskip2.5ex + \text{k}_{ \text{10} }^{\text{on}}  \left[ \text{(ps:TFPI:e$_{5\text{h}}^{\text{m}}$):e$_{10}$} \right]  \left[ \text{p}_{ \text{10} }^{\text{avail}} \right] - \text{k}_{ \text{10} }^{\text{off}}  \left[ \text{(ps:TFPI:e$_{5\text{h}}^{\text{m}}$):e$_{10}^{\text{m}}$} \right] \\
 &\hskip2.5ex + \text{k}_{ \text{(ps:TFPI:e$_{5\text{h}}^{\text{m}}$):e$_{10}^{\text{m}}$} }^{+}  \left[ \text{ps:TFPI:e$_{5\text{h}}^{\text{m}}$} \right]  \left[ \text{e$_{10}^{\text{m}}$} \right] - \text{k}_{ \text{(ps:TFPI:e$_{5\text{h}}^{\text{m}}$):e$_{10}^{\text{m}}$} }^{-}  \left[ \text{(ps:TFPI:e$_{5\text{h}}^{\text{m}}$):e$_{10}^{\text{m}}$} \right] &
 \end{flalign*} 

 \begin{flalign*} 
 \frac{d}{dt}\left[ \text{  ps$^{\text{m}}$:TFPI:e$_{5\text{h}}^{\text{m}}$:e$_{10}^{\text{m}}$  } \right]&=  \text{k}_{ \text{ps} }^{\text{on}}  \left[ \text{(ps:TFPI:e$_{5\text{h}}^{\text{m}}$):e$_{10}^{\text{m}}$} \right]  \left[ \text{p}_{ \text{ps} }^{\text{avail}} \right] - \text{k}_{ \text{ps} }^{\text{off}}  \left[ \text{(ps$^{\text{m}}$:TFPI:e$_{5\text{h}}^{\text{m}}$):e$_{10}^{\text{m}}$} \right]\\
 &\hskip2.5ex + \text{k}_{\text{5}}^{\text{on,psc}}  \left[ \text{(ps$^{\text{m}}$:TFPI:e$_{\text{5h}}$):e$_{10}^{\text{m}}$} \right]  \left[ \text{p}_{ \text{5} }^{\text{avail}}  \right]- \text{k}_{\text{5}}^{\text{off,psc}}  \left[ \text{(ps$^{\text{m}}$:TFPI:e$_{5\text{h}}^{\text{m}}$):e$_{10}^{\text{m}}$} \right] \\
 &\hskip2.5ex + \text{k}_{ \text{10} }^{\text{on}}  \left[ \text{(ps$^{\text{m}}$:TFPI:e$_{5\text{h}}^{\text{m}}$):e$_{10}$} \right]  \left[ \text{p}_{ \text{10} }^{\text{avail}} \right] - \text{k}_{ \text{10} }^{\text{off}}  \left[ \text{(ps$^{\text{m}}$:TFPI:e$_{5\text{h}}^{\text{m}}$):e$_{10}^{\text{m}}$} \right] \\
 &\hskip2.5ex + \text{k}_{ \text{(ps$^{\text{m}}$:TFPI:e$_{5\text{h}}^{\text{m}}$):e$_{10}^{\text{m}}$} }^{+}  \left[ \text{ps$^{\text{m}}$:TFPI:e$_{5\text{h}}^{\text{m}}$} \right]  \left[ \text{e$_{10}^{\text{m}}$} \right] - \text{k}_{ \text{(ps$^{\text{m}}$:TFPI:e$_{5\text{h}}^{\text{m}}$):e$_{10}^{\text{m}}$} }^{-}  \left[ \text{(ps$^{\text{m}}$:TFPI:e$_{5\text{h}}^{\text{m}}$):e$_{10}^{\text{m}}$} \right] &
 \end{flalign*} 

 \begin{flalign*} 
 \frac{d}{dt}\left[ \text{  ((ps:TFPI:e$_{\text{5h}}$):e$_{10}$):e$_7^{\text{m}}$  } \right]&= + \text{k}_{ \text{((ps:TFPI:e$_{\text{5h}}$):e$_{10}$):e$_7^{\text{m}}$} }^{+}  \left[ \text{(ps:TFPI:e$_{\text{5h}}$):e$_{10}$} \right]  \left[ \text{e$_7^{\text{m}}$} \right] - \text{k}_{ \text{((ps:TFPI:e$_{\text{5h}}$):e$_{10}$):e$_7^{\text{m}}$} }^{-}  \left[ \text{((ps:TFPI:e$_{\text{5h}}$):e$_{10}$):e$_7^{\text{m}}$} \right]\\
 &\hskip2.5ex - \left[ \text{((ps:TFPI:e$_{\text{5h}}$):e$_{10}$):e$_7^{\text{m}}$} \right]  \frac{d}{dt} \left( [\text{PL}^{\text{as}}] \right) /\text{(P$_\text{{max}}$ - [PL$^{\text{as}}$])} &
 \end{flalign*} 

 \begin{flalign*} 
 \frac{d}{dt}\left[ \text{  ps\_fr  } \right]&= \text{k}_{\text{flow}}  ( \text{ [ps\_fr]$^{\text{up}}$}  - \left[ \text{ps\_fr} \right] ) \\
 &\hskip2.5ex - \text{k}_{ \text{ps} }^{\text{on}}  \left[ \text{ps\_fr} \right]  \left[ \text{p}_{ \text{ps} }^{\text{avail}} \right] + \text{k}_{ \text{ps} }^{\text{off}}  \left[ \text{ps$^{\text{m}}$\_fr} \right] \\
 &\hskip2.5ex + \text{n}_\text{ps}  \left[ \frac{d}{dt} \left( [\text{PL}^{\text{as}}] + [\text{PL}^{\text{av}}] \right) \right]  \\
 &\hskip2.5ex - \text{k}_{ \text{ps\_fr:e$_{10}$} }^{+}  \left[ \text{ps\_fr} \right]  \left[ \text{e$_{10}$} \right] + \text{k}_{ \text{ps\_fr:e$_{10}$} }^{-}  \left[ \text{ps\_fr:e$_{10}$} \right] \\
 &\hskip2.5ex - \text{k}_{ \text{ps\_fr:e$_{10}^{\text{m}}$} }^{+}  \left[ \text{ps\_fr} \right]  \left[ \text{e$_{10}^{\text{m}}$} \right] + \text{k}_{ \text{ps\_fr:e$_{10}^{\text{m}}$} }^{-}  \left[ \text{ps\_fr:e$_{10}^{\text{m}}$} \right] &
 \end{flalign*} 

 \begin{flalign*} 
 \frac{d}{dt}\left[ \text{  ps$^{\text{m}}$\_fr  } \right]&= \text{k}_{ \text{ps} }^{\text{on}}  \left[ \text{ps\_fr} \right]  \left[ \text{p}_{ \text{ps} }^{\text{avail}} \right] - \text{k}_{ \text{ps} }^{\text{off}}  \left[ \text{ps$^{\text{m}}$\_fr} \right]\\
 &\hskip2.5ex - \text{k}_{ \text{ps$^{\text{m}}$\_fr:e$_{10}$} }^{+}  \left[ \text{ps$^{\text{m}}$\_fr} \right]  \left[ \text{e$_{10}$} \right] + \text{k}_{ \text{ps$^{\text{m}}$\_fr:e$_{10}$} }^{-}  \left[ \text{ps$^{\text{m}}$\_fr:e$_{10}$} \right] \\
 &\hskip2.5ex - \text{k}_{ \text{ps$^{\text{m}}$\_fr:e$_{10}^{\text{m}}$} }^{+}  \left[ \text{ps$^{\text{m}}$\_fr} \right]  \left[ \text{e$_{10}^{\text{m}}$} \right] + \text{k}_{ \text{ps$^{\text{m}}$\_fr:e$_{10}^{\text{m}}$} }^{-}  \left[ \text{ps$^{\text{m}}$\_fr:e$_{10}^{\text{m}}$} \right] &
 \end{flalign*} 

 \begin{flalign*} 
 \frac{d}{dt}\left[ \text{  ps\_fr:e$_{10}$  } \right]&= \text{k}_{\text{flow}}  ( \text{ [ps\_fr:e$_{10}$]$^{\text{up}}$}  - \left[ \text{ps\_fr:e$_{10}$} \right] ) \\
 &\hskip2.5ex - \text{k}_{ \text{ps} }^{\text{on}}  \left[ \text{ps\_fr:e$_{10}$} \right]  \left[ \text{p}_{ \text{ps} }^{\text{avail}} \right] + \text{k}_{ \text{ps} }^{\text{off}}  \left[ \text{ps$^{\text{m}}$\_fr:e$_{10}$} \right] \\
 &\hskip2.5ex - \text{k}_{ \text{10} }^{\text{on}}  \left[ \text{ps\_fr:e$_{10}$} \right]  \left[ \text{p}_{ \text{10} }^{\text{avail}} \right] + \text{k}_{ \text{10} }^{\text{off}}  \left[ \text{ps\_fr:e$_{10}^{\text{m}}$} \right] \\
 &\hskip2.5ex + \text{k}_{ \text{ps\_fr:e$_{10}$} }^{+}  \left[ \text{ps\_fr} \right]  \left[ \text{e$_{10}$} \right] - \text{k}_{ \text{ps\_fr:e$_{10}$} }^{-}  \left[ \text{ps\_fr:e$_{10}$} \right]&
 \end{flalign*} 

 \begin{flalign*} 
 \frac{d}{dt}\left[ \text{  ps$^{\text{m}}$\_fr:e$_{10}$  } \right]&= \text{k}_{ \text{ps} }^{\text{on}}  \left[ \text{ps\_fr:e$_{10}$} \right]  \left[ \text{p}_{ \text{ps} }^{\text{avail}} \right] - \text{k}_{ \text{ps} }^{\text{off}}  \left[ \text{ps$^{\text{m}}$\_fr:e$_{10}$} \right]\\
 &\hskip2.5ex - \text{k}_{ \text{10} }^{\text{on}}  \left[ \text{ps$^{\text{m}}$\_fr:e$_{10}$} \right]  \left[ \text{p}_{ \text{10} }^{\text{avail}} \right] + \text{k}_{ \text{10} }^{\text{off}}  \left[ \text{ps$^{\text{m}}$\_fr:e$_{10}^{\text{m}}$} \right] \\
 &\hskip2.5ex + \text{k}_{ \text{ps$^{\text{m}}$\_fr:e$_{10}$} }^{+}  \left[ \text{ps$^{\text{m}}$\_fr} \right]  \left[ \text{e$_{10}$} \right] - \text{k}_{ \text{ps$^{\text{m}}$\_fr:e$_{10}$} }^{-}  \left[ \text{ps$^{\text{m}}$\_fr:e$_{10}$} \right]&
 \end{flalign*} 

 \begin{flalign*} 
 \frac{d}{dt}\left[ \text{  ps\_fr:e$_{10}^{\text{m}}$  } \right]&= \text{k}_{ \text{10} }^{\text{on}}  \left[ \text{ps\_fr:e$_{10}$} \right]  \left[ \text{p}_{ \text{10} }^{\text{avail}} \right] - \text{k}_{ \text{10} }^{\text{off}}  \left[ \text{ps\_fr:e$_{10}^{\text{m}}$} \right]\\
 &\hskip2.5ex - \text{k}_{ \text{ps} }^{\text{on}}  \left[ \text{ps\_fr:e$_{10}^{\text{m}}$} \right]  \left[ \text{p}_{ \text{ps} }^{\text{avail}} \right] + \text{k}_{ \text{ps} }^{\text{off}}  \left[ \text{ps$^{\text{m}}$\_fr:e$_{10}^{\text{m}}$} \right] \\
 &\hskip2.5ex + \text{k}_{ \text{ps\_fr:e$_{10}^{\text{m}}$} }^{+}  \left[ \text{ps\_fr} \right]  \left[ \text{e$_{10}^{\text{m}}$} \right] - \text{k}_{ \text{ps\_fr:e$_{10}^{\text{m}}$} }^{-}  \left[ \text{ps\_fr:e$_{10}^{\text{m}}$} \right]&
 \end{flalign*}

 \begin{flalign*} 
 \frac{d}{dt}\left[ \text{  ps$^{\text{m}}$\_fr:e$_{10}^{\text{m}}$  } \right]&= \text{k}_{ \text{ps} }^{\text{on}}  \left[ \text{ps\_fr:e$_{10}^{\text{m}}$} \right]  \left[ \text{p}_{ \text{ps} }^{\text{avail}} \right] - \text{k}_{ \text{ps} }^{\text{off}}  \left[ \text{ps$^{\text{m}}$\_fr:e$_{10}^{\text{m}}$} \right] \\
 &\hskip2.5ex + \text{k}_{ \text{10} }^{\text{on}}  \left[ \text{ps$^{\text{m}}$\_fr:e$_{10}$} \right]  \left[ \text{p}_{ \text{10} }^{\text{avail}} \right] - \text{k}_{ \text{10} }^{\text{off}}  \left[ \text{ps$^{\text{m}}$\_fr:e$_{10}^{\text{m}}$} \right] \\
 &\hskip2.5ex + \text{k}_{ \text{ps$^{\text{m}}$\_fr:e$_{10}^{\text{m}}$} }^{+}  \left[ \text{ps$^{\text{m}}$\_fr} \right]  \left[ \text{e$_{10}^{\text{m}}$} \right] - \text{k}_{ \text{ps$^{\text{m}}$\_fr:e$_{10}^{\text{m}}$} }^{-}  \left[ \text{ps$^{\text{m}}$\_fr:e$_{10}^{\text{m}}$} \right]&
 \end{flalign*}

	\end{document}